\documentclass[nofootinbib,twocolumn]{revtex4}
\usepackage{graphicx,natbib,multirow}
\usepackage{url}
\usepackage{color}
\usepackage{rotating}
\usepackage{ulem}
\usepackage{amssymb,amsmath}

\newcommand{\Mpc}{{\rm Mpc}}

\newcommand{\alm}{a_{\ell m}}
\newcommand{\Ylm}{Y_{\ell m}}

\newcommand{\NoverN}{\frac{\delta N}{\bar N}({\bf \hat n})}
\newcommand{\nhat}{{\bf \hat n}}

\newcommand{\la}{\left \langle}
\newcommand{\ra}{\right \rangle}

\newcommand{\dbest}{d_{\rm best}}
\newcommand{\dbestbold}{{\bf d_{best}}}
\newcommand{\ddotn}{\hat {\bf d} \cdot \hat {\bf n}}

\newcommand{\bcut}{b_{\rm cut}}
\newcommand{\zmin}{z_{\rm min}}
\newcommand{\zmax}{z_{\rm max}}
\newcommand{\zbar}{\bar z}
\newcommand{\zstar}{z_*}
\newcommand{\beff}{b_{\rm eff}}
\newcommand{\Cl}{C_{\ell}}
\newcommand{\Clobs}{C_{\ell}^{\rm obs}}
\newcommand{\Clth}{C_{\ell}^{\rm th}}
\newcommand{\Coneobs}{C_1^{\rm obs}}
\newcommand{\Coneth}{C_1^{\rm th}}

\newcommand{\fsky}{f_{\rm sky}}
\newcommand{\fsources}{f_{\rm sources}}
\newcommand{\blteight}{|b|<8^\circ}
\newcommand{\bltfifteen}{|b|<15^\circ}
\newcommand{\blttwenty}{|b|<20^\circ}
\newcommand{\bltthirty}{|b|<30^\circ}

\newcommand{\sgblt}{|{\rm SGB}|<}
\newcommand{\sgbge}{|{\rm SGB}| \ge}
\newcommand{\sgbgt}{|{\rm SGB}| >}
\newcommand{\declt}{{\rm dec} <}
\newcommand{\decgt}{{\rm dec} >}

\def\mpcinv{{\rm Mpc}^{-1}}

\everymath{\displaystyle}

\begin{document}

\title{Dipoles in the sky}

\author{Cameron Gibelyou}
\email{gibelyou@umich.edu}
\affiliation{Department of Physics, University of Michigan, 
450 Church St, Ann Arbor, MI 48109-1040}

\author{Dragan Huterer}
\email{huterer@umich.edu}
\affiliation{Department of Physics, University of Michigan, 
450 Church St, Ann Arbor, MI 48109-1040}

\begin{abstract}

We perform observational tests of statistical isotropy using data from
large-scale-structure surveys spanning a wide range of wavelengths. Using data
from 2MASS, 2MRS, and NVSS galaxies, and BATSE gamma-ray bursts, we constrain
the amplitude and direction of dipolar modulations in the number count of
sources projected along the line of sight. We pay particular attention to the
treatment of systematic errors and selection effects, and carefully
distinguish between different sources of dipole signal previously considered
in the literature. Dipole signals detected in these surveys are
consistent with the standard, statistically isotropic expectation,
except for the NVSS result, which is likely biased
  by remaining systematics in
  the data. We place constraints on the amplitude of any intrinsic
dipole driven by novel physics in the early universe.

\end{abstract}

\maketitle

\section{Introduction}\label{sec:intro}

The cosmological principle holds that the universe is homogeneous and
isotropic on its largest observable scales. While the cosmological principle
is a crucial ingredient in obtaining many important results in quantitative
cosmology, there is no fundamental reason why our universe \textit{must} obey
it. The most general aim of the research we present here has been to directly test the
cosmological principle using data from recent astrophysical observations.

Observations do bear out that the cosmological principle is a reasonable
approximation for most purposes. However, few rigorous observational tests
have been applied to test homogeneity and isotropy. The work presented here is
directed toward performing tests of the statistical isotropy of the universe
using large-scale structure (LSS). The goal of this work is,
fundamentally, to bring statistical isotropy more fully out of the realm of
assumption and into the realm of observation. Any observed violations of
statistical isotropy could have far-reaching implications for our
understanding of the universe's earliest moments, and violations of isotropy
would also invalidate basic assumptions that serve as prerequisites to typical
methods of data analysis in observational cosmology.

To define statistical isotropy, consider a fluctuating field on the sky
$T(\hat {\bf n})$ (the same arguments will apply for any field, including the
cosmic microwave background (CMB) temperature field, the galaxy density field,
etc.). The field is statistically isotropic if the two-point correlation
function depends \textit{only} on the separation between points:
\begin{equation}
\left \langle T({\bf \hat n}) T({\bf \hat n '}) \right \rangle = C({\bf \hat n} \cdot {\bf \hat n '})
\end{equation} 
while in the case where statistical isotropy is violated, the right-hand side
would read $C(\hat {\bf n}, {\bf \hat n '})$.

Alternatively, we could expand the field in spherical harmonics $T(\hat {\bf
  n}) \equiv \sum_{\ell =0}^\infty T_{\ell} \equiv \sum_{\ell=0}^\infty
\sum_{m= - \ell}^{\ell} \alm\Ylm(\nhat)$, where the $a_{\ell m}$ are the
coefficients of the expansion. If statistical isotropy is assumed, the angular
power spectrum $C_{\ell}$ may be defined via $\left \langle a_{\ell m}^*
a_{\ell ' m '} \right \rangle = C_{\ell} \delta_{\ell \ell '} \delta_{m m '}$,
where $C_{\ell}$ does not depend on the rotational degrees of freedom
$m$. Hence, it is meaningful to calculate $C_{\ell}$ by averaging over the $(2
\ell + 1)$ (on a full sky) ``samples" corresponding to the $(2 \ell + 1)$
values of $m$ for each $\ell$. However, without the assumption of statistical
isotropy, we have in general
\begin{equation}
\left \langle a_{\ell m}^* a_{\ell ' m '} \right \rangle = \tilde C_{\ell m \ell ' m '},
\end{equation}
which is much more difficult to measure than $C_{\ell}$ since we get only one
sample of $\tilde C_{\ell m \ell ' m '}$ for each $(\ell, m, \ell ', m ')$ in
our universe. This is a big reason why statistical isotropy is such a crucial
assumption in cosmology: it is much easier to work with $\Cl$ than the much
more complicated quantity above.

\subsection{Point of Departure: Large-Scale Structure as Probe of CMB Anomalies}
\label{sec2.1.1}

While measurements of the angular power spectrum of the CMB by the Wilkinson
Microwave Anisotropy Probe (WMAP) experiment have provided strong support for
the inflationary Hot Big Bang model and allowed for unprecedentedly precise
determination of cosmological parameters, these successes have come along with
various unexpected ``anomalies" in the data (for reviews, see
  \citet{Copi_AdvAstro} and \citet{wmap7_anomalies}).

There are already tantalizing hints of violations of statistical
isotropy in CMB data; in fact, several observed anomalies, especially those
indicating correlations between patterns in the CMB of (apparently)
cosmological origin (e.g., the quadrupole and octopole) and the geometry and
motion of the Solar System, can be understood as indications that statistical
isotropy fails to hold. In addition, the southern ecliptic hemisphere has
significantly more power than the northern ecliptic hemisphere on scales of
about 3 degrees and larger (multipoles $\lesssim 60$) (\citet{Hoftuft}). This
so-called ``hemispherical power anomaly" is a dipole modulation (see the
beginning of Sec.~\ref{sec3.1} for precise discussion of what this entails
mathematically) of CMB power (direction in Galactic coordinates is
$(l,b)=(224^\circ,-22^\circ)$) that is completely distinct from the measured
CMB dipole $C_1$ (which is due to our motion with respect to the CMB rest
frame), and is yet another possible indication of breaking of statistical
isotropy in the CMB\footnote{However, see \citet{hanson2009estimators} for the
updated result that the significance of the hemispherical power anomaly
decreases when smaller scales are taken into account: previous analyses had
been done for $\ell \lesssim 60$; when analysis is extended out to $\ell \sim
500$, the effect becomes less than $3\sigma$-anomalous. The fact that greater
resolution reduces the significance of the signal calls into question the
authenticity of the signal as a genuine cosmological effect.}. Finally, the low
power at large angles in $C(\theta)$ may itself be an indication that
statistical isotropy is violated: it seems that a conspiracy of low-$\ell$
multipoles in the angular power spectrum $C_{\ell}$ is responsible for
creating the suppressed $C(\theta)$ \cite{wmap12345}, and correlations
between different multipoles could be the result of a lack of statistical
isotropy.

\subsection{Goals of This Work}
\label{sec2.1.2}

We take these considerations as general motivation for the work performed in
the rest of this paper, which concerns testing statistical isotropy with
large-scale structure, and will analyze the specific issue of dipole patterns
in various surveys being used as probes of the statistical isotropy of the
universe. This work also complements the (surprisingly few!) studies
  of tests of statistical isotropy with current or future LSS observations
  \cite{ellis1984expected,Hirata_hemispherical,Pullen_Hirata,Gibelyou,Zunckel}. 

The paper is organized as follows.  In Sec.~\ref{sec2.2}, we point out the
various effects that are expected to contribute to a dipole signal in a
large-scale-structure survey. In Sec.~\ref{sec3.1}, we outline the formalism
used in this paper to detect such dipole signals. We then apply that formalism
to several surveys: the 2MASS Redshift Survey (Sec.~\ref{sec2mrs}), the Two
Micron All-Sky Survey (2MASS) as a whole (Sec.~\ref{sec4.3}), the Burst And
Transient Source Experiment (BATSE) of the Compton Gamma-Ray Observatory
(Sec.~\ref{sec5.2}), and the NRAO VLA Sky Survey (NVSS; Sec.~\ref{sec5.3}). In
Sec.~\ref{sec5.1}, we also examine issues surrounding searching for dipole
signals in high-redshift objects and review work that has searched for dipoles
in X-ray surveys. We conclude in Sec.~\ref{secconclusion}, summarizing our
results in Table \ref{tab6.1} in that section.

\section{Types of Dipoles: Specific Theoretical Considerations}
\label{sec2.2}

It is completely expected that a dipole will be present in any survey of
objects that trace large-scale structure. Both of the following effects
contribute to the dipole: (a) there are local anisotropies since the universe
is not homogeneous and isotropic except on its very largest scales, and (b)
the Earth has a total motion relative to the large-scale-structure rest frame
that is the sum of several vector contributions (Earth moves around the Sun,
the Sun moves around the center of the Milky Way, the Milky Way moves with
respect to the Local Group barycenter, and the Local Group barycenter moves
with respect to the structure around it and, ultimately, the
large-scale-structure rest frame). That motion produces dipole anisotropy due
to two effects, the Doppler effect and relativistic aberration of angles
(\citet{itoh2010dipole}; see Sec.~\ref{sec2.2.3} for mathematical
details). The Doppler effect is relevant because it changes how magnitude
varies with frequency, and since LSS surveys invariably operate within limited
frequency ranges, the Doppler effect may shift certain objects into or out of
a magnitude-limited sample. Since frequencies will increase in the direction
of motion and decrease in the opposite direction, this produces a small dipole
in the number of objects detected. Meanwhile, relativistic aberration causes
the measured positions of galaxies to be displaced toward our direction of
motion. This effect is on the order of $v/c \sim 10^{-3}$, relevant for our
purposes.

We expect that as we go from smaller to larger survey volumes, the measured
value of the dipole amplitude should converge to that of the CMB dipole. This
is because with larger survey volumes, we average over more and more
structure, and the universe approaches homogeneity and isotropy. Any dipole
left over should be due only to our motion, a kinematic dipole (with amplitude
on the order of $10^{-3}$) just as in the CMB. There are several reasons why
the dipole might {\it not} converge to that of the CMB:
\begin{enumerate}
\item the rest frame of the CMB may not be the same as the rest frame of the
  LSS: novel horizon-scale physics (explored below) could induce a relative
  velocity between the CMB and LSS, so that galaxies would have a nonzero
  average streaming velocity with respect to the CMB rest frame;
\item there is also the possibility that there is genuinely more mass (and
  therefore more galaxies/objects that trace the mass distribution) in one
  direction, corresponding to modulation of primordial curvature perturbations
  due to the physics of inflation. For example, isocurvature perturbations can
  produce such an effect (and explain the CMB hemispherical power anomaly)
  \cite{Erickcek_iso,Erickcek_hem,langlois1996entropy,langlois1996double}. 
\end{enumerate}
Careful measurement of dipoles in various surveys, such as those we perform
here, help zone in on these possibilities, which correspond to a violation of
statistical isotropy.

The rest of this section will flesh  out the details of the discussion in the
preceding two paragraphs.

\subsection{Flux-Weighted Dipole vs.\ 2D-Projected Dipole}
\label{sec2.2.1}

One very commonly computed type of dipole is not, strictly speaking, a dipole
at all, but is frequently referred to as such. This is the flux-weighted
``dipole," where instead of calculating a (genuine) dipole based only on the
two-dimensional projected positions of objects on the sky, some radial
information is preserved by weighting each object by the flux we receive from
it. We follow \citet{bilicki2011two} in the following explanation of how the
flux-weighted dipole is calculated.

%
%
%

The flux-weighted dipole, as typically computed, is a measure of the
acceleration due to gravity on the Local Group. In linear theory, the peculiar
velocity \textbf{v(r)} at position \textbf{r} is proportional to the peculiar
acceleration vector \textbf{g(r)} induced by the matter distribution around
position \textbf{r} \citep{erdogdu2006dipole,bilicki2011two}:
\begin{equation}\label{eq:v_and_g}
{\bf v(r)}=\frac{H_0 f(\Omega_M)}{4 \pi G \bar \rho}\;{\bf g(r)} =
\frac{2 f(\Omega_M)}{3 H_0 \Omega_M}\;{\bf g(r)}\;.  
\end{equation}
Here $H_0=100\,h\,{\rm km/s} / \mathrm{Mpc}$ is the Hubble constant,
$\Omega_M$ is the matter density divided by the critical density, and
$f(\Omega_M)\equiv(d\ln D/ d\ln z)\vert_{z=0}$ (where $D$ is
the growth factor). In the $\Lambda$CDM model, the factor $f(\Omega_M) \approx
\Omega_M^{0.55}$ and is only weakly dependent on the cosmological constant
(\citet{lahav1991dynamical}). The acceleration vector itself is given by
\begin{equation}\label{eq:g_acc}
{\bf g}({\bf r})=G \bar \rho \int \delta_M({\bf r}')\,
\frac{{\bf r} '-{\bf r} }{|{\bf r} '-{\bf r} |^3}\,d^3{\bf r}'\;,   
\end{equation}
where $\delta_M({\bf r})=\left[\rho_M({\bf r})-\bar \rho\right]/ \bar \rho$ is
the density contrast of the mass perturbations at the point ${\bf r}$; note
that $\rho_g$ is the mass density of galaxies, and $b$ is the bias factor that
relates mass density of galaxies to that of matter, $b\equiv \rho_g/\rho_M$
(assuming constant, scale- and time-independent bias). The bias is usually
packaged with the factor $f(\Omega_M)$ into the parameter $\beta\equiv
f(\Omega_M)\slash b$. Comparing Eqs.\ (\ref{eq:v_and_g}) and (\ref{eq:g_acc}),
we get the proportionality valid in linear theory:
\begin{equation}
{\bf v} \propto  \beta\, {\bf g}\;.
\label{eq:beta}
\end{equation}
Comparison of the peculiar velocity ${\bf v}$ (determined from either the LSS
surveys or using the CMB dipole) and acceleration ${\bf g}$ of the Local Group
serves as a tool to estimate the $\beta$ parameter. Independent knowledge of
biasing allows one to estimate the cosmological matter density $\Omega_M$. The
programme of measuring the matter density in this way has been ongoing for over
three decades
(e.g.\ \cite{Yahil_1980,Davis_Huchra_1982,Yahil_1986,Lahav_Lynden_Rowan_1987,Hudson_1993,lauer1994motion,Lavaux,
  nusser2011cosmological, davis2011local, nusser2011bulk}).

The challenging part in this procedure is evaluating the acceleration in
Eq.~(\ref{eq:g_acc}). This is where the flux-weighted-dipole approach comes
in, where position vectors of the objects in a survey are weighted according
to their fluxes (used as a rough proxy for mass, since both gravity and flux
go as $1/r^2$), and added together. By preserving some radial information in
this way, the flux-weighted dipole allows one to obtain a measure of the
direction and strength of the acceleration of the Local Group due to the
Newtonian gravitational attraction from objects in the survey. The
flux-weighted dipole from 2MASS has previously been found to be in the
direction $(l,b)=(264.5^\circ,43.5^\circ) \pm (2.0^\circ, 4.0^\circ)$
(\citet{maller2003clustering}; note the rather serious discrepancy between the
published result and the arXiv version, and see \citet{bilicki2011two} for
detailed discussion of convergence of the flux-weighted dipole and possible
shortcomings of Maller et al.); the flux-weighted dipole from 2MRS is in the
direction $(l,b)=(251^\circ,38^\circ)$ (\citet{erdogdu2006dipole}) in the
Local Group frame, or $(245^\circ,39^\circ)$ in the CMB frame.

Just to be clear, we do {\it not} consider further the flux-weighted
  dipole in this work; we henceforth study the 2D-projected dipole described
  below.

\subsection{2D-Projected Dipole: Local-Structure Dipole}
\label{sec2.2.2}

For the rest of this section, we focus on what we term the 2D-projected
dipole, which is the quantity that is usually indicated by the isolated use of
the word ``dipole." This quantity relies on objects at any given redshift
being projected on the celestial sphere (hence ``2D-projected") with no
weighting scheme.

For a survey with very large (hundreds of Mpc- to Gpc-scale) volume, the
universe is at least close to homogeneous and isotropic on the scales relevant
for the survey. We naturally expect that any dipole signal in such a
large-volume survey will be strongly suppressed. However, on much smaller
scales, where the universe is not at all homogeneous and isotropic, dipole
signals should naturally emerge in any survey of objects that trace
large-scale structure at all, and certainly in any galaxy survey. To take a
particularly simple example, there is a large dipole in the galaxy
distribution if we survey only objects within the Local Group.

But even if a survey encompasses structure on scales of tens of Mpc, we fully
expect that given the non-uniformity of nearby structure, there should be a
dipole component in the pattern of galaxies observed on the sky. This dipole
component will be strongest for the smallest surveys, and should die away
monotonically (at least in statistical average) for larger and larger
surveys. The effect turns out to be on the order of $10^{-1}$ for scales of
tens to a couple hundred of Mpc, meaning that the fluctuations (contributing
to the dipole) in the number of galaxies, as a function of position on the
sky, are on the order of 1/10 the size of the mean number of galaxies across
the sky.

More rigorously, we make predictions for the full angular power spectrum
$C_{\ell}$ of large-scale structure as a function of maximum redshift of a
survey. The angular power spectrum of density fluctuations of halos is usually
expressed within the Limber approximation, where the contribution of modes
parallel to the line of sight is ignored. In this approximation, the angular
power spectrum is given by
\begin{equation}
C_\ell  =  {2\pi^2\over \ell^3}\int_0^\infty dz \, {W^2(z)\over r(z)^2
  H(z)} \Delta^2\left (k={\ell\over r(z)}, z\right ),
\end{equation}
where $\Delta^2(k)\equiv k^3P(k)/(2\pi^2)$ is the dimensionless power
spectrum, $r(z)$ is the comoving distance, and $H(z)$ is the Hubble
parameter. The weight $W(z)$ is given by 
\begin{equation}
W(z)=\frac{\displaystyle b(z)\, {dN\over dz}}{\displaystyle\int_{\zmin}^{\zmax} 
\left ({dN\over dz}\right ) dz}, \label{eq:Wz}
\end{equation}
where $\zmin$ and $\zmax$ are the lower and upper end of the redshift range,
and $dN/dz (z)$ is the number of galaxies per unit redshift.\footnote{Note
  that a sometimes-used alternative definition of $n(z)$ refers to the spatial
  density of galaxies (e.g.~\citet{hu2004joint}); it is related to the quantity we use via $dN/dz =
    n(z)\,\Omega\,r^2(z)/H(z)$, where $\Omega$ is the solid
    angle spanned by the survey, and $r$ and $H$ are the comoving distance and
    Hubble parameter,
    respectively. Note also that our $W(z)$ is equivalent to the quantity
  $f(z)$ from \citet{ho2008correlation}.} We adopt the tabulated values, or
else functional form, of $N(z)$ directly from the respective surveys that we
study.

\begin{figure}[]
\begin{center}
\includegraphics[width=.45\textwidth]{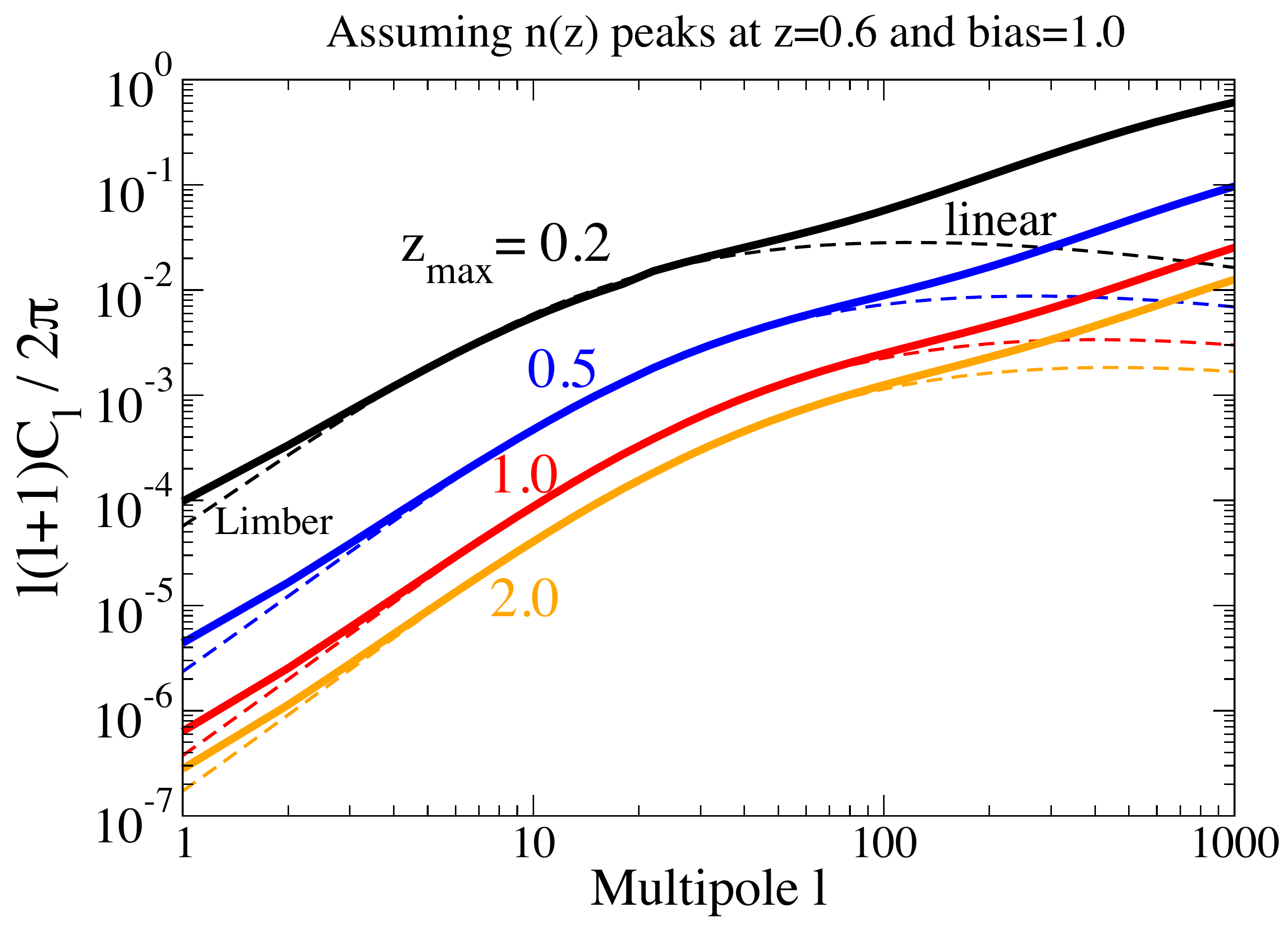}
\caption{A plot of the angular power spectrum $\Cl$ predicted for a galaxy
  survey with a peak in the galaxy redshift distribution at $z=0.6$ and the
  given maximum depth $\zmax$. In the remainder of this paper we will focus
  attention on the dipole, $\ell=1$. For this particular redshift
  distribution, the local-structure dipole becomes subdominant to the
  kinematic dipole for around $\zmax \sim 1.0$. For higher $\zmax$, we should
  get convergence to the kinematic dipole plus any intrinsic dipole that might
  be present. The dotted curves correspond to the power spectrum within the
  Limber approximation (low $\ell$) and assuming linearity (high $\ell$); the
  solid curves correspond to the more accurate set of assumptions where the
  Limber approximation is relaxed and nonlinearity is taken into account.}
\label{fig:Clpredictions}
\end{center}
\end{figure}

However, we are interested in the dipole $\ell=1$ where the Limber
approximation is not accurate anymore (it is accurate at $\ell \gtrsim 10$); see
Fig.~\ref{fig:Clpredictions}. Therefore, we adopt the exact expression for the
power spectrum; using notation from (e.g.)~\citet{hearin2011testing}, this is
\begin{eqnarray}
\label{eq:Cl_noLimber}
C_\ell &=& 4\pi\int_0^\infty d\ln k\, \Delta^2\left (k, z=0\right )I^2(k)\\[0.2cm]
I(k) &\equiv & \int_0^\infty dz\, W(z) D(z)\, j_\ell(k\chi(z))
\label{eq:Ik}
\end{eqnarray}
where $\chi(r)$ is the radial distance, and $\chi(z)=r(z)$ in a flat universe,
the case that we consider. Here $D(z)$ is the linear growth function of
density fluctuations, so that $\delta(z)=D(z)\delta(0)$, where $D(0)=1$. Note
that, over the shallow range for 2MASS we can assume that $b(z)$ is constant,
and factor it outside of Eqs.~(\ref{eq:Cl_noLimber}) and (\ref{eq:Ik}), but
over the much deeper range for NVSS the bias may vary with redshift, and we
adopt the expression for $W(z)$ from \citet{ho2008correlation} that implicitly
integrates bias and number density as per Eq.~(\ref{eq:Wz}). This is explained
in detail in Sec.~\ref{sec5.3.2}.

To produce the fiducial theoretical predictions, we consider the standard
cosmological model with the following parameter values: matter density
relative to critical $\Omega_M=0.25$, equation of state parameter $w=-1$,
spectral index $n=0.96$, and amplitude of the matter power spectrum $\ln A$
where $A=2.3\times 10^{-9}$ (corresponding to $\sigma_8=0.8$) defined at scale
$k=0.002~\mpcinv$.  The power spectrum $\Delta^2(k, z)\equiv k^3
P(k)/(2\pi^2)$ is calculated using the transfer function output by CAMB. We do
not vary the values of cosmological parameters, since they are measured to
sufficient accuracy that any shifts in predicted dipole amplitude that could
occur due to realistic changes in cosmological parameters are tiny in
comparison with cosmic variance given the finite sky coverage and relative
shallowness of the surveys we employ (as we have explicitly verified).

The power spectra are shown in Fig.~\ref{fig:Clpredictions}. Note the
substantial, order-unity, difference between the exact and approximate
(Limber) expressions at $\ell=1$; in the remainder of this work we use the
exact, double-integral expression.

\begin{table*}[t]
\caption{Motions that give rise to a kinematic dipole in the CMB and
  large-scale structure.}
\label{tab2.1}
\centering
\begin{tabular}{| c | c | c |}
\hline
\rule[-2mm]{0mm}{6mm} Motion & Approximate Speed (km/s) & Direction \\
\hline

\rule[-2mm]{0mm}{6mm} Earth around Sun & $\sim 30$ km/s & annually varying \\ \hline
\rule[-2mm]{0mm}{6mm} Sun wrt Local Group & $\sim 306$ km/s & $(l,b) = (99,-4) \pm (5,4)$ \\ \hline
\rule[-2mm]{0mm}{6mm} Local Group wrt CMB & $\sim 622$ km/s & $(l,b) = (272,28)$ \\ \hline
\rule[-2mm]{0mm}{6mm} Overall CMB kinematic dipole & $\sim 370$ km/s & $(l,b) = (264.4,48.4) \pm (0.3, 0.5)$ \\ \hline

\end{tabular}
\end{table*}

\subsection{2D-Projected Dipole: Kinematic Dipole}
\label{sec2.2.3}

A dipole pattern may also arise due to motion of the Earth with respect to the
astrophysical objects or structure being measured. This is what produces the
dipole in the cosmic microwave background, and it also contributes to the
total dipole in a measurement of the large-scale structure.

\subsubsection{Kinematic Dipole in the CMB}
\label{sec2.2.3.1}

Probably the best-known dipole in all of cosmology is the dipole measured in
the CMB temperature distribution. This dipole, which has an amplitude on the
order of $10^{-3}$ times the amplitude of the CMB monopole, arises due to the
motion of the Solar System with respect to the CMB rest frame. This motion is
the vector sum of several different motions, summarized in Table \ref{tab2.1}.

Values of the kinematic dipole in the CMB are cited with the contribution from
the Earth's motion around the Sun subtracted out, so that the dipole is due
only to the Sun's velocity with respect to the CMB
(\citet{kogut1993dipole}). The value of the Local Group's peculiar velocity
with respect to the CMB is from \citet{maller2003clustering} and was computed
using the value of the Sun's velocity with respect to the Local Group in
\citet{courteau1999solar}. (When the velocity of the Local Group with respect
to the CMB rest frame is inferred from the measurement of the CMB dipole, the
direction becomes $(l,b)=(276,30) \pm (3,3)$ \cite{kogut1993dipole}. Compare
also values determined in \citet{tully2008our} and \citet{bilicki2011two}.)
The peculiar velocity predicted from linear-theory $\Lambda$CDM is $\sim 470$
km/s \citep{bilicki2012motion}. Note that the speed of the Sun with respect to
the CMB rest frame would be considerably greater if not for the fact that the
Sun's velocity vector with respect to the Local Group points in a direction
nearly opposite that of the Local Group's velocity vector with respect to the
CMB. Also note that the dominant contribution to the Sun's motion with respect
to the Local Group is the Sun's motion around the center of the Galaxy, which
has speed $\sim 220$ km/s, and is composed of the Sun's motion with respect to
the Local Standard of Rest and the LSR's motion with respect to the Galactic
Center \cite{itoh2010dipole,courteau1999solar}.

\subsubsection{Kinematic Dipole in LSS}
\label{sec2.2.3.2}

The kinematic dipole in the CMB, which is due to the Sun's motion with respect
to the CMB rest frame, is observed as a Doppler shifting of the CMB
photons. The effect that gives rise to a kinematic dipole in the large-scale
structure is not quite as direct. Rather, it includes contributions both from
the Doppler effect and relativistic aberration. We derive the relevant
equations in Appendix \ref{appB}.

\subsection{2D-Projected Dipole: Intrinsic Dipole}
\label{sec2.2.4}

In the CMB, the intrinsic dipole corresponding to adiabatic perturbations is
zero (\citet{Erickcek_iso}). When we switch over from talking about the CMB to
talking about large-scale structure, we expect that there may be an intrinsic
dipole in the LSS. Below, we explore possible reasons why an intrinsic dipole
might compete with or even (conceivably) dominate the LSS kinematic dipole.

\citet{Erickcek_iso} propose a scenario in which the curvaton (particle
mediating a scalar field that may generate fluctuations during inflation
without actually driving inflation) has a large-scale spatial gradient, which
in turn causes variation in the amplitude of the primordial curvature
perturbations, modulating $\Delta_R$ across the
sky. \citet{Hirata_hemispherical} shows how this modulation due to
isocurvature perturbations would transfer to the CMB and large-scale
structure, in the latter case causing a dipolar variation in the abundance of
massive haloes (and objects that occupy them). This inflationary scenario is
one scenario that invokes the physics of the early universe to explain why
there might be an intrinsic dipole in the large-scale structure above and
beyond what we naturally expect to be present from typical scale-invariant
fluctuations/adiabatic perturbations laid down in the simplest inflationary
scenarios. While the simplest curvaton-gradient model has been ruled out by
Hirata's analysis of constraints on the dipole in SDSS quasars, and
corresponding constraints on dipolar modulation of the primordial power
spectrum,\footnote{According to Hirata, any smooth gradient in the amplitude
  of the primordial curvature perturbations is no more than 2.7 percent per
  present-day horizon radius (99 percent confidence); cf.~the 11 percent
  variation required in the Erickcek et al.~model needed for consistency with
  the CMB hemispherical power anomaly.} similar but more complicated scenarios
are still possible.

Note that Hirata's constraints on the primordial dipole amplitude using SDSS
quasars are on the order of $2 \times 10^{-2}$, which corresponds to
constraints on the amplitude of the dipole in the quasars themselves roughly
an order of magnitude higher. Hence current constraints on this particular
intrinsic-dipole scenario are not down to the level associated with the
kinematic dipole, though this was not a problem in Hirata's analysis since
that analysis looked specifically for a dipole effect that would accompany the
primordial conditions needed to explain the CMB hemispherical power anomaly
given the curvaton-gradient model, and that would have required a $10^{-1}$
dipole.

Another possibility for generating an intrinsic dipole is that the CMB rest
frame is not the same as the large-scale-structure rest frame. This happens,
for example, in models with a large ($\sim$ Gpc radius) underdense void, in
which we are located close to the center. The observed bulk flow is then equal
to the difference between the Hubble parameters inside and outside the void
multiplied by our distance away from the center\footnote{Therefore,
    the CMB frame and the LSS frame {\it inside the void} (but not the global
    LSS frame) are different in this case. Most measurements of bulk flows,
    being much shallower than the size of the void, would interpret this as
    a legitimate difference between the CMB and LSS frames.}  \cite{Tomita:1999rw}.
Requiring that the intrinsic dipole thus measured is consistent with
observations of the CMB dipole ($v/c\simeq O(10^{-3})$) requires that we live
very close ($\lesssim 15\,\Mpc$) to the center of the void
\cite{Alnes:2006pf}, making these models rather fine-tuned.

An alternate mechanism for how a CMB-LSS rest-frame disagreement may arise is
provided by the Grishchuk-Zel'dovich effect (\citet{grishchuk1978long};
\citet{erickcek2008superhorizon}; see also \citet{gunn1988deviations}). As
\citet{turner1991tilted} explains, if inflation lasts only a little longer
than necessary to solve the flatness and horizon problems, scales that were
superhorizon-sized at the onset of inflation -- these scales cannot be
affected by events during or later than the inflationary epoch, and thus
contain imprints of the pre-inflationary universe -- may not be much larger
than our present horizon, and thus may have some effect in the current
universe. In particular, he proposes that large density fluctuations with
wavelengths slightly larger than the Hubble radius (modes that are ``just
barely" superhorizon-sized) may exist, and would appear to us as a density
gradient in a particular direction. Such a density gradient could produce a
``tilted universe": a universe in which all the matter within the Hubble
volume gains a peculiar velocity due to the greater gravitational attraction
from one ``side" of the universe than the opposite. The effect is equivalent
to saying that the rest frame of the CMB is not the same as the rest frame of
large-scale structure: from the rest frame of the CMB, all matter would have a
nonzero average streaming velocity. This would produce an intrinsic dipole due
to relativistic aberration and the Doppler effect (or, equivalently, it would
produce an additional kinematic dipole on top of that expected from analysis
of the CMB) \cite{itoh2010dipole}. In order for this ``tilting" effect to be
observable, isocurvature modes must be present, even in the presence of
late-time acceleration \citep{zibin2008gauging}.

Given the presence of isocurvature modes, the Grishchuk-Zel'dovich effect
would also produce an (additional) intrinsic dipole due to the simple fact of
the superhorizon-scale density perturbation. As another example of physical
mechanisms that would produce an intrinsic dipole, \citet{Gordon2005} examine
a scenario in which there are spatial perturbations in the density of dark
energy from a quintessence field: that is, a late-time effect produces
horizon-scale fluctuations. More generally, Gordon et al.~examine a class of
models in which the full fundamental theory is homogeneous and statistically
isotropic, but statistical isotropy is broken from a given observer's position
because of superhorizon-scale perturbations that appear as a gradient in
density across the sky on the largest observable scales.  Any theory that
generates such a variation in density would give rise to what we have termed
an intrinsic large-scale-structure dipole, and the appearance of the breaking
of statistical isotropy. These density variations could, at least
theoretically, exist on essentially any order of magnitude in $\delta
\rho/\rho$.

There is some reason to take the idea of a tilted universe
seriously. \citet{kashlinsky2008measurement} investigate the bulk motion of
galaxies in the universe out $\sim300$ Mpc/$h$ and find, somewhat
controversially (see, e.g.,~\cite{keisler2009statistical,Osborne,Mody}), that
there is a coherent bulk flow in their sample. The evidence they develop for
this claim comes from attempting to detect the kinetic Sunyaev-Zel'dovich
effect by computing the dipole of the CMB temperature field evaluated at the
positions of galaxy clusters. This dipole, evaluated as it is in a small
number of pixels, does not receive appreciable contributions from our own
motion (i.e., from the CMB kinematic dipole due to the Sun's motion with
respect to the CMB rest frame), but does receive contributions from instrument
noise, the thermal SZ effect, the \textit{intrinsic} CMB dipole, and
foreground components. However, contributions other than the kinematic SZ
effect are, they argue, accounted for in their analysis, with the thermal SZ
effect in particular canceling out/integrating down when averaged over a large
number of clusters. Their conclusion is that the dipole in CMB temperature
evaluated at cluster positions is due to the kinetic SZ effect due to the bulk
flow of the cluster sample. If this effect is authentic, then it fits well
with the tilted-universe scenario: the bulk motion is detectable in
large-scale structure but does not generate a primordial dipole CMB component.

Other studies that use measurements of peculiar velocities of local
neighborhood galaxies --- typically determined by combining the measurements
of their distances and redshifts --- find larger-than expected bulk flows
\cite{Watkins_2009,Feld_Watk_Hudson_2010}. These flows are of order $400$ km/s
and seem to be showing no signs of convergence out to the probed distance
$R\sim 60~h\mpcinv$; they are estimated to be $\sim 1$\% likely in the
standard $\Lambda$CDM cosmological model for the given observed volume. The
relation of these larger-than-expected bulk-flow measurements to findings by
\citet{kashlinsky2008measurement} is unclear at this time
\cite{Feld_Watk_Hudson_2010}, especially given that bulk flows
  inferred from distances obtained from type Ia supernovae indicate somewhat
  lower bulk flows that are therefore in better agreement with $\Lambda$CDM
  \cite{Colin,Dai,Turnbull}.
  

\begin{table*}[]
  \renewcommand\multirowsetup{\centering} 
\begin{normalsize}
\caption{Table reviewing the sources of dipole signal in a
  large-scale-structure survey. }
\label{orgtable}
\centering
\begin{tabular}{| c | c | c | c |}
\hline\hline 
\rule[-3mm]{0mm}{8mm} & Local-Structure Dipole & \quad Kinematic Dipole \quad
& Intrinsic Dipole \\
\hline\hline 

\rule[-3mm]{0mm}{8mm} Typical size & $\sim 0.1$--$10^{-5}$ depending on $\zmax$ 
& $\sim 10^{-3}$ & $\gtrsim 10^{-5}$ \\\hline \hline 
\rule[-2mm]{0mm}{8mm} Flux-Weighted & Local Group's
& \multirow{2}{100pt}{N/A} & \multirow{2}{100pt}{N/A} \\[-0.1cm] 
\rule[-3mm]{0mm}{6mm} \quad method probes: \quad  & acceleration; Sec.~\ref{sec2.2.1} & &  \\\hline
\rule[-2mm]{0mm}{8mm} 2D-Projected & small-scale departures from
& our motion; & \quad many possible theoretical \quad  \\[-0.1cm] 
\rule[-3mm]{0mm}{6mm} method probes: & \quad statistical isotropy;
Sec.~\ref{sec2.2.2} \quad &  Sec.~\ref{sec2.2.3} & explanations; Sec.~\ref{sec2.2.4} 
\\ \hline
\hline
\end{tabular}
\end{normalsize}
\end{table*}

\subsection{Types of Dipoles: Review}
\label{sec2.2.5}

In summary, when we observe some tracer of large-scale structure (galaxies,
quasars, gamma-ray bursts, etc.), we may observe a dipole in counts. If the
dipole we are observing is what we have called the 2D-projected dipole -- that
is, a dipole in surface density of the object -- then contributions to that
dipole may come from (1) the local-structure dipole, (2) the kinematic dipole
(which is due to the Doppler effect and relativistic aberration), and (3) an
intrinsic LSS dipole (which is really the $z\ggg 1$ limit of the
  local-structure dipole). There are a couple observations to make about these
effects:

\begin{itemize}
\item In the limit of very high redshift, the local-structure dipole goes to
  amplitudes on the order of $10^{-5}$ (as we have explicitly verified), the
  kinematic dipole is expected to go to amplitude $\sim v/c \sim 10^{-3}$ and
  align with the direction of the CMB dipole, and the intrinsic dipole could
  take on a wide variety of values depending on certain theoretical
  considerations.
\item For structures/galaxies at relatively small redshifts, the
  local-structure dipole amplitude $>>$ the kinematic dipole
  amplitude. However, even though the kinematic dipole is swamped by the
  local-structure dipole, we expect that these two dipoles should point in
  somewhere close to the same direction. While no particular level of
  agreement is guaranteed, the fact remains that local structure is what
  accelerates us in the direction that the kinematic dipole points. This is
  why, in linear theory, the velocity of the Local Group is proportional to
  its acceleration due to gravity. However, since the 2D-projected dipole
  takes no radial information into account, it is not a true measure of
  gravitational attraction or acceleration, but only a partially reliable
  proxy.
\end{itemize}

In Table \ref{orgtable}, we show the types of dipoles considered in
  this work, and the section in which they are introduced.

\section{Formalism for Detecting Dipoles}
\label{sec3.1}

Some forays have already been made into tests of statistical isotropy, and
dipoles in particular, using measurements of large-scale structure. Many
estimators for the dipole have been employed, some of which do better jobs
than others at naturally incorporating sky cuts, allowing for systematic
effects to be accounted for, etc. Here we adopt the estimator used by
\citet{Hirata_hemispherical} to test the WMAP hemispherical power anomaly
using quasars detected by the Sloan Digital Sky Survey. This is
the best unbiased estimator for determining the amplitude
and direction of a dipole in counts of objects on the sky
under conditions of cut skies and in the presence of systematics. We now
describe this formalism.

\subsection{Obtaining the Direction of the Dipole}

Consider a dipolar modulation on the sky with some amplitude $A$ in a (unit)
direction $\hat {\bf d}$. We may write the observed density field $N$ of the
objects in question as a function of direction $\hat {\bf n}$ as
\begin{equation}
N(\hat {\bf n}) = [1+A(\hat {\bf d} \cdot \hat {\bf n})] \bar N + \epsilon(\hat {\bf n})
\label{eqn:introdipole}
\end{equation}
where $\bar N$ is the intrinsic statistically isotropic field and $\epsilon$
combines random and instrumental noise. If we momentarily drop the $\epsilon$
term, we can write
%
$\delta N/\bar N = A(\ddotn)$.
%
Reinstating a term corresponding to systematic errors, the fluctuations in
density as a function of direction can be written as the sum of contributions
from a dipole, fluctuations due to systematics, and a mean offset
\cite{Hirata_hemispherical}
\begin{equation}
\NoverN = A {\bf \hat d} \cdot {\bf \hat n} + \sum_i k_i t_i({\bf \hat n}) + C.
\label{eqn:masterequation}
\end{equation}
Here $t_i({\bf \hat n})$ are possible systematics templates in the sky map
(such as an extinction map), the coefficients $k_i$ give the amplitudes of the
contributions of these systematics to the observed density field, and the
presence of the monopole term, $C$, allows us to account for covariance
between the monopole and other estimated parameters, especially covariance
between the monopole and any systematic templates.

It is then straightforward to write down the best linear unbiased estimator of
the combination ({\bf d}, $k_i$, $C$) with corresponding errors. The procedure
is as follows: First, we rewrite the above equation as
\begin{equation}
\NoverN = {\bf x} \cdot {\bf T(\hat n)}
\end{equation}
where ${\bf x} = (d_x, d_y, d_z, k_1, ..., k_N, C)$, ${\bf T(\hat n)} = (n_x, n_y, n_z, t_1(\hat n), ..., t_N(\hat n), 1)$, and $n_x^2 + n_y^2 + n_z^2 = 1$.

The best linear unbiased estimator of {\bf x} is
\begin{equation}
{\bf \hat x} = F^{-1} g
\end{equation}
where the components of the vector $g$ are
\begin{equation}
g_i = \int T_i(\nhat) \delta N^{\Omega}(\nhat) d^2 \nhat
\label{eq:geq}
\end{equation}
and the Fisher matrix $F$ is given by
\begin{equation}
F_{ij} = \bar N^{\Omega} \int T_i(\nhat) T_j(\nhat) d^2 \nhat,
\end{equation}
where $N^{\Omega} \equiv dN/d\Omega$ is the number of galaxies per
steradian. To actually compute these quantities with discretized data, it is
convenient to work with a data map and a random map, the latter of which is
simply a set of randomly chosen directions/points $\nhat_R$ on the unit
sphere:
\begin{equation}
g_i = \sum_D T_i(\nhat_D) - \frac{N_D}{N_R} \sum_R T_i(\nhat_R)
\end{equation}
\begin{equation}
F_{ij} = \frac{N_D}{N_R} \sum_R T_i(\nhat_R) T_j(\nhat_R)
\end{equation}
where $N_D$ and $N_R$ represent galaxy counts rather than the number of galaxies per steradian as in the continuous case.

Note that the component of $g$ corresponding to the monopole term in Equation
(\ref{eqn:masterequation}), which we will refer to as $g_C$, must be zero,
even if the sky is cut. This can be seen in the analytic formula for $g$ by
noting that we are integrating fluctuations relative to the mean, where the
mean is determined from whatever portion of the sky is being integrated
over. In the formulation where we discretize the celestial sphere, $g_C = N_D
- (N_D/N_R)(N_R) = 0$; $g_C$ represents the monopole of the fluctuations from
the mean on the cut sky, which must be zero. Hence, the only way the monopole
term $C$ in Equation (\ref{eqn:masterequation}) can be nonzero is by picking
up on the covariances between variables.

To show explicitly how we calculate the Fisher matrix $F_{ij}$ in the discrete
formalism, we take $F_{zz}$ as an example:
\begin{equation}
F_{zz} = \frac{N_D}{N_R} \sum_{i=1}^{N_R} z_i^2 = \left(\frac{N_D}{N_R}\right)(N_R)\la z^2 \ra = N_D \la z^2 \ra
\end{equation}
where we have used $z$ to designate the $z$-coordinate of the vector pointing
to the center of the pixel in which count $i$ is found. Since $\la z^2 \ra =
1/3$ over the entire sphere, we have $F_{zz} = N_D/3$ (for the entire
celestial sphere) in the limit of sufficiently large number of counts in the
random map to have suppressed Poisson noise.

\subsection{Errors in Estimated Dipole Direction}

The matrix $F_{ij}$ is the Fisher matrix for the full parameter set
  $p_i = \{d_x, d_y, d_z, k_1, ..., k_N, C\}$, and hence the covariance
matrix is Cov$(p_i, p_j) = (F^{-1})_{ij}$. By the Cram\'{e}r-Rao inequality, the best-case marginalized
errors on the parameters are
\begin{equation}
\sigma_{\rm marg}(p_i) = \sqrt{(F^{-1})_{ii}};
\label{eqn:marg}
\end{equation}
inverting $F$ automatically mixes all the elements together and takes into
account how they covary. Meanwhile, the best-case unmarginalized errors are
\begin{equation}
\sigma_{\rm unmarg}(p_i) = 1/\sqrt{F_{ii}}.
\label{eqn:unmarg}
\end{equation}
Note that the errors on our estimates of the dipole are based on the shape of
the sky cut, the input systematic templates, and the number of data points
$N_D$.

As a side note, the correlation between parameters $p_i$ and $p_j$ is
\begin{equation}
\rho_{ij} = \frac{F^{-1}_{ij}}{\sqrt{(F^{-1}_{ii} F^{-1}_{jj})}}.
\end{equation}

\subsection{Estimating the Amplitude of the Dipole}

In this formalism, we need only to acquire the components of the dipole $(d_x,
d_y, d_z)$, and the associated errors $(\sigma_x, \sigma_y,
\sigma_z)$. Combining the components by squaring, summing, and taking the
square root of the sum would create a biased estimator of the dipole amplitude
$A$, so we never do this. Instead, once we have the best-fit dipole
$\dbestbold \equiv A {\bf \hat d}$, we can construct a marginalized likelihood
function for the amplitude $A$ \cite{Hirata_hemispherical}:
\begin{equation}
\mathcal{L}(A) \propto \int \exp \left \lbrack -\frac{1}{2}(A \nhat - \dbestbold) {\rm Cov}^{-1}(A \nhat - \dbestbold) \right \rbrack d^2 \nhat
\label{eq:like_A}
\end{equation}
where $d^2 \nhat$ indicates integration over all possible directions on the
sphere.\footnote{Even though $\dbest$ itself is a biased estimator of the
  amplitude $A$, our likelihood as written in Eq.~(\ref{eq:like_A}) returns an
  unbiased estimate. The reason is that the likelihood effectively compares
  theory to measurement in each component (x, y, or z) separately, as can be
  seen by diagonalizing and rewriting the exponent in that expression, and the
  components themselves are unbiased. Numerical evidence that the estimate of
  $A$ is entirely unbiased is shown in Fig.~\ref{fig:dipquadoct}, in Appendix
  \ref{appB}.} In this equation, for a given amplitude
$A$, we take the best-fit dipole $\dbestbold$ as a given, and then compare
each direction $\nhat$ (with the given amplitude) with the best-fit
dipole. The likelihood function is a Gaussian in $A$ for fixed direction
$\nhat$, by construction, but may not be Gaussian when marginalized over
direction. That marginalization occurs in the equation above when we integrate
over all $\nhat$, giving us the likelihood of a particular amplitude $A$
marginalized over all directions, given the best-fit dipole $\dbestbold$
``selected" by the data. Posterior analysis will then show where 95 percent of
the weight lies.

Given that we ultimately work discretely, with a celestial sphere that is
pixellized using HEALPix \cite{gorski2005healpix}, the likelihood turns into a
sum over pixels:
\begin{equation}
\mathcal{L}(A) \propto \sum \exp \left \lbrack -\frac{1}{2}(A \nhat - \dbestbold) {\rm Cov}^{-1}(A \nhat - \dbestbold) \right \rbrack \Delta Area
\end{equation}
The factor $\Delta Area$ will come out of the summation since all pixel areas
are equal in HEALPix and all that matters is the ratio of likelihoods rather
than the absolute values of likelihoods, so we literally sum over all the
pixels in order to get the marginalized $\mathcal{L}(A)$. We drop the
prefactor on the likelihood that includes covariance since the covariance does
not depend on parameters.

Finding the likelihood distribution as a function of direction, $\mathcal{L}(\nhat)$, follows from an exactly analogous procedure, but we sum over all possible amplitudes associated with a given pixel rather than over all possible pixels associated with a given amplitude.

\subsection{Converting From Dipole Amplitude $A$ to Angular Power Spectrum $C_1$}

Here we show that there is a simple relationship between the dipole power
$C_1$, the $\ell=1$ mode of the angular power spectrum familiar from several
areas of cosmology, and the amplitude $A$ of the dipole computed
above. Without loss of generality, we assume that the dipole points in the
positive $z$ direction and write the fluctuation in counts in two ways:
\begin{equation}
{\delta N\over N}(\nhat) = A \nhat \cdot {\bf \hat{d}} = A\, \cos\theta =
a_{10}Y_{10}(\bf \hat{n})
\end{equation}
where $Y_{10}=\sqrt{3/(4\pi)}\cos \theta$. Therefore, 
%
$a_{10} = \sqrt{4\pi/3}A$.
%
Using the usual relationship $\Cl=\sum_{m=-\ell}^{\ell} |\alm|^2/(2\ell+1)$, the power spectrum
$C_1$ contribution is then given by
\begin{equation}
C_1 = {4\pi\over 9}A^2.
\end{equation}
For the purposes of order-of-magnitude calculations, the rule of the thumb is $C_1 \simeq A^2$.

\subsection{Commentary on the Formalism}
\label{sec3.2}

It is straightforward to show that this estimator is either precisely or
approximately equivalent to similar estimators used by recent authors.
Nevertheless the formalism of Hirata that we presently use has several
practical advantages. First, the real-space estimator employed here is
more convenient to implement than multipole-space estimators
employed in previous analyses (e.g., \citet{frith20052},
\citet{blake2002detection}, \citet{baleisis1998searching}, etc.). Many
analyses use pseudo-$C_{\ell}$ to deal with sky cuts, while sky cuts are very
straightforward to deal with in this formalism (see Sec.~\ref{sec3.2} for
further details). Finally, estimating the coefficients $k_i$ allows one to
very naturally incorporate any systematics templates one suspects might be
relevant and ensure that they do not interfere with estimation of the
dipole. This form of component separation allows one to isolate the different
contributions to the observed fluctuations in counts, and separate those
contributions into actual dipole plus systematic effects. Any pattern put into
this formalism as a systematic template will be marginalized over in the
determination of dipole amplitude $A$ and direction ${\bf \hat d}$.

Moreover, this formalism has several other noteworthy features:

\begin{itemize}
\item{It allows very naturally for arbitrary sky cuts: all that is necessary
  is to remove pixels from both the data map and the random map when performing the dipole
  analysis. When the sky is cut, the dipole becomes coupled to other
  multipoles, and the errors on the detection of $d_x$, $d_y$, and $d_z$
  derived from the Fisher matrix correspondingly increase to account for
  this.}
\item{It allows for straightforward incorporation of arbitrary
  pixellization. The scale of the pixellization should not matter, because
  Poisson noise is on the scale of the pixellization, which is much smaller
  than the scale of the dipole, and Poisson noise in larger pixels means a
  smaller effect (goes as $1/\sqrt{N}$), so the Poisson noise cancels
  out. However, different pixellization schemes can affect the size, shape,
  and nature of a mask if there is any sky cut, and when discrepancies appear
  between dipole results using different pixellization schemes, it is
  virtually always traceable to this.}
\item{It allows for the possibility of $A>1$. This may seem counterintuitive
  since it implies negative counts in some pixels. However, even though it is
  true that counts cannot go negative in real data, it is still possible for a
  \textit{model} in which some pixels have negative counts to be the best fit
  to the data.}
\end{itemize}

Note that the level of degeneracy between the systematic template $t_i({\bf \hat n})$ and the dipole ${\bf\hat d}$ depends
on structure of the former, and on the relative orientation of the two. We have
performed extensive tests to verify and build our intuition about this
degeneracy. 
For example, if a pure-dipole map is created with
the dipole pointing in the z-direction, and this map is used as a template
$t(\nhat)$, the effect is that the detection of the dipole in the z-direction,
$d_z$, gives an unreliable number, while the error bar in the z-direction,
$\sigma_z$, blows up to a much larger number than $d_z$, so that $(S/N)_z =
d_z/\sigma_z \ll 1$. Also, the correlation $\rho_{zk}$ between $d_z$ and the
template coefficient $k$ becomes 1.0. In other words, a dipole template in a given direction
takes out any component of the dipole detection in that direction.

Finally, we performed tests with varying sky coverage and survey
depth, and the number of (mock) galaxies available to verify that the input
dipole is successfully recovered within the estimator's reported error.

We proceed to apply this estimator to data from several surveys. We
select surveys with very wide sky coverage, ideally almost full-sky coverage,
because for a fixed depth, the number of modes available scales as the
fraction of the sky covered, $\fsky$. This is useful especially for beating
down cosmic variance in theoretical predictions, which as we will see, is the
dominant source of uncertainty in our comparisons of observations with theory
(observational results tend to be much more tightly constrained than
theoretical predictions, since we are working at very low $\ell$).

We also find
that higher multipoles do not contribute appreciably to the recovered signal,
or even strongly affect the error bars on the dipole signal, as long as more
than roughly half the sky is probed in the survey. See Appendix \ref{appB} for
details of how we use maps of $\ell=2$ and $\ell=3$ modes as systematics
templates to detect the presence of coupling among multipoles.

Previous research on dipoles in similar surveys will be profiled as different
types of surveys are brought up.

\section{Dipole in 2MRS}
\label{sec2mrs}

\subsection{Introduction to Dipole Signals in 2MASS}
\label{sec2mrsintro}

The Two Micron All Sky Survey (2MASS), which imaged 99.998 percent of the
celestial sphere (\citet{skrutskie2006two}), provides an excellent starting
point in testing for dipoles in tracers of large-scale structure. This survey
includes two main catalogs, the point-source catalog (PSC) and extended-source
catalog (XSC). The latter is of interest here, since it includes roughly 1.6
million sources, nearly all of which are extragalactic.

2MASS used two 1.3-m equatorial Cassegrain telescopes, one in the Northern
Hemisphere and one in the Southern Hemisphere (Mt.~Hopkins, Arizona; Cerro
Tololo, Chile), to observe in the $J$, $H$, and $K_s$ bands, corresponding to
wavelengths of 1.25, 1.65, and 2.16 $\mu$m, respectively. The XSC contains
sources that are extended with respect to the instantaneous point-spread
function \cite{skrutskie2006two}, including galaxies and Galactic
nebulae. The S/N$=$10 sensitivity limits are met by sources as bright or
brighter than 15.0, 14.3, and 13.5 mag in the $J$, $H$, and $K_s$ bands,
respectively, and (very importantly for our dipole-related considerations)
exhibit a mean color difference of less than 0.01 mag between hemispheres,
meaning that the photometry is highly uniform between hemispheres. The
reliability (corresponding to the ratio of the number of genuine extended
sources to the total number of sources, spurious or genuinely extended, in the
dataset) of the XSC is greater than 99 percent for Galactic latitude $|b| >
20^\circ$. Some extended sources in the catalog are not extragalactic, though
these can be easily removed with the right color cuts (as detailed later in
Sec.~\ref{sec4.3}).

A small subset of the 1.6 million extended sources in the 2MASS XSC were
assigned redshifts in the 2MASS Redshift Survey (2MRS), a catalog which
includes position and redshift information for over 40,000 galaxies present in
the original 2MASS sample. In this section and the next, we apply the
dipole-detecting formalism outlined in Sec.~\ref{sec3.1} to the entire 2MRS
catalog, as well as to appropriate subsets of the 2MASS XSC sources.

\citet{erdogdu2006dipole}, \citet{maller2003clustering}, and
\citet{bilicki2011two}, for example, have calculated the flux-weighted dipoles
(see Sec.~\ref{sec2.2.1}) for 2MRS (a 23,000-galaxy subset thereof, actually,
with $K_s<11.25$ -- a preliminary version of the catalog) and 2MASS,
respectively. This stands in a longer tradition of attempting to calculate
flux-weighted dipoles from near-infrared surveys, since near-infrared light
closely traces the mass distribution of large-scale structure. For instance,
\citet{rowan2000iras} calculate a flux-weighted dipole from the IRAS PSCz
Redshift Survey, which had redshifts for over 15,000 IRAS galaxies (at 60
$\mu$m; cf.~the wavelengths of 2MASS, over an order of magnitude shorter). The
IRAS PSCz ($\zmax \sim 0.1$), 2MRS ($\zmax \sim 0.1$), and 2MASS XSC ($\zbar >
0.07$) studies all find tolerably small discrepancies between the direction of
the flux-weighted dipole (and thus the acceleration of the Local Group) and
the CMB velocity dipole that partially results from that acceleration
(velocity of the Local Group being proportional to acceleration of the Local
Group in linear theory). (As noted before, the motion of the Sun with respect
to the Local Group also contributes to the kinematic dipole, but the direction
is nearly opposite the direction of the Local Group's motion with respect to
the CMB rest frame, and hence changes the magnitude of the velocity vector but
does not substantially change its direction.)

The number-weighted dipole of \citet{erdogdu2006dipole} comes closer than the
flux-weighted dipole to mimicking the quantity that we calculate here, but the
number-weighted dipole, like the closely related flux-weighted dipole, is
another quantity that seeks to measure the acceleration of the Local Group due
to surrounding large-scale structure, but instead of using flux as a proxy for
mass as in the flux-weighted dipole, the number density of galaxies in a given
direction on the sky is assumed to serve as a good proxy for mass. Our goals
and aims, as well as the precise quantity we calculate, are different: We seek
to measure not an acceleration, but rather the simple 2D-projected dipole,
which in the case of the relatively nearby survey 2MASS (and 2MRS, as a subset
of 2MASS), is dominated completely by the contributions from the
local-structure dipole (see Sec.~\ref{sec2.2.2}).

The entire power spectrum of 2MASS has been calculated by \citet{frith20052},
which means that at least one measure of the 2D-projected dipole that we
explore here has already been obtained. We compare our results to this
previous result later in this section. However, we do not regard our result as
a simple replication of the previous result (and, in fact, we note substantial
disagreement): we compute not just the amplitude of the dipole, but also its
direction; we account for systematics in a direct and natural way; and we
place the 2MASS dipole into a larger context of exploring the various
contributions to dipoles, and testing observational results against
theoretical predictions, in a wide variety of surveys.

\subsection{2MRS Profiled}

We begin with the 2MASS Redshift Survey, the densest all-sky redshift survey
to date. The 2MRS team (\citet{huchra20112mass})
measured redshifts of 43,533 bright ($K_s<11.75$) sources with $E(B-V) \le 1$
mag and $|b| \ge 5^\circ$ (for $30^\circ \le l \le 330^\circ$; $|b| \ge
8^\circ$ otherwise). Sources were carefully screened to ensure that all were
genuinely extragalactic sources and do not have compromised photometry. As
explained below, we err on the conservative side and make a symmetric cut at
$\blteight$ for all Galactic longitudes $l$, which eliminates roughly 1700 of
the galaxies in the survey. Nearly all 2MRS galaxies are within the range
$0<z<0.1$. Previous tests have worked with the flux-weighted dipole in 2MRS
\cite{erdogdu2006dipole}, but have not explored the various contributions
to the 2D-projected dipole.

The results for the dipole amplitude in this survey are strongly expected to
agree with theoretical predictions. Given the relatively small volumes
surveyed (and the fact that we are dealing with the very large-scale $\ell=1$
mode), cosmic variance is very large, and so a statistically significant
discrepancy between theory and observation would require highly anomalous
disagreement between the two. We find, even with relatively cursory checks,
that there are no serious discrepancies between theory and observation for
2MRS. Nevertheless, in the coming sections, we profile the various tests
performed on the data, and the results for both dipole direction and
amplitude, in the interests of presenting results for this survey as something
of a model (in addition to being an important test in its own right): this is
a dataset with well-controlled systematics and very little chance of giving
anomalous results, where we can test out several different types of systematic
checks, to gain intuition for what results should look like when we perform
similar tests on surveys at higher redshifts and/or with less well-understood
systematics.

\begin{figure}[]
\begin{center}
\includegraphics[width=.48\textwidth]{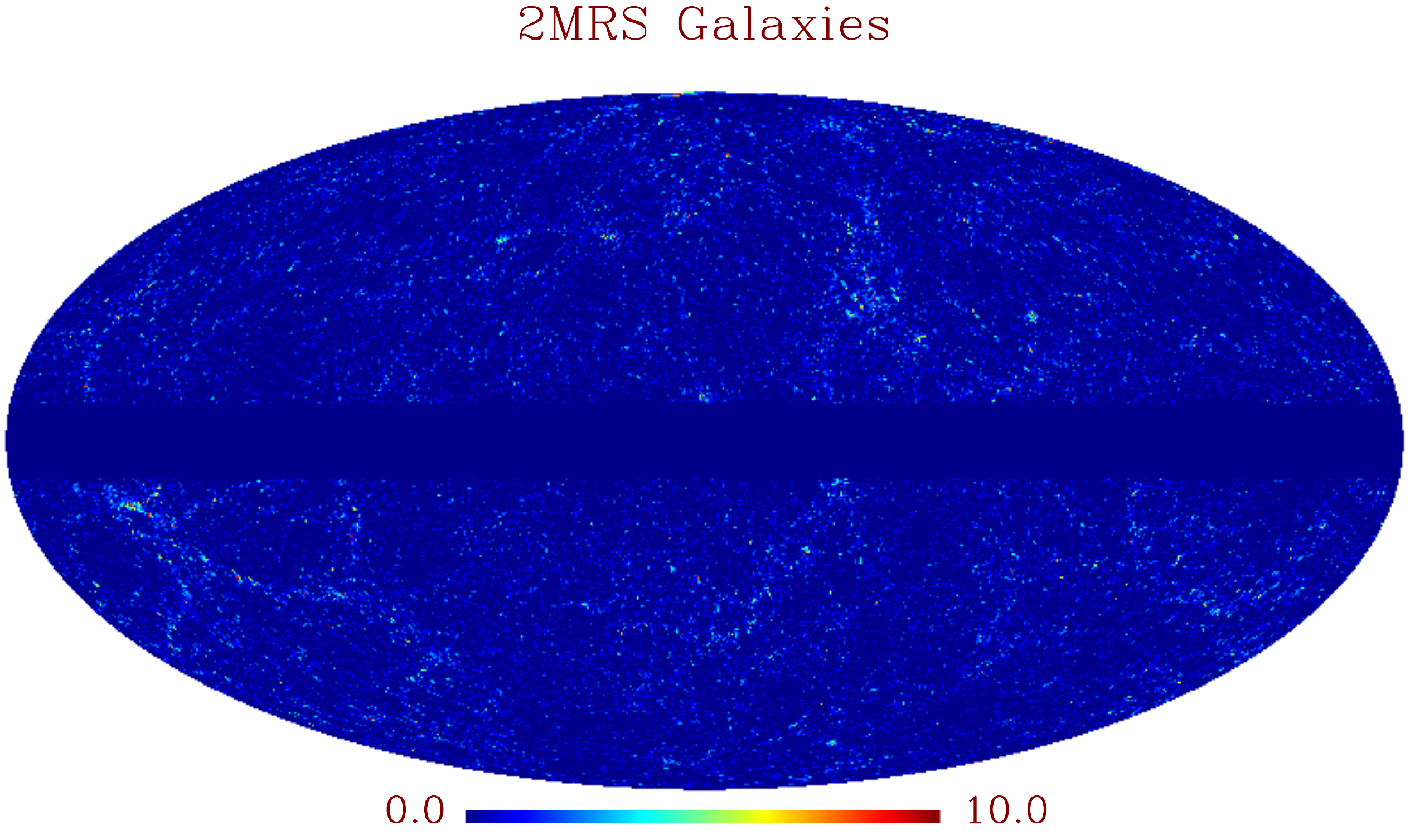}
\includegraphics[width=.48\textwidth]{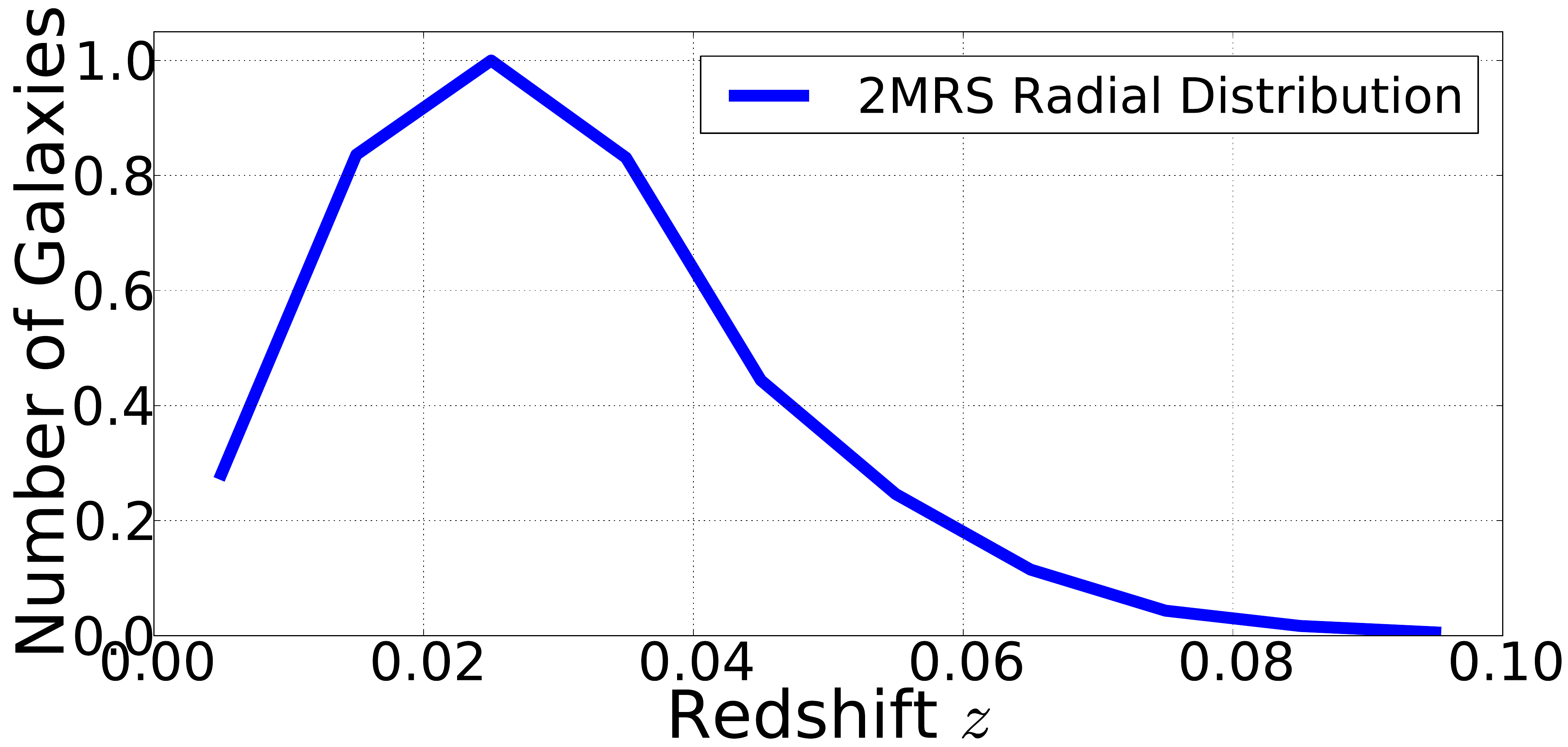}
\caption{\textit{Top panel:} All sources in the 2MASS Redshift Survey that
  escape the $\blteight$ cut. The mean redshift in the survey is approximately
  $\zbar = 0.028$. Even by eye, it is clear that the dipole due to local
  structure has not died away at these scales. In particular, the
  supergalactic plane is still fairly clearly visible in the data (see,
  e.g.,~\citet{maller2003clustering}). (Note that the dynamic range of this
  plot has been limited so that structures outside the supergalactic plane are
  also visible.) \textit{Bottom panel:} A plot of the radial distribution of 2MRS galaxies. The data are
  put in redshift bins [0, 0.01), [0.01, 0.02), ..., [0.09, 0.10), where the
        plot shows the number of galaxies in each bin as a function of bin
        center.}
\label{fig:2mrsallsources}
\end{center}
\end{figure}

This survey is essentially ready to be analyzed ``straight out of the box,"
meaning that major systematic errors have already been addressed (especially
Galactic extinction), and we already have a sample of extragalactic sources
with uniform sky coverage outside the Galactic plane. (The latter is important
since a lack of uniform completeness across the sky could, if not properly
accounted for, mimic the effect of a dipole.) More careful attention must be
paid to these matters in the 2MASS sample as a whole, and in other surveys,
but 2MRS requires only that we cut out pixels (in both the data map and the
random map to which it is compared; see Sec.~\ref{sec3.1}) within 8 degrees of
the Galactic equator, $\blteight$. Note that pixels whose \textit{centers} are above 8
degrees, and thus escape a straightforward cut of pixels with centers below 8
degrees, still may have area below 8 degrees, especially if the pixelization
is coarse, as in the cases of HEALPix resolution\footnote{Resolution at some NSIDE roughly
    corresponds to pixel size $\theta = 1^\circ \times $(60/NSIDE).} NSIDE$=$8 or
16. We adopt NSIDE$=$128 for the rest of this paper, except where otherwise
noted, and also cut conservatively so that pixels with any area at all with
$\blteight$ are cut.

\subsection{Observational Constraints on Dipole Amplitude as a Function of Redshift}
\label{sec4.2.1}

\begin{figure}[]
\begin{center}
\includegraphics[width=.48\textwidth]{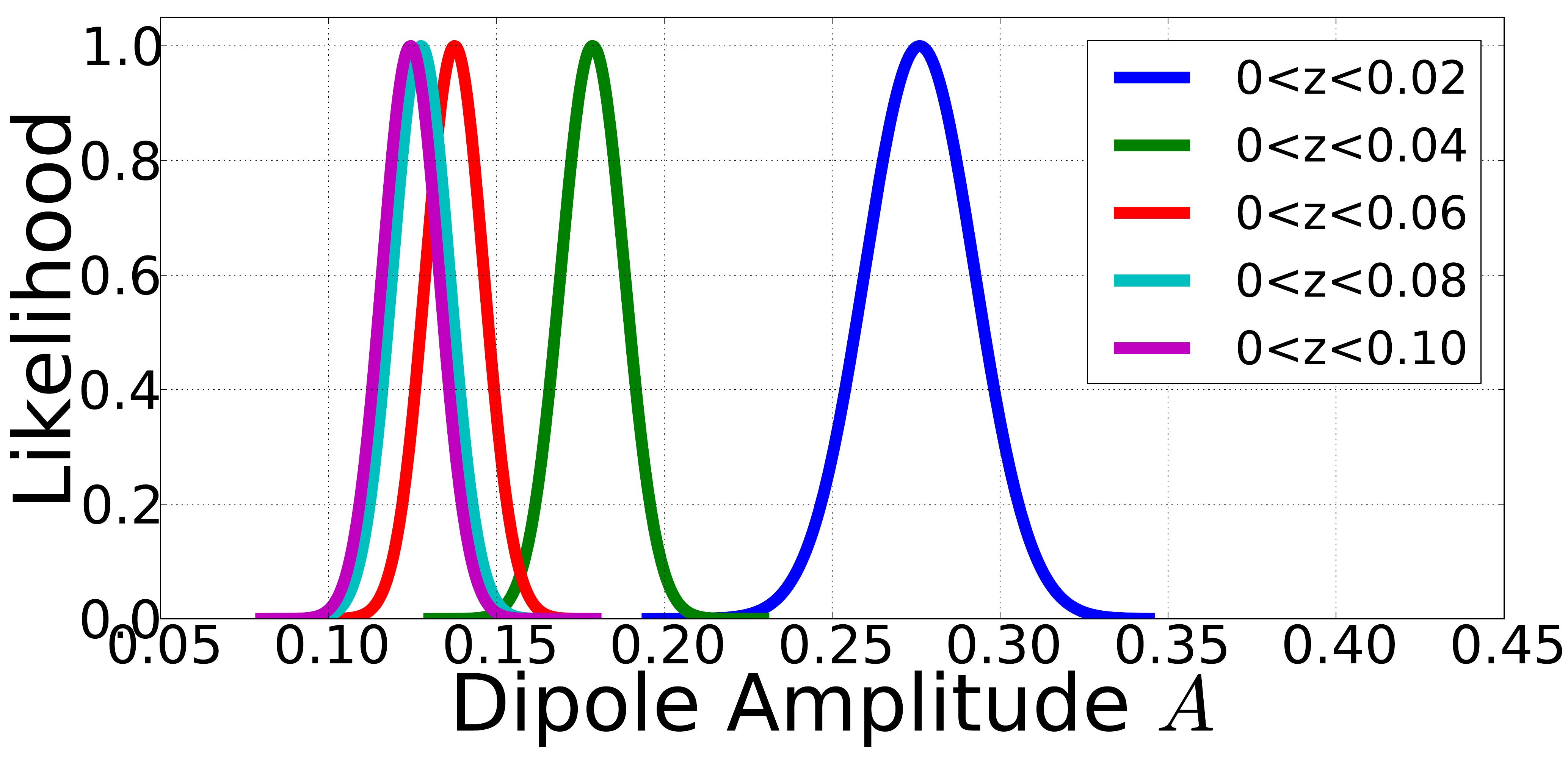}
\caption{Likelihood curves for different maximum redshifts in 2MRS. Any
  galaxies with $\blteight$ are removed from the sample.}
\label{fig:likelihoodfunctionofzmax}
\end{center}
\end{figure}

With $\blteight$ excised from the map (see Fig.~\ref{fig:2mrsallsources}), we
can apply the formalism outlined in Sec.~\ref{sec3.1} directly to different
subsets of the survey. We pixelize the data using HEALPix, meaning that we
take the Galactic coordinates given in survey data and assign the given galaxy
to the pixel corresponding to those
coordinates. Fig.~\ref{fig:likelihoodfunctionofzmax} gives the likelihood of
the dipole amplitude, $\mathcal{L}(A)$, for different subsets of the entire
survey, reaching out to $\zmax = 0.02, 0.04, 0.06, 0.08,$ and 0.10 (where the
last value in that list represents the entire survey except for 25 sources at
an assortment of higher redshifts).

Note that the behavior is as expected in several regards:

\begin{itemize}
\item{The dipole amplitude $A$ starts off larger and grows smaller as we go out further in redshift.}
\item{$A$ converges to a certain value. This should happen simply because we
  run out of sources as we go to higher and higher redshift (e.g., 41,446
  sources are at $z<0.06$, while the total number of sources with $z<0.10$ is
  43,506.)}
\item{Because the redshifts in this sample are relatively low in cosmological
  terms, going out only to $z \sim 0.10$, we expect that the dipole amplitude
  should remain on the order of $10^{-1}$, and it does.}
\end{itemize}

Although 2MRS should be relatively systematic-free, we proceed to perform
straightforward tests for two types of systematic effects: Galactic extinction
as characterized by the maps of \citet{schlegel1998maps}, and star-galaxy
confusion or other systematics that vary as a function of Galactic latitude
$b$.

\subsection{Comparison of Dipole Parameters With and Without Extinction Template}
\label{sec4.2.2}

If we wish to take Galactic extinction into account, the formalism we are
employing to find dipole amplitudes and directions allows for very
straightforward incorporation of extinction due to dust as a systematic
template. The 2MRS maps are already extinction-corrected, so this is much more
of a sanity check than anything else.

We use the SFD map \citep{schlegel1998maps}, an intensity map of the sky at 100
microns and a reprocessed composite of the COBE/DIRBE and IRAS/ISSA maps. The resolution is on the order of
a few arcminutes, and so the SFD map can in general be used to reliably derive
extinction due to dust, assuming a standard reddening law. This works best
away from the plane of the Galaxy, since within the Galactic plane dust
conditions tend to fluctuate much more strongly on small scales than they do
away from the Galactic plane (with the possibility of multiple dust
temperature distributions, variable grain sizes, etc.). However, since we work
almost exclusively well away from the Galactic equator, we do not expect this
to be an issue.

The SFD map is nearly parity-even in Galactic coordinates, as the Galaxy is
itself nearly parity-even, so when the sky is symmetrically cut, extinction is
not likely to contribute to, or diminish, a dipole-like (parity-odd) signal,
at least not in the $z$-direction. However, we still include it as a
precaution against a known potential source of systematic error.

The results are best presented in the form of a direct comparison, in Table \ref{tab4.1}. Entries in this table take into account
  the entire 2MRS sample, 43,506 galaxies (before the symmetric cut in
  Galactic latitude) with $0.00 < z < 0.10$.

\begin{table*}[]
\footnotesize
\caption{Comparison of dipole parameters with systematics templates
  vs.~without templates, for 2MRS. The first column gives the Galactic
  latitude $b$ of the cut; the second column identifies any systematic
  template present; the third gives the HEALPix NSIDE parameter; the fourth
  gives the number of sources that were still available after the cut was
  made; the fifth gives the dipole amplitude with highest likelihood; the
  sixth and seventh give $l$ and $b$ of the best-fit dipole; and the eighth
  and ninth give the 68 and 95 percent confidence intervals on the amplitude
  of the dipole. The first row gives results when we include no systematics
  templates; the second when we include only the SFD dust map as a template,
  and the third when we include SFD and quadrupole and octopole modes as
  templates.}
\label{tab4.1}
\centering
\setlength{\tabcolsep}{1em} 
\begin{tabular}{| c | c | c | c | c | c | c | c | c |}
\hline
\rule[-2mm]{0mm}{6mm} $|b| \ge$ & Systematics & NSIDE & $N$ & $A_{\rm peak}$ & $l$ & $b$ & 68 percent CI & 95 percent CI \\
\hline

\rule[-2mm]{0mm}{6mm} 8.0 & none & 128 &   41834 & 0.124 & 228.0 &  38.7 & 0.116 - 0.132 & 0.108 - 0.141 \\ \hline
\rule[-2mm]{0mm}{6mm} 8.0 & SFD Dust & 128 &   41834 & 0.118 & 222.3 &  38.3 & 0.110 - 0.126 & 0.102 - 0.135 \\ \hline
\rule[-2mm]{0mm}{6mm} 8.0 & SFD Dust + Quad + Oct & 128 &   41834 & 0.120 & 213.8 &  35.2 & 0.111 - 0.128 & 0.103 - 0.137 \\ \hline
 
\hline
\end{tabular}
\end{table*}

The results change only slightly when the SFD template is added, and are
statistically consistent with the no-dust results. The fact that the dipole
amplitude drops slightly with addition of the template is an indication that a
very small amount of the dipole power in the 2MRS map can be attributed to the
pattern set up by the distribution of dust in our Galaxy.

The final row of Table \ref{tab4.1} gives the results on dipole amplitude and
direction when we include not only the SFD map but also the five $\ell=2$ and
seven $\ell=3$ modes as templates. This ensures, as discussed earlier and
elaborated upon in Appendix \ref{appB}, that we are detecting only dipole
signal and are not receiving contributions from higher multipoles. The
contributions from higher multipoles are almost negligible in this case,
though they do slightly affect the direction of the detected dipole signal.

\subsection{Dipole Parameters as a Function of Sky Cut}
\label{sec4.2.3}

With the SFD extinction template in place, and again using the entire 2MRS
sample out to $z=0.10$, we may also vary the location where the cut in
Galactic latitude is placed. Verifying that the placement of the cut (as long
as it is kept at least as aggressive as the $\blteight$ cut) does not affect
the results beyond widening our error bars serves as a check for any source of
systematic error that varies as a function of Galactic latitude. Most notably,
any star-galaxy confusion that might creep into the survey (very unlikely in
the case of this particular survey) would vary strongly as a function of
Galactic latitude, with the density of stars dropping precipitously as one
moves away from the Galactic equator, and so this test serves to verify that
star-galaxy confusion is not a major contributor to the detection of a
dipole. It also helps to guard against the possibility that variations in sky
coverage (see Sec.~\ref{sec4.2.8}) affect the dipole signal. (Sky coverage is
better at higher Galactic latitudes, since extended sources cannot be observed
very close to very bright stars -- less than 2 percent of the area of a
typical high-latitude sky is masked by stars, as noted by
\citet{erdogdu2006dipole}.)

Note that as the sky cut becomes more and more aggressive, we expect the error
bars on the observed value of $A$ to become wider (simply because we have less
data and therefore looser constraints), but we expect the
best-fit/peak-likelihood value of $A$ itself to remain consistent with values
found given less aggressive sky cuts. If $A$ shifts in such a way that more
aggressive sky cuts yield amplitudes inconsistent with amplitudes from maps
with less aggressive sky cuts, this is a fairly good indication that
star-galaxy confusion or another systematic effect that varies with Galactic
latitude is at play. While the angular dimensions (and thus angular cuts) are
not what determine the amplitude of the dipole, for very aggressive cuts we
are left with far less data than we are with minimal cuts, and this means that
(a) the ``measurement error" on the observation becomes greater, and, even
more importantly, (b) the cosmic variance associated with the theoretical
prediction becomes much greater. Due to (b) in particular, a much wider range
of peak-likelihood values for the amplitude becomes consistent with theory, as
we cut the sky more aggressively.

\begin{figure}[]
\begin{center}
\includegraphics[width=.48\textwidth]{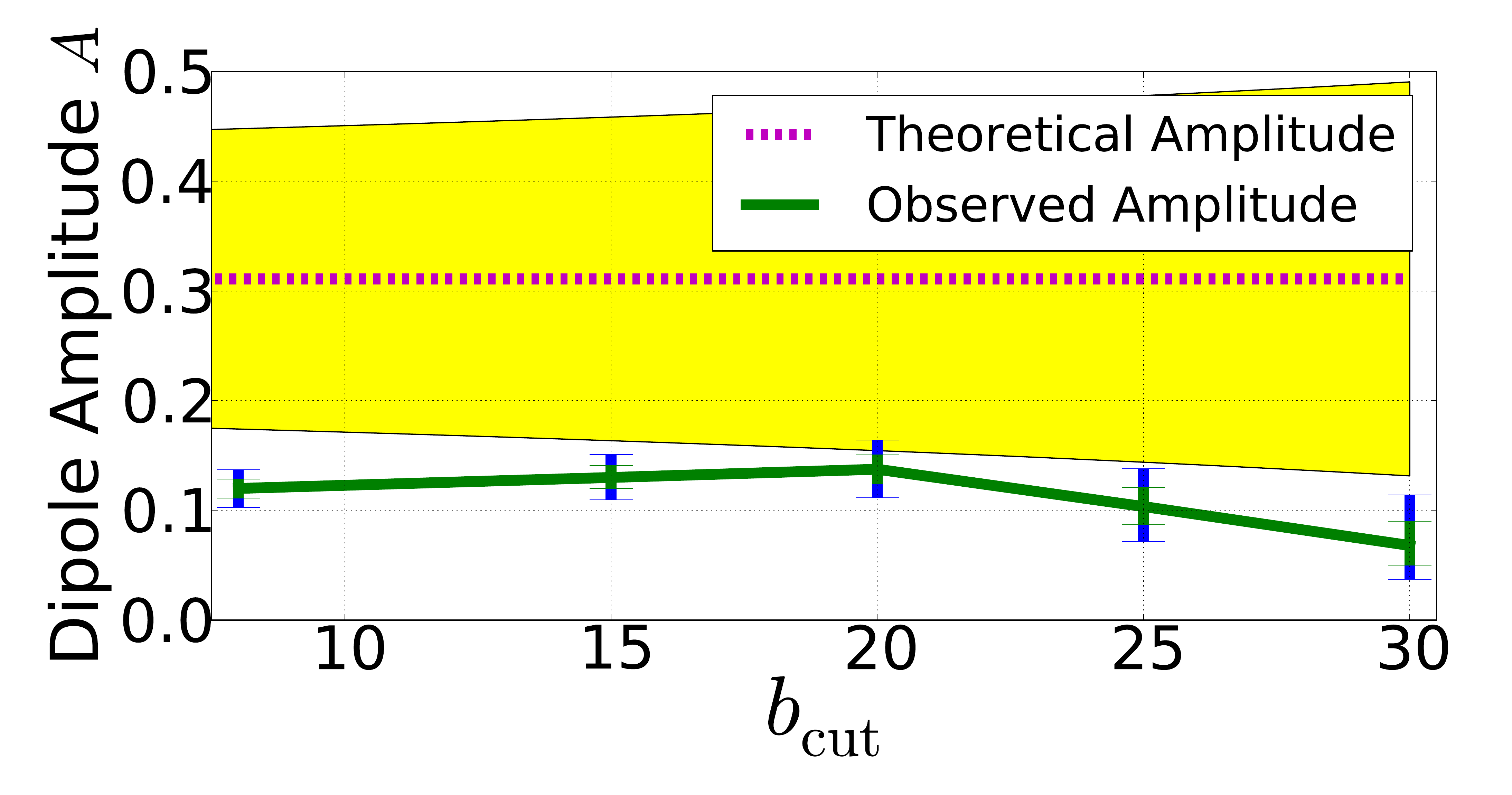}
\caption{Results for the dipole amplitude as a function of cut in Galactic
  latitude $b$. Here $\bcut$ indicates that $|b|<\bcut$ was cut out of the
  map. Notice the very wide cosmic-variance band (yellow shaded region) around
  the theoretically predicted value for the dipole amplitude, and the
  consistency of all observed values within cosmic-variance limits. The small
  green (blue) error bars indicate 68\% (95\%) measurement
  errors. The measurement errors are
    tiny in comparison with cosmic variance. Possible contamination from the
    quadrupole and octopole, which becomes more important as more of the sky
    is cut, is taken into account by including all $\ell=2$ and $\ell=3$ modes
    as systematics templates in the analysis.}
\label{fig:2mrsfunctionofbcut}
\end{center}
\end{figure}

Results for different sky cuts are found in
Fig.~\ref{fig:2mrsfunctionofbcut}. The results are all consistent with one
another, and with theory (all measurements being just outside the one-sigma
cosmic-variance band). Note that we have included quadrupole and octopole
modes as systematics templates (see Appendix B) in creating this plot since
$\ell=2$ and $\ell=3$ modes affect the error bars somewhat for more aggressive
cuts.

\subsection{Dipole Amplitude: Theory vs. Observation}
\label{sec4.2.4}

Now that we have established basic consistency among dipole determinations
with different sky cuts, we go back to the least restrictive cut, at $\blteight$, and keep
the SFD template in place. We proceed to compare theory and observation in
dipole amplitude as a function of redshift in 2MRS.

As noted earlier, it is important to take the bias of tracers of the LSS into
account when producing theoretical expectations for the clustering of these
objects. \citet{frith20052} find the bias in the 2MASS $K_s$ band to be $1.39
\pm 0.12$, employing a technique that uses constraints on the galaxy
power-spectrum normalization as well as previous constraints on $\sigma_8$. We
therefore adopt 1.4 as the value of the bias for both 2MRS and 2MASS in
general. The qualitative conclusions drawn from these surveys do not depend
strongly on the precise value adopted for the bias. Note that, for constant
bias $b$, dipole amplitude is proportional to the bias, $A \propto b$.

\begin{figure}[]
\begin{center}
\includegraphics[width=.48\textwidth]{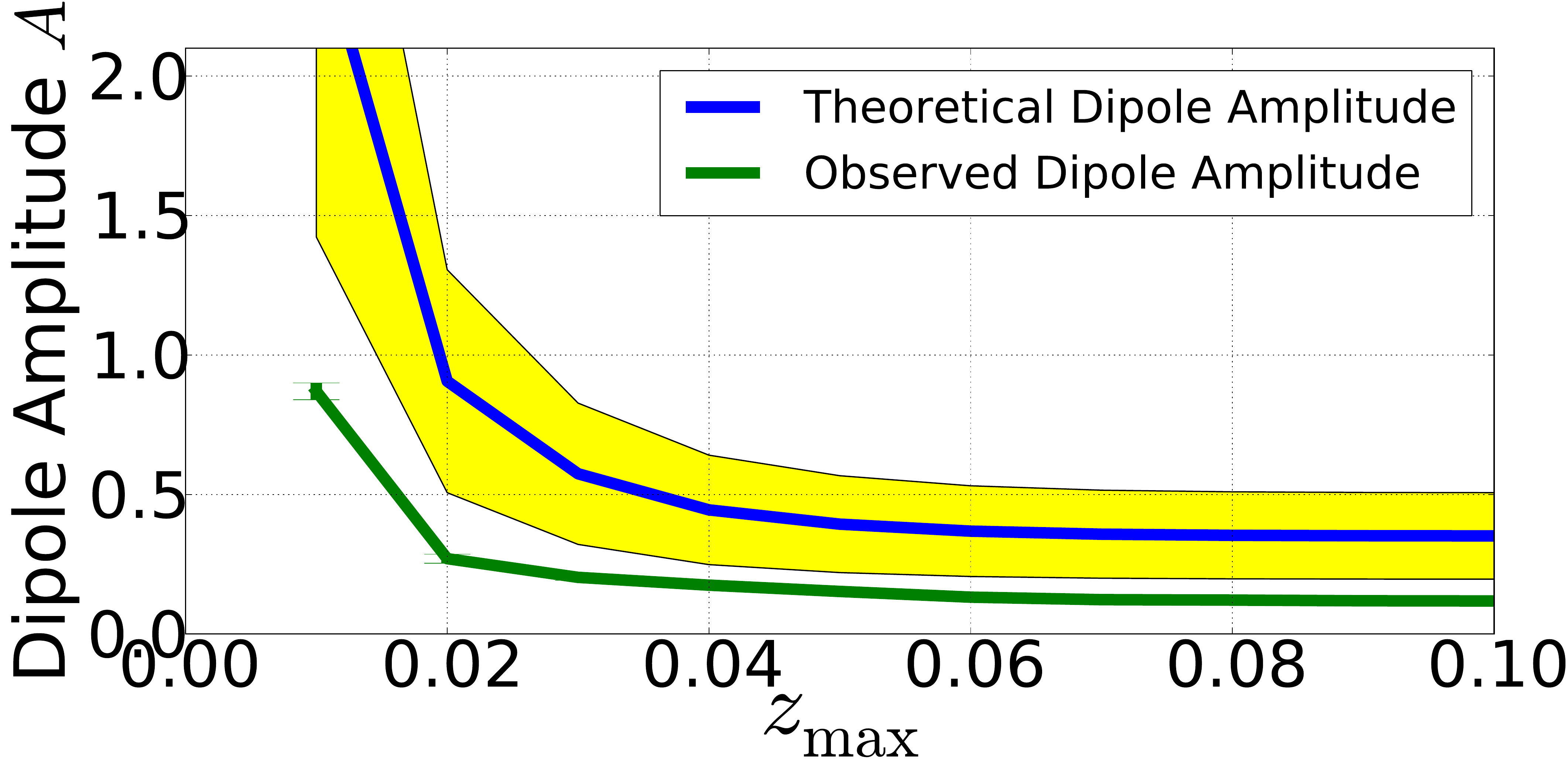}
\includegraphics[width=.48\textwidth]{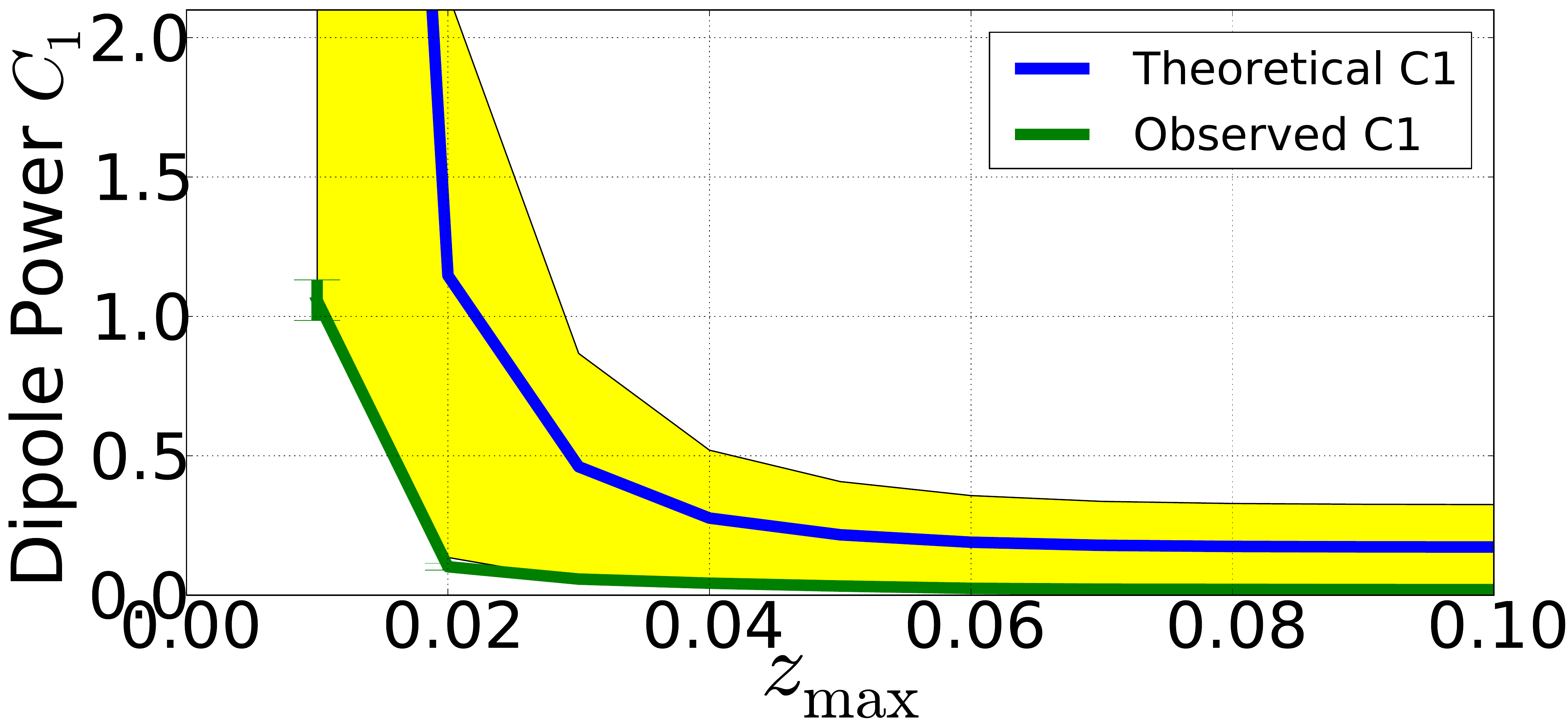}
\caption{\textit{Top panel:} Comparison of observations with theory for the
  dipole amplitude, as a function of how much of the 2MRS sample is included
  ($0.00 < z < \zmax$). All observed values are found including the SFD
  template, and with a cut at $\blteight$. For the purposes of calculating
  theoretical predictions, we take $\fsky = 0.86$, corresponding to the
  $\blteight$ cut.) \textit{Bottom panel:} The same results, only with the dipole
  power $C_1$ rather than dipole amplitude $A$. In all cases, 68 percent error
  bars on observations are shown, but are either invisible or nearly invisible
  due to how tiny they are. See Sec.~\ref{sec3.1} for details of the procedure
  to convert between $A$ and $C_1$.}
\label{fig:Avsz2mrs}
\end{center}
\end{figure}

There are cosmic-variance errors on all theoretical predictions:
\begin{equation}
\delta \Cl = \sqrt{\frac{2}{(2 \ell + 1)\fsky}} \Cl.
\end{equation}
We can relate $\Cl$ to the amplitude $A$ via
%
$C_1 = (4\pi/9) A^2$
%
and so doing error propagation to get the cosmic-variance error on the amplitude, we have
\begin{equation}
\delta A = \frac{1}{2} \sqrt{\frac{2}{3 \fsky}} A.
\end{equation}
This allows us to plot cosmic-variance uncertainties in both $C_1$ and
$A$. The basic result of doing so is shown in Fig.~\ref{fig:Avsz2mrs}. Note
that the errors on the observations are very small in comparison with cosmic
variance. Therefore, for the rest of this section, we consider measurement
errors negligible and focus only on how observational results compare with
theory within the bounds of cosmic variance.

The dipole amplitude, both theory and measurement, decreases as redshift
increases, exactly as it should given our previous arguments that averaging
over more structure at larger distances yields lower values of the
dipole. Whether the amplitude is expressed as $A$ or as $C_1$, the
observational results are consistently lower than the theoretical
expectations. If these measurements for different $\zmax$ were all
independent, there would be a highly significant inconsistency between theory
and observation, but the measured values are highly correlated since samples
with higher $\zmax$ contain all samples with lower $\zmax$.

More specifically, correlations between a bin $i$ going out to $z_{\rm max,1}$
and another bin $j$ going out to $z_{\rm max,2}$ are calculated as
\begin{equation}
{\rm Cov} \left [C_{ii}, C_{jj} \right ] = \frac{2}{2l+1} C_{ij}^2,
\end{equation}
where we have made the usual assumption that the galaxy overdensity is
  a Gaussian random variable, so that Wick's theorem can be applied to obtain
  the expression above. Then the correlation coefficient is
\begin{eqnarray}
\rho &=& \frac{{\rm Cov}(C_{ii}, C_{jj})}{\sqrt{{\rm Cov}(C_{ii}, C_{ii}) {\rm
    Cov}(C_{jj}, C_{jj})}}\nonumber  \\[0.12cm]
&=& \frac{C_{ij}^2}{(C_{ii} C_{jj})} 
\end{eqnarray}
Correlations between 2MRS samples range from 0.42 (between the full ($0.00 <
z < 0.10$) sample and the smallest ($0.00 < z < 0.01$) sample), to 0.81
(between the full sample and the second-smallest ($0.00 < z < 0.02$) sample),
to well over 0.99 for many combinations of samples. (This is also the reason
why all bins have similar significance as compared with one another.)
Therefore, rather than being a 10-sigma inconsistency between theory and
observation, Fig.~\ref{fig:Avsz2mrs} represents only slightly more than a
1-sigma discrepancy. The next section will find the precise ``discrepancy"
rigorously.

\subsection{Comparison of Theory and Observation for Dipole Amplitude}
\label{sec4.2.5}

\begin{figure}[]
\begin{center}
\includegraphics[width=.49\textwidth]{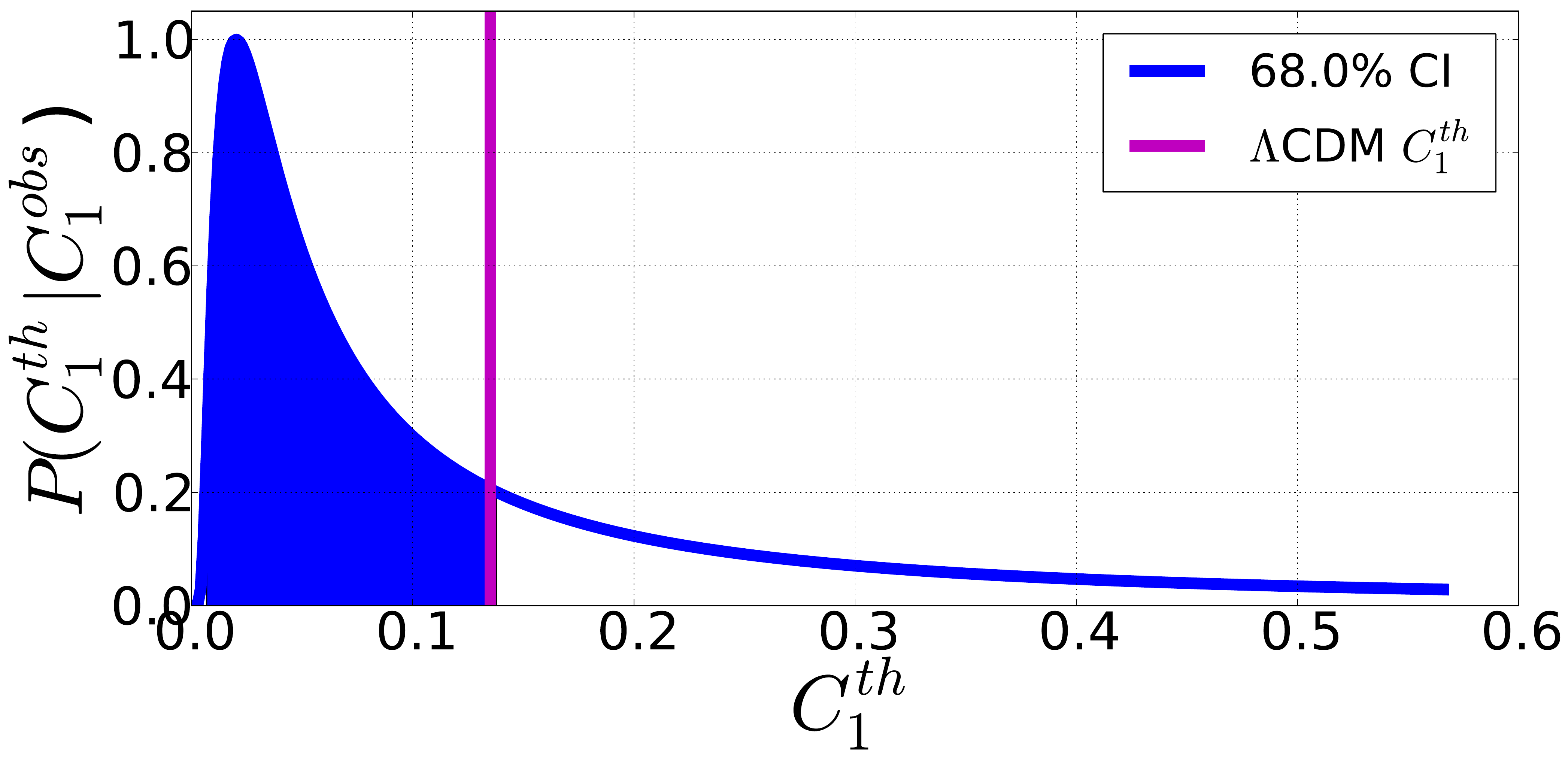}
\caption{Posterior probability of theoretical $C_1$ given observed $C_1$ as a
  function of the theoretical value, for the full sample of 2MRS galaxies. The
  observed $C_1$ determines the likelihood distribution for $\Coneth$, and we
  can then compare the $\Lambda$CDM value (vertical magenta line) for
  $\Coneth$ to that distribution. The $\Lambda$CDM value is clearly consistent with the
  observation.}
\label{fig:probthgivenobs}
\end{center}
\end{figure}

Given a Gaussian field on the celestial sphere with observed angular power
spectrum $\Clobs$, the power is $\chi^2$-distributed, and the likelihood of a
given theoretical value $\Clth$ is
\begin{equation}
\ln P(\Clth | \Clobs) = \sum^\infty_{\ell=0} \frac{2 \ell + 1}{2} \left \lbrack -\frac{\Clobs}{\Clth} + \ln \frac{\Clobs}{\Clth} \right \rbrack - \ln \Clobs.
\end{equation}
(see, for example, \citet{chu2005cosmological}).\footnote{This expression can
  be derived by noting that the random variable $Y = (2 \ell + 1)
  \frac{\Clobs}{\Clth}$ is $\chi^2$-distributed with $2 \ell + 1$ degrees of
  freedom. Inserting this expression for $Y$ into the general expression for a
  $\chi^2$ distribution, and then using the fact that $P(Y) dY = P(\Cl) d
  \Cl$, it is relatively straightforward to show that the proportionality for
  $P(\Cl)$ given in \citet{chu2005cosmological} holds, and from there the
  expression for the log-likelihood given above immediately follows.}  Here
the observed quantity $\Clobs$ is treated as a realization of the theoretical
value $\Clth$. For $\ell=1$, this simplifies to
\begin{equation}
\ln P(\Coneth | \Coneobs) = \frac{3}{2} \left \lbrack -\frac{\Coneobs}{\Coneth} + 
\ln \frac{\Coneobs}{\Coneth} \right \rbrack - \ln \Coneobs
\end{equation}
Again, we treat this as a likelihood, so that $P$ is a function of the
theoretical model $\Coneth$, with the observed quantity held fixed. Then, as
usual, we can plot the likelihood of a parameter value (in this case,
theoretical $\Coneth$) and see where our ``actual" theoretical $\Coneth$ falls
with respect to that distribution. In each redshift bin, we could generate a
different likelihood distribution based on the observation, and then compare
to the actual $\Coneth$ in each case. However, because of the very high
correlations between redshift samples, we gain very little by doing tomography
in this way, and so we only perform this analysis on the full 2MRS sample,
$0.00 < z < 0.10$. The results are shown in Fig.~\ref{fig:probthgivenobs}.

Whether we calculate a simple signal/noise ratio to compute the significance
of $\Coneth$ results given $\Coneobs$, as in Fig.~\ref{fig:Avsz2mrs}, or
whether we use the more detailed comparison of
Fig.~\ref{fig:probthgivenobs}, the qualitative conclusion is the same: the
$\Lambda$CDM prediction matches observations within appropriate
cosmic-variance limits. Note that the reason why these two strategies do not
match up quantitatively is that the signal/noise strategy assumes cosmic
variance is symmetric, while using the $P(\Coneth | \Coneobs)$ distribution
takes into account the asymmetry of cosmic variance, particularly at the very
low $\ell$ at which we are working. This is also why significances in $C_1$
and $A$ do not match up with one another exactly.

\subsection{Observational Constraints on Dipole Direction as a Function of Redshift and Sky Cut}
\label{sec4.2.6}

Up to this point, we have been focusing on the dipole amplitude and comparing
theoretical and observed amplitudes. The direction of the dipole, however, is
also a quantity of considerable interest.

As discussed in Sec.~\ref{sec2.2}, there are three major types of dipoles that
could contribute to any detected dipole in objects that trace large-scale
structure: the local-structure dipole, the kinematic dipole, and the intrinsic
dipole. At the scales probed by 2MRS, we expect the local-structure dipole to
completely dominate other contributions, since it is on the order of $10^{-1}$
while the kinematic dipole falls two orders of magnitude below this and the
intrinsic dipole may very well fall even further below that. Therefore, there
is no reason to expect that the direction of the 2MRS dipole should align with
the direction of the CMB dipole, as we would expect it to do if the kinematic
dipole were dominant at these scales.

That said, it should not be at all surprising if the 2MRS local-structure
dipole points somewhere near the CMB kinematic dipole. The reason for this has
to do with what generates the motion that gives rise to the kinematic
dipole. As discussed in Sec.~\ref{sec2.2}, the total velocity of the Sun with
respect to the Local Group is directed along almost the same line as the
velocity of the Local Group with respect to the CMB rest frame, but in the
opposite direction. So the direction of the Sun's total motion with respect to
the CMB rest frame is essentially the same as the direction of the Local
Group's motion with respect to the CMB rest frame, but the speed is lower than
that of the Local Group since the contribution of the Sun's motion with
respect to the Local Group gets subtracted off. The Local Group moves in a
certain direction with respect to the CMB rest frame because of the
gravitational pull of structure in the relatively nearby universe. The
acceleration due to gravity of the Local Group, as determined via
flux-weighted dipole measurements, is directed less than $20^\circ$ away on
the sky from the direction of the velocity of the Local Group
\cite{maller2003clustering}. Therefore, insofar as the local-structure dipole
gives information about the clustering of local structure and the direction of
the acceleration due to gravity of the Local Group, it is expected that it
should point in at least the same general direction as the CMB kinematic
dipole, which is generated in part by the velocity of the Local Group induced
by its acceleration due to gravity. Since the local-structure dipole is a
2D-projected quantity rather than one that preserves radial information, it is
not a perfect indicator of where gravitational pulls on the Local Group are
coming from. But we do expect the direction of the local-structure dipole to
feel some influence from the direction of the CMB kinematic dipole.

\begin{figure}[]
\begin{center}
\includegraphics[width=.48\textwidth]{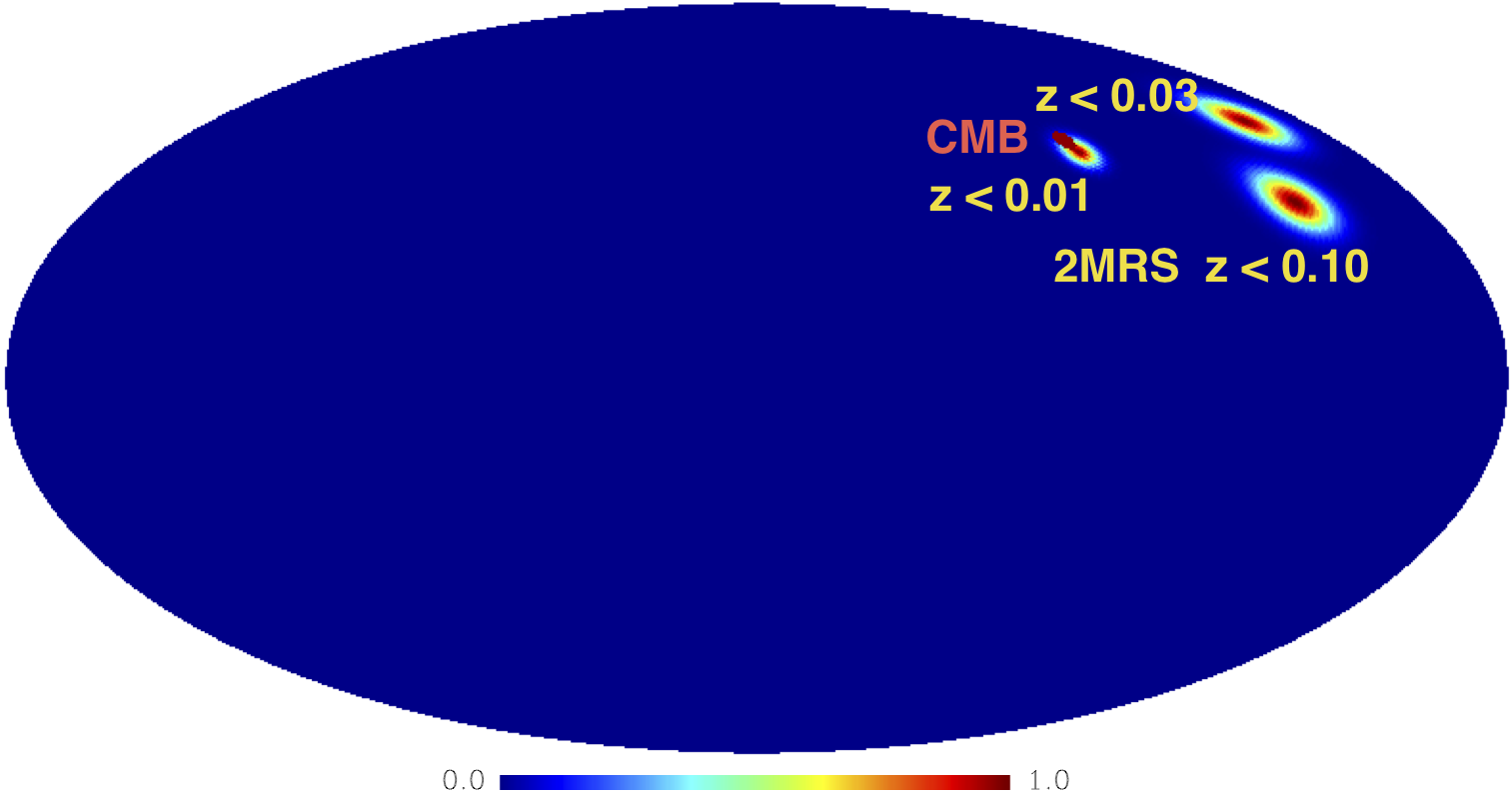}
\includegraphics[width=.48\textwidth]{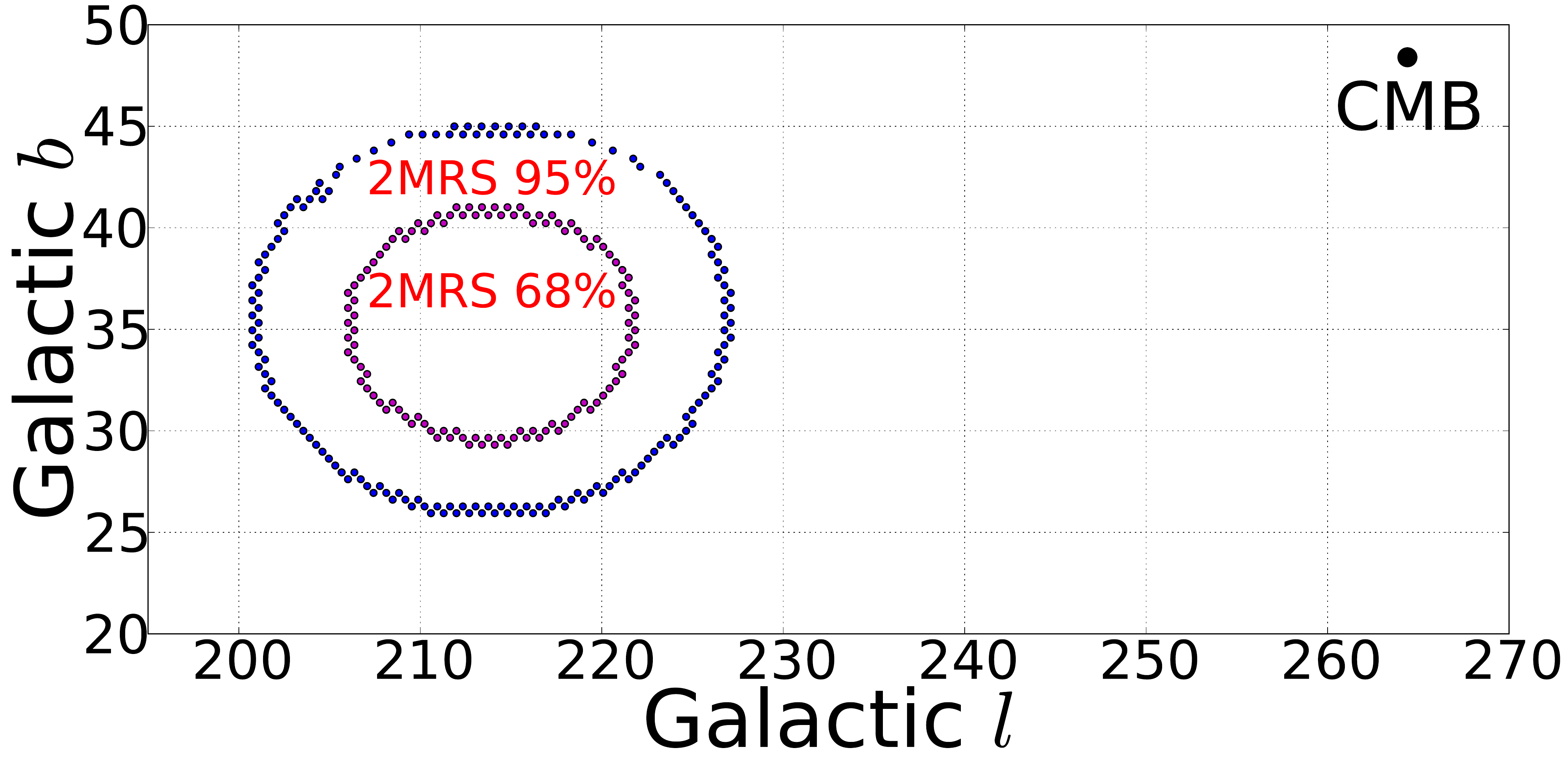}
\caption{\textit{Top panel:} Likelihood associated with each dipole direction
  on the sky, marginalized over amplitude, shown for 2MRS redshift shells
  $0.00 < z < 0.01$ (leftmost multicolored oval), $0.00 < z < 0.03$ (uppermost
  multicolored oval), and $0.00 < z < 0.10$ (rightmost multicolored oval). We
  assume a $\blteight$ cut, and incorporate the SFD dust systematic template
  and quadrupole and octopole templates. The color scale represents normalized
  likelihood as a function of direction. The single-colored disc that overlaps
  with one of the multicolored likelihood ovals represents the direction of
  the CMB kinematic dipole, with error bars exaggerated to a circle of 2
  degrees in order to make the position clearly visible on the map.
  \textit{Bottom panel:} Confidence intervals for the direction of the dipole
  in the full 2MRS survey, with the position of the CMB dipole
  shown. Agreement was not expected, but it is reassuring that the 2MRS
  projected dipole does lie in the same general region of sky as the CMB
  dipole.}
\label{fig:2mrsdirection}
\end{center}
\end{figure}

\begin{table*}[th]
\footnotesize
\caption{Key patterns in cutting in supergalactic coordinates, for
  2MRS.}
\label{tab4.4}
\centering
\setlength{\tabcolsep}{0.7em} 
\begin{tabular}{| c | c | c | c | c | c || c | c | c | c | c | c |}
\hline
\rule[-3mm]{0mm}{8mm} $\sgbge$ & $\fsky$ & $N$ & $\fsources$ & $\frac{\fsources}{\fsky}$ & $A_{\rm peak}$ & $\sgblt$ & $\fsky$ & $N$ & $\fsources$ & $\frac{\fsources}{\fsky}$ & $A_{\rm peak}$ \\
\hline

\rule[-2mm]{0mm}{6mm} 0.0 & 0.86 & 41834 & 1.00 & 1.17 & 0.12 & --- & --- & --- & --- & --- & --- \\ \hline
\rule[-2mm]{0mm}{6mm} 2.0 & 0.82 & 39964 & 0.96 & 1.16 & 0.12 & 74.82 & 0.84 & 41234 & 0.99 & 1.17 & 0.12 \\ \hline
\rule[-2mm]{0mm}{6mm} 5.0 & 0.78 & 37124 & 0.89 & 1.14 & 0.12 & 65.90 & 0.81 & 39867 & 0.95 & 1.18 & 0.11 \\ \hline
\rule[-2mm]{0mm}{6mm} 10.0 & 0.70 & 32673 & 0.78 & 1.11 & 0.12 & 55.73 & 0.74 & 36882 & 0.88 & 1.20 & 0.09 \\ \hline
\rule[-2mm]{0mm}{6mm} 20.0 & 0.55 & 24799 & 0.59 & 1.08 & 0.11 & 41.15 & 0.59 & 30321 & 0.72 & 1.22 & 0.05 \\ \hline

\hline
\end{tabular}
\end{table*}

The observational results for the direction of the dipole are displayed in
Fig.~\ref{fig:2mrsdirection}. The results align with the qualitative
expectations detailed above. It turns out that the 2MRS dipole direction is
indeed not consistent with the direction of the CMB kinematic dipole, but
still within the same basic region of sky.

Note that the constraints on the dipole direction are actually tighter for the
very small $0.00 < z < 0.01$ sample than for the full sample of 2MRS galaxies,
despite the fact that the number of sources is an order of magnitude
smaller. This is not anomalous since the higher-redshift sources actually
decrease the prominence of the dipole in local structure, producing a result
with roughly half the total signal/noise as in the case where we take the
$\zmax = 0.01$ subsample.

The best-fit direction for the 2MRS dipole is $(l,b) = (228.0^\circ, 38.7^\circ)$. \citet{erdogdu2006dipole} find that the 2MRS number-weighted dipole (the quantity they analyze that is ``closest" to our dipole) is at $(l,b) = (218^\circ,33^\circ)$ in the CMB frame,
in close agreement with our results.



\subsection{Cutting the Supergalactic Plane}
\label{sec4.2.7}

As a final check, we wish to know how much of the dipole signal in 2MRS is
coming from the vicinity of the supergalactic plane (SGP), a planar structure
in the local galaxy distribution (\citet{lahav2000supergalactic}). We
therefore progressively excise more and more of the SGP and see how much the
amplitude of the dipole dies away. We compare this to the effect of excising
\textit{similar areas} from the vicinity of the supergalactic poles. We expect
that there should be more sources near the supergalactic plane, and that the
dipole should die away much more quickly when the supergalactic plane is
excised than when similar areas around the supergalactic poles are
excised. 

This check will become more important as we proceed to perform our analysis on
surveys that probe much larger radial distances than does 2MRS, as the
structure associated with the supergalactic plane will only contribute to the
dipole on relatively nearby scales, and the effect should diminish as we probe
to larger and larger redshifts.

\begin{figure*}[t]
\begin{center}
\includegraphics[width=.48\textwidth]{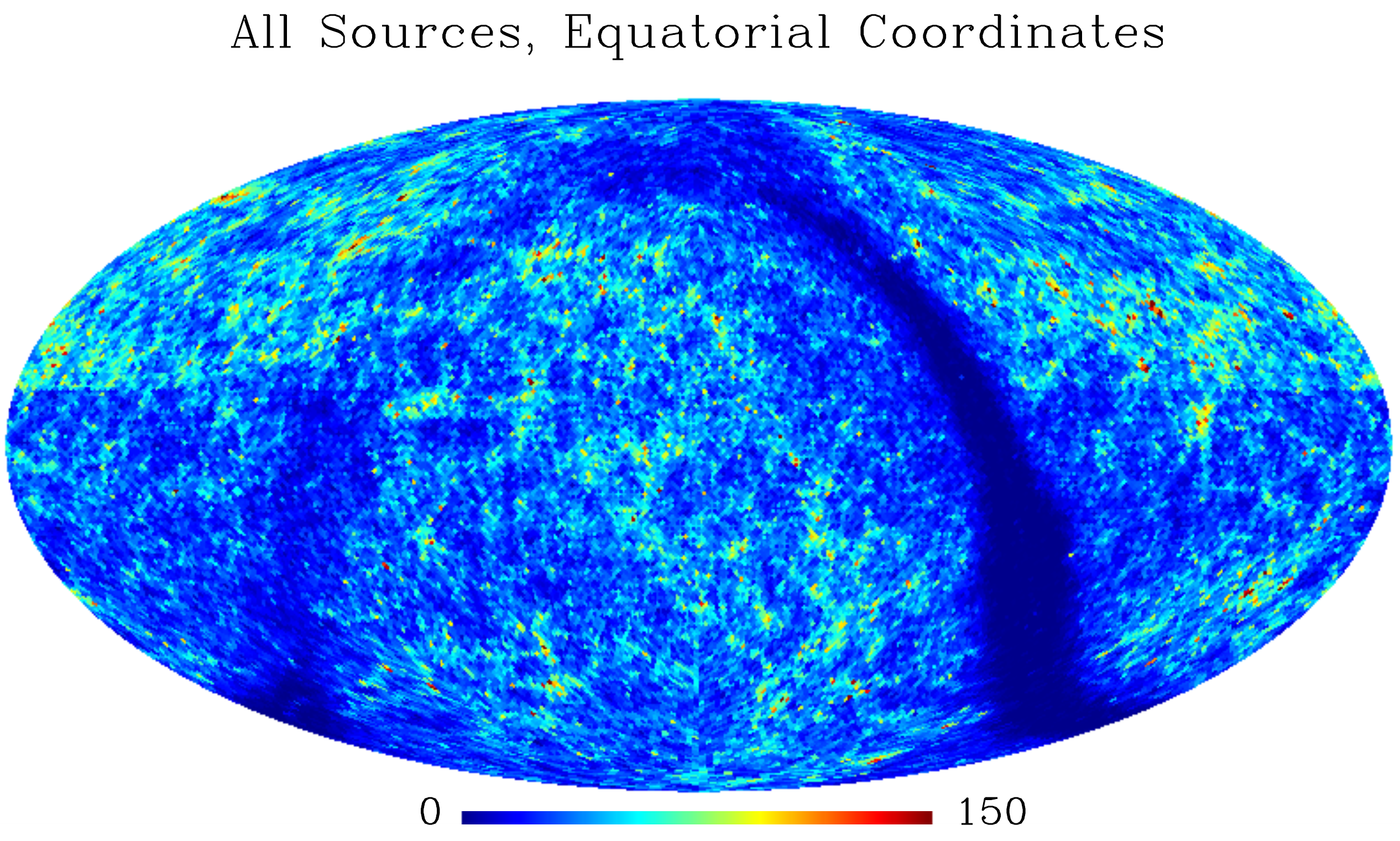}
\includegraphics[width=.48\textwidth]{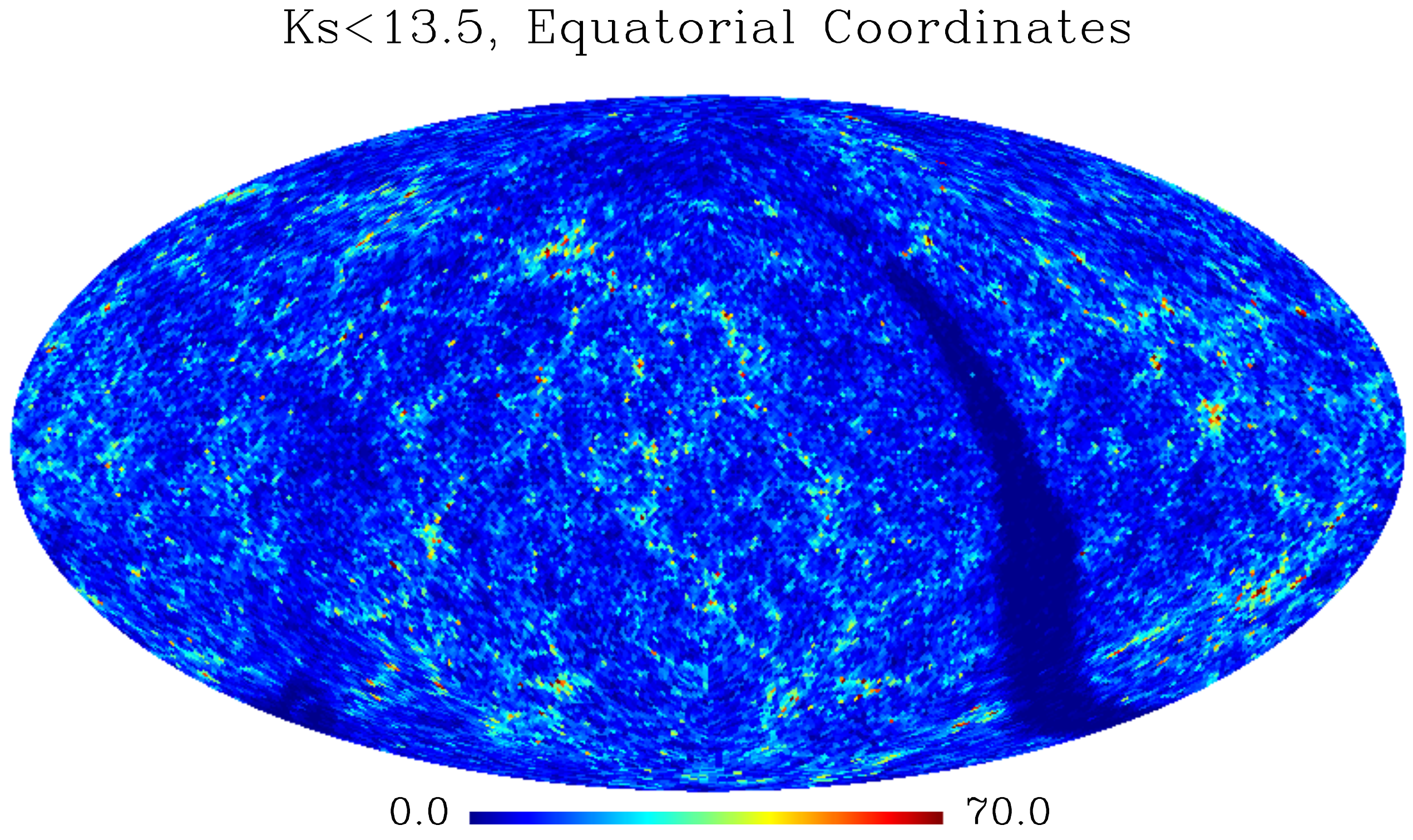}\\[0.5cm]
\includegraphics[width=.48\textwidth]{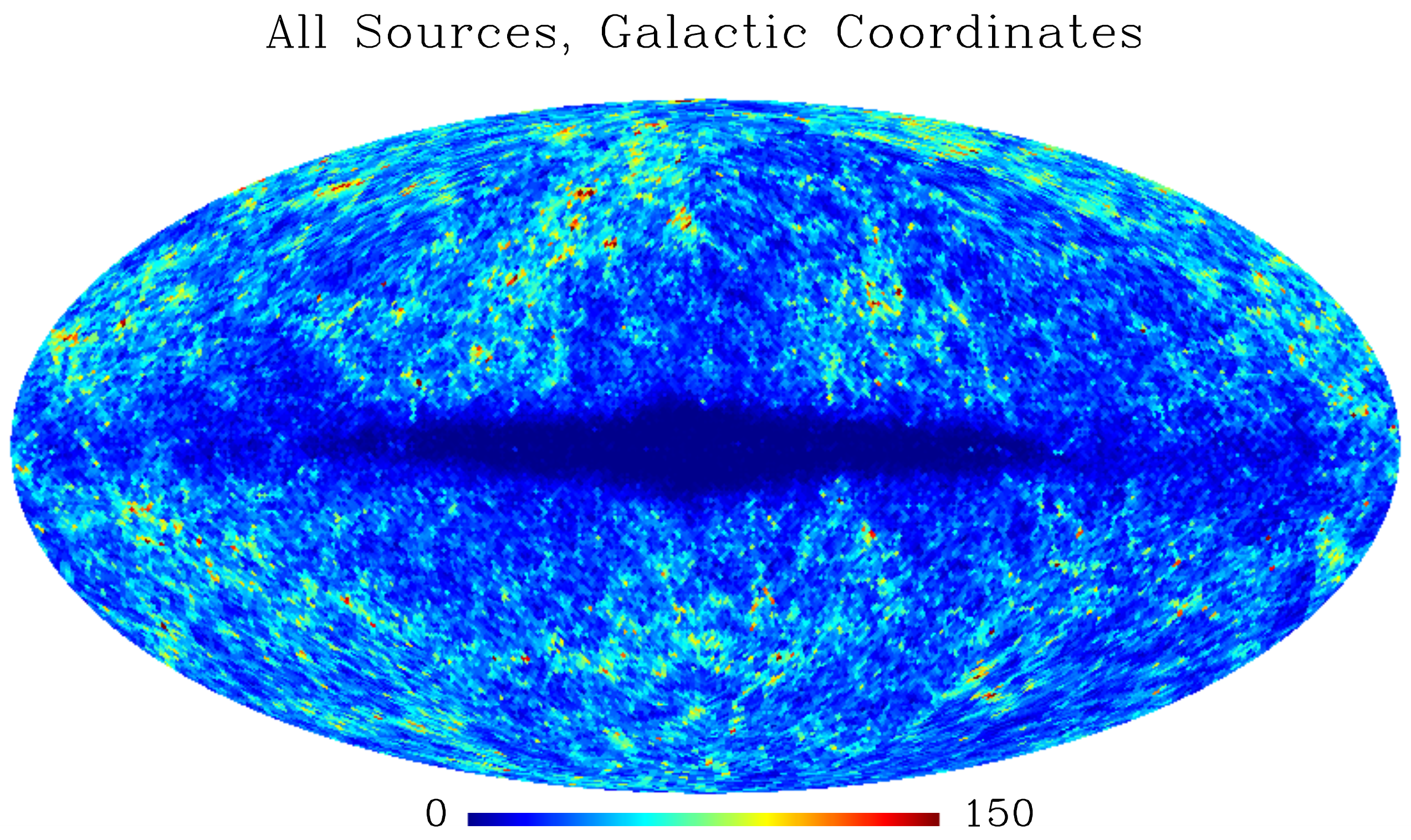}
\includegraphics[width=.48\textwidth]{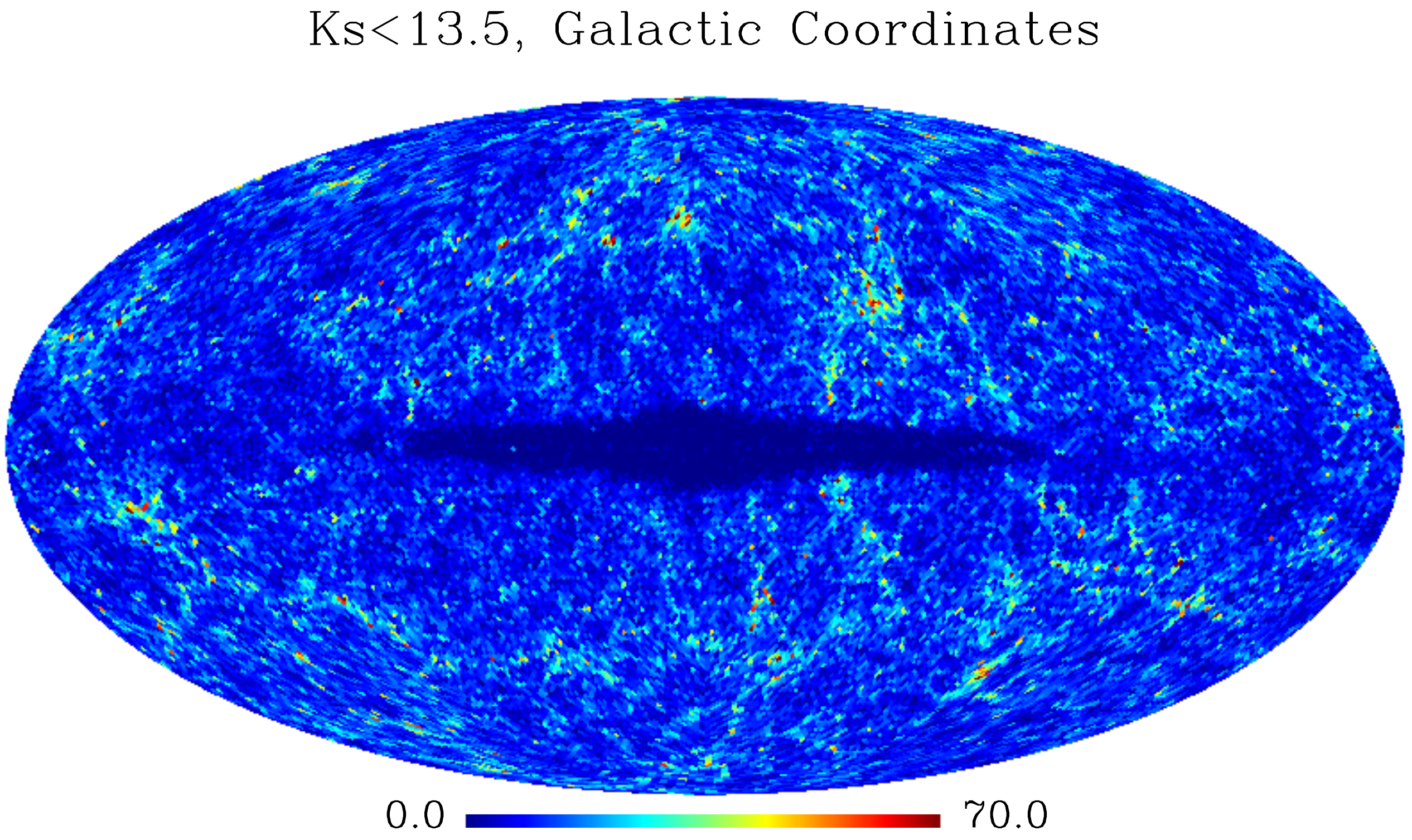}
\caption{\textit{Left side:} All 2MASS sources, in equatorial and Galactic
  coordinates. Note the very strong discontinuity in the selection function,
  visible in both images (especially in the top image, where it appears as a
  horizontal line), at declination around 20 degrees. \textit{Right side:}
  2MASS sources with $K_s$-band magnitude less than 13.5. The survey has nearly uniform completeness when this criterion is
  imposed. }
\label{fig:2massgalaxies}
\end{center}
\end{figure*}

We find the results in Table~\ref{tab4.4} when we use the usual $\blteight$
Galactic cut and also include a cut around the supergalactic plane ($\sgbge$
in the first column\footnote{Supergalactic coordinates SGB and SGL are defined
  in analogy to Galactic $b$ and $l$, where SGL is the azimuthal coordinate,
  and SGB$=$0 corresponds to the middle of the supergalactic plane, so we care
  only about making cuts in SGB.}) or a cut around the supergalactic poles
($\sgblt$ in the seventh column). The table is aligned so that cuts around the
supergalactic plane are in the same row as cuts of similar area around the
supergalactic poles. The $\fsky$ columns give the fraction of the sky that
remains when we perform the given cut. We also calculate $\fsources$, the
fraction of the total number $N$ of sources that remain when we perform the
given cut. The cuts in SGB (less than 2.0, 5.0, 10.0, 20.0 degrees; greater
than 74.82, 65.90, 55.73, 41.15 degrees) were chosen so that equal areas
around the plane and around the poles would be cut if there were no cut in
Galactic $b$ (as in our tests on the BATSE catalog; Sec.~\ref{sec5.2}). Since
there is a cut in Galactic latitude here, $\fsky$ does not match up exactly
between the cuts around the supergalactic plane and poles, but the values are
still close, and in any case they are normalized in the $\fsources/\fsky$
calculation; we employ the same cuts for all surveys tested in this paper.

From Table
  \ref{tab4.4}, we learn the following:

\begin{itemize}
\item{The fraction of sources associated with the supergalactic plane is
  greater than the fraction of sources associated with the supergalactic poles
  for every $\fsky$. This is our first indication that a greater-than-random
  portion of the dipole signal comes from the vicinity of the supergalactic
  plane.}
\item{The ratio $\fsources/\fsky$, which gives a measure of how overdense the
  uncut portion of the sky is, dwindles steadily as we cut more and more of
  the SGP, but increases as we cut more and more area around the supergalactic
  poles. Therefore, as expected, there are more sources near the supergalactic
  plane than near the supergalactic poles, and this is true essentially
  regardless of how much of the area around the plane/poles is cut.}
\end{itemize}

In general, we conclude that more of the dipole signal comes from the area of
the supergalactic plane than from the vicinity of the supergalactic
poles. This serves as a good check that the source of the dipole signal in the
relatively local structure surveyed by 2MRS is generally where we expect it to
be. When we perform analyses of higher-redshift objects in Sec.~\ref{sec5.2}
and \ref{sec5.3}, the supergalactic plane should not ``show up" as it has
here.

\subsection{2MRS: Conclusion}
\label{sec4.2.8}

We conclude our analysis of the dipole signal in 2MRS by pointing out that all
results are consonant with theoretical expectations. This comes as no surprise
given that 2MRS was the most well-controlled and well-understood of the
surveys we analyze here, and was being treated exhaustively as something of a
model for our other analyses.

It should also be noted that we have verified that the results above do not
change appreciably when corrections are made for the 2MASS coverage map (see
Sec.~\ref{sec4.3}).

We now proceed to apply our dipole analysis to the full 2MASS dataset.

\section{Dipole in 2MASS}
\label{sec4.3}

We now analyze the full 2MASS survey in a manner similar to how we analyzed
the very well-characterized 2MRS subsample. The challenges associated with
analyzing 2MASS as a whole are greater, in part because the full sample of
2MASS galaxies (1.6 million extended sources) does not have uniform
completeness across the entire sampled sky.

It should be noted that some previous results concerning the 2MASS galaxy
distribution stand in some tension with $\Lambda$CDM predictions. For example,
\citet{frith20052mass} point out that the angular correlation function and
angular power spectrum of 2MASS galaxies (under cuts reasonably similar to the
ones we perform here) display fluctuations that are 3-5 sigma out of line with
$\Lambda$CDM predictions. We focus attention here on the dipole alone, which
of course sacrifices a certain amount of information with respect to what
could be gained from analysis of the entire power spectrum, but also lends
itself to much better and more detailed analysis of contributions to the
signal at this one multipole.

\subsection{Selection Cuts}

We make several cuts to the sample of 2MASS galaxies in order to ensure
uniformity of the sample:

\begin{enumerate}

\item{As shown in Fig.~\ref{fig:2massgalaxies}, the biggest issue in connection with
survey completeness is that the selection function has a sharp discontinuity
for galaxies with $K_s$-band magnitude greater than roughly 13.5. We therefore
cut out all these sources, roughly 2/3 of the sample, at the outset, and
consider only that portion of the survey with nearly uniform completeness over
the entire sky (with the exception of the Galactic plane), that is, sources
with $K_s<13.5$.}

\item{We must make a more aggressive Galactic cut than we did for 2MRS in
  order to ensure that star-galaxy confusion does not come into
  play. \citet{maller2005galaxy} and \citet{skrutskie2006two} note that the
  2MASS XSC is highly reliable and complete for $|b| > 20^\circ$ (more than 98
  percent galaxies rather than stars at these latitudes), but that star-galaxy
  confusion is an increasingly large problem at lower latitudes: the XSC is 10
  percent stars for $5^\circ < |b| < 20^\circ$; and within the Galactic plane,
  $|b| < 5^\circ$, there is additional contamination by artifacts (10 to 20
  percent) and Galactic extended sources ($\sim 40$ percent) including
  globular clusters, open clusters, planetary nebulae, and giant molecular
  clouds (\citet{jarrett20002mass2}). In particular, Maller et
  al.~cross-correlate the 2MASS stellar density $n_{\rm star}$ with the XSC
  galaxy density as a function of the latitude of a symmetric (in Galactic
  coordinates) cut and find that including XSC objects with $\bltfifteen$
  gives a galaxy-star cross-correlation that is higher in amplitude than the
  galaxy-galaxy autocorrelation, suggesting the presence of multiple-star
  systems mistakenly identified as extended sources. However, this excess
  signal goes away for a cut of $\blttwenty$. Cutting at $\blttwenty$ ensures
  less than 2 percent contamination from Galactic sources
  \cite{maller2005galaxy,frith20052}.}

\item{We use the XSC confusion flag ({\tt cc\_flg}) to eliminate known
artifacts (diffraction spikes, meteor streaks, infrared airglow, etc.).}

\item{Again following \citet{frith20052}, as well as \citet{maller2005galaxy},
  we also cut out bright ($K_s<12.0$) objects with $(J-K_s)$ colors that are
  outside the range $[0.7, 1.4]$ (see \citet{bilicki2012motion} for
  alternative choices of which ranges to cut). This is a conservative measure
  designed to get rid of a final set of objects which are in the 2MASS XSC but
  which are not extragalactic sources. This removes a few thousand sources.}

\item{As explained later, we take the 2MASS sky coverage into
  account. The XSC does not have completely uniform sky coverage given the
  presence of bright stars and other foreground objects that make it more
  difficult for the telescopes to detect extended sources in particular
  directions. Although the pattern of sky coverage is parity-even (following
  the shape of the Galaxy) and unlikely to mimic a dipole in any way (as
  implied by the 2MRS analysis), we still take this into account in the
  present analysis.}

\item{K-corrections (corrections to magnitudes in a given
  passband that are made necessary by the fact that light can redshift into or
  out of a given range of wavelengths)
for the $K_s$ band can make a non-negligible difference in the calculation of
a flux-weighted dipole or other quantity that depends on specifics of
photometry. However, in this case
we do not need to take them into account, because they are actually accounted
for in our predictions: K-corrections are tied to the same (pseudo-)Doppler
effect that helps to generate the kinematic dipole (see Sec.~\ref{sec2.2}),
and so accounting for them in observational results as well would amount to
double-counting.}

\end{enumerate}

All photometric cuts are applied to 2MASS isophotal magnitudes -- not total
magnitudes, which are an extrapolated quantity and viewed as less reliable for
the purposes of this kind of analysis. While many analyses which use 2MASS
data actually use the extrapolated magnitudes (since, according to
\citet{jarrett20032mass}, the isophotal magnitudes underestimate total
luminosity by 10 percent for early-type and 20 percent for late-type
galaxies), we stick here with the more conservative isophotal magnitudes,
especially since the cut at $K_s<13.5$ is much surer to accomplish its purpose
if the more conservative magnitude estimates are used. It is worth noting that
the 2MRS team used isophotal magnitudes in their sample selection
(\citet{huchra20112mass}).

\begin{figure}[t]
\begin{center}
\includegraphics[width=.48\textwidth]{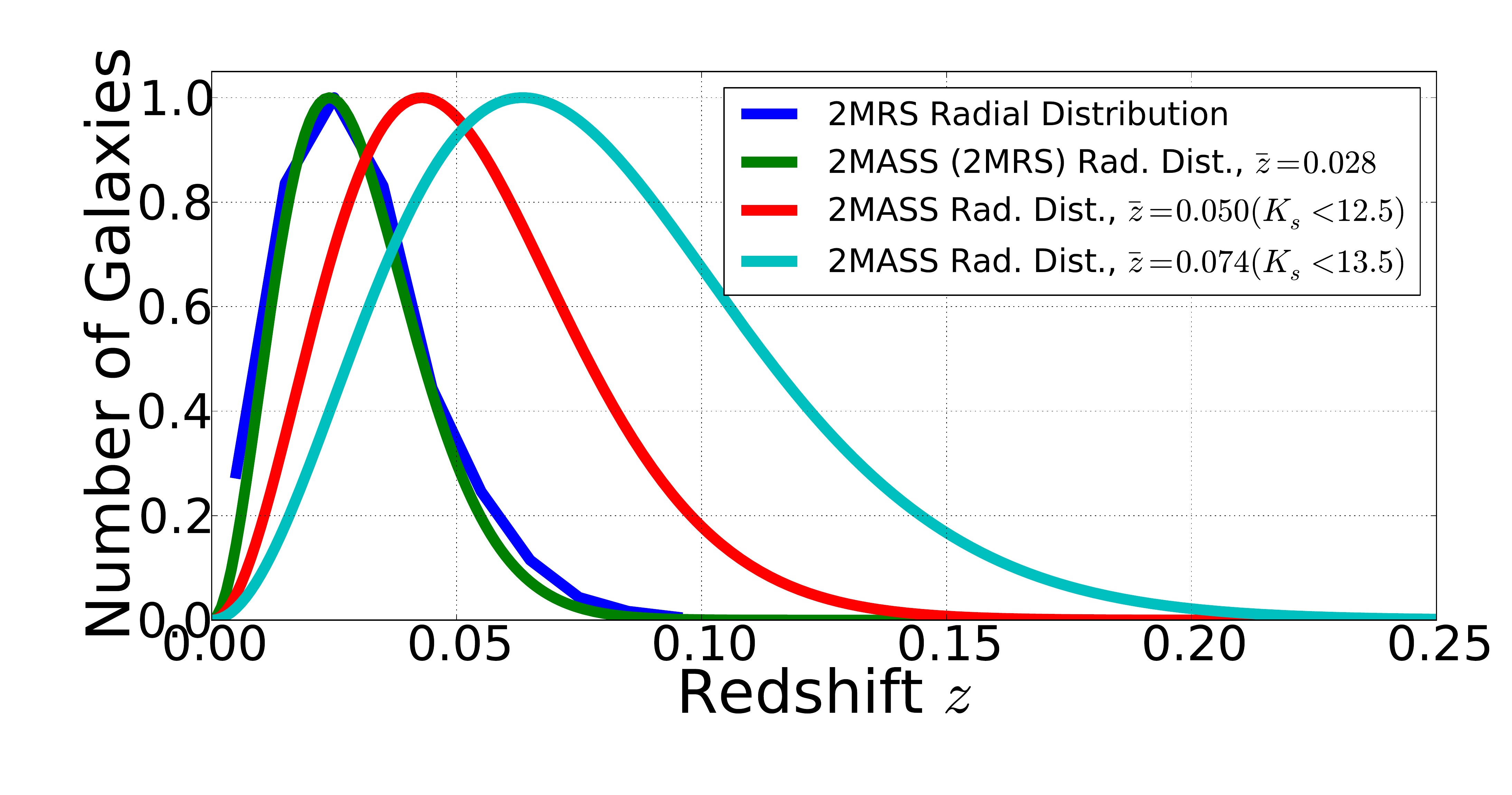}
\caption{Plot of the
  radial distribution of 2MASS galaxies as a function of redshift. Three
  different mean redshifts are shown, one which corresponds to the 2MRS
  distribution (the actual 2MRS $dN/dz$ is plotted in blue), and the other two
  of which correspond to photometric cuts in the full 2MASS survey of
  $K_s<12.5$ and $K_s<13.5$.}
\label{fig:2massdirectionFOS}
\end{center}
\end{figure}

\begin{table*}[t]
\caption{Comparison of dipole parameters without any templates, with SFD template, and with SFD,
    quadrupole and octopole templates, for 2MASS, for two different limiting $K$-band
magnitudes.\vspace{0.1cm}}
\label{tab4.5}
\centering
\setlength{\tabcolsep}{0.7em} 
\begin{tabular}{| c | c | c | c | c | c | c | c | c |}
\hline
\multicolumn{9}{|c|}{\rule[-3mm]{0mm}{8mm} For objects with $K_s<13.5$:}\\
\hline
\rule[-2mm]{0mm}{6mm} $|b| \ge$ & Template & NSIDE & $N$ & $A_{\rm peak}$ & $l$ & $b$ & 68
percent CI & 95 percent CI \\
\hline
\rule[-2mm]{0mm}{6mm} 20.0 & none & 128 &  386008 & 0.089 & 303.4 &   7.3 & 0.086 - 0.092
& 0.083 - 0.095 \\ \hline
\rule[-2mm]{0mm}{6mm} 20.0 & SFD  & 128 &  386008 & 0.088 & 305.0 &   4.5 & 0.085 - 0.091
& 0.082 - 0.094 \\ \hline
\rule[-2mm]{0mm}{6mm} 20.0 & SFD + Quad + Oct & 128 &  386008 & 0.104 & 268.4 &   0.0 & 0.100 - 0.108 & 0.096 - 0.112 \\ \hline

\multicolumn{9}{|c|}{\rule[-3mm]{0mm}{8mm} For objects with $K_s<12.5$:}\\
\hline
\rule[-2mm]{0mm}{6mm} $|b| \ge$ & Template & NSIDE & $N$ & $A_{\rm peak}$ & $l$ & $b$ & 68
percent CI & 95 percent CI \\
\hline

\rule[-2mm]{0mm}{6mm} 20.0 & none & 128 &   91008 & 0.0848 & 275.0 &  28.2 & 0.078 - 0.091
& 0.072 - 0.097 \\ \hline
\rule[-2mm]{0mm}{6mm} 20.0 & SFD & 128 &   91008 & 0.0812 & 276.3 &  25.9 & 0.075 - 0.088
& 0.069 - 0.094 \\ \hline
\rule[-2mm]{0mm}{6mm} 20.0 & SFD + Quad + Oct & 128 &   91008 & 0.134 & 267.3 &   8.5 & 0.126 - 0.142 & 0.117 - 0.150 \\ \hline

\hline
\end{tabular}
\end{table*}

\subsection{Radial Selection Function}

With these photometric cuts applied, we are ready to proceed with the
analysis. While no spectroscopic redshifts are available for the 2MASS XSC as
a whole, and the photometric redshifts that do exist are not particularly
reliable, considerable information is available about the overall radial
distribution of 2MASS galaxies. In particular, \citet{frith20052} and others
give the 2MASS radial selection function as 
\begin{equation}
\frac{dN}{dz}(z) = \frac{3 z^2}{2(\zbar/1.412)^3} \exp \left ( - \left ( \frac{1.412
  z}{\zbar} \right )^{3/2} \right )
\label{eq:Nz_2MASS}
\end{equation}
with $\zbar = 0.074$ for $K_s<13.5$ and $\zbar = 0.050$ for $K_s<12.5$. With
these values of $\zbar$, we can determine theoretical predictions for the
local-structure dipole (which is still dominant by two orders of magnitude
over other contributions to the dipole at these scales) for these two
photometric cuts. Combined with the 2MRS sample, which follows this same form
for the selection function quite closely (see
Fig.~\ref{fig:2massdirectionFOS}) and corresponds to approximately $\zbar =
0.028$, we can perform a comparison of theory and observation for multiple
subsamples of the entire 2MASS catalog.

For $K_s<12.5$ without (with) the $\blttwenty$ cut, there are 127,030 (89,980) galaxies. For
$K_s<13.5$ without (with) the cut, there are 542,201 (381,586) galaxies.

\subsection{Systematic Checks: Extinction}
\label{sec4.3.1}

In 2MRS, extinction corrections were already applied to the magnitudes of the
galaxies, but in 2MASS, the catalog values for the magnitudes are not
corrected for extinction. This means that it becomes much more important in
this case to make sure that we have adequately controlled for the effects of
extinction. Knowing the magnitudes is
  important to determine which objects get into the sample in the first
place. We find that several thousand galaxies that do not make the $K_s<13.5$
cut before extinction correction do make the cut when magnitudes are corrected
for extinction.\footnote{Note that extinction
corrections always bring magnitudes down, since sources appear dimmer due to
extinction, and so are assigned higher brightness/lower magnitude when
corrected for extinction.}

We have performed various extinction corrections, experimenting with slightly
different extinction coefficients $R = A_V/E(B-V)$ for the 2MASS $K_s$ band
(0.367 from \citet{ho2008correlation}; 0.302 from the analogous UKIRT value in
\citet{schlafly2010measuring}; cf. 0.35, which is used by
\citet{erdogdu2006dipole}, following \citet{cardelli1989relationship}). We
find that the results when extinction corrections are applied directly are
essentially identical to the results obtained when the SFD dust systematic
template is applied as it was to the 2MRS maps, so we explicitly present only
the SFD-template results here.


Table~\ref{tab4.5} shows the effects of the inclusion of the SFD template for
the $K_s<13.5$ and $K_s<12.5$ maps. As should be clearly visible from a quick
glance at the values in the table, very little changes when the SFD template
is included, and there is substantial overlap even of the 68 percent
confidence intervals for each of the no-template cases with each of the
corresponding SFD-template cases. We conclude that although the results shift
slightly, and therefore it is worth keeping the SFD template in our analysis,
extinction does not have a substantial impact on the dipole results. This is
as expected based on considerations of how extinction affects 2MASS coverage,
as outlined in \citet{jarrett20002mass1} and \citet{jarrett20002mass2}: the
completeness of the 2MASS XSC is much more adversely affected by source
confusion than by extinction.

In the same table, we also show the effect of including templates for the
quadrupole and octopole modes, and note that including these templates shifts
the peak dipole amplitude by more than 10 percent for $K_s<13.5$ and more than
50 percent for $K_s<12.5$. (Note also the changes in direction in each case,
which actually serve to bring into agreement the dipole directions for the two
different photometric cuts.) Based on the results presented in Appendix
\ref{appB}, it is not surprising that inclusion of these templates has more of
an effect than it did for 2MRS, since our sky cut is significantly more
aggressive here.



\subsection{Systematic Checks: Coverage Map}

Uniform completeness in an infrared survey like 2MASS is impossible due to the
presence of foreground stars. In some directions, the presence of foreground
stars makes observation of distant background galaxies impossible. Sky
coverage, which ranges from 0 to 1 within a given pixel, tends to be well
above 0.98 for the high-Galactic-latitude sky. Data products from 2MASS
include coverage maps\footnote{See
  \url{http://www.ipac.caltech.edu/2mass/releases/allsky/doc/sec2_6f.html}.}
that indicate coverage as a function of direction.

\begin{figure}[]
\begin{center}
\includegraphics[width=.48\textwidth]{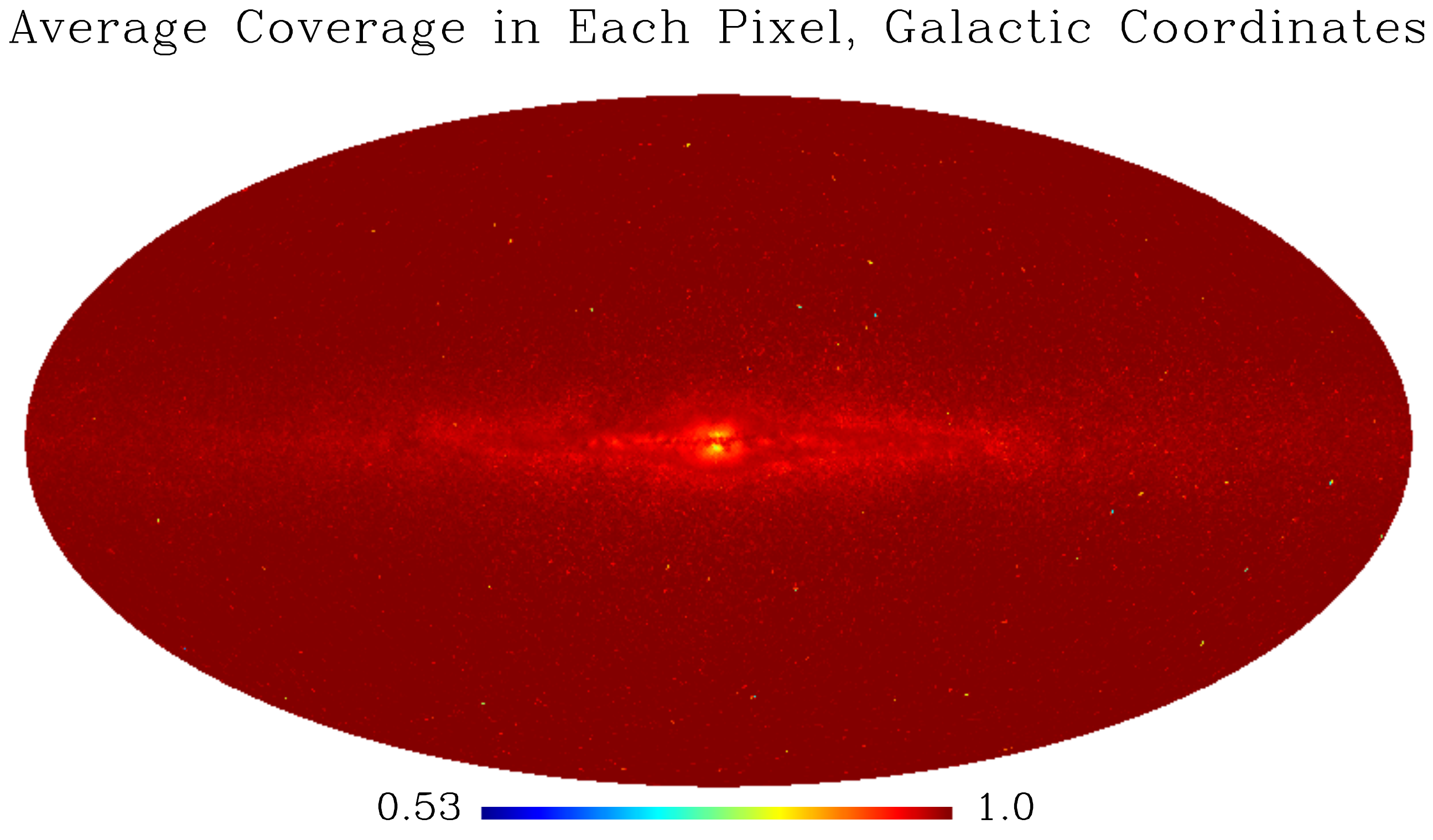}
\caption{2MASS sky coverage as a function of direction for the entire sky. Coverage can be zero (full
    masking) or one (no masking), or anywhere between these extremes. We use this
  map as a systematic template in the dipole formalism.}
\label{fig:coverage}
\end{center}
\end{figure}

We convert these maps into the same HEALPix pixelization scheme we use to
pixelize all the surveys in this paper, including 2MASS itself. Each HEALPix
pixel contains at least 4, and up to 19, ``subpixels" associated with the
pixelization of the 2MASS coverage maps, so
resolution is not an issue. See Fig.~\ref{fig:coverage}.

There are several ways in which we could take these coverage maps into account
in our analysis. First, we could mask out all HEALPix pixels that have an
average coverage less than some threshold (the threshold is usually chosen as
0.98 in the literature; see, e.g.,~\citet{ho2008correlation}). Second, we
could mask out all pixels that have \textit{any} subpixel with coverage less
than some threshold. Third, we could use the entire coverage map as a
systematic template. We have not found a case in which it makes anything even
close to a statistically significant difference which of these strategies we
choose, so we choose the option that is simplest and arguably best: we use the
entire coverage map as a systematic template. This has the advantage of not
privileging any particular threshold, but rather taking the variation in
coverage over the entire sky into account evenhandedly. This accounts for the
actual pattern of observations on the sky and weights them accordingly
(cf.~our treatment of the BATSE exposure function in Sec.~\ref{sec5.2}). In
any case, given that the coverage map closely follows the shape of the Galaxy
(foreground stars are, after all, our primary concern), and the Galaxy is
nearly parity-even, we do not expect coverage to contribute significantly to
the (parity-odd) dipole anyway; results are very much in accord with these
expectations.

In summary: We apply the coverage map throughout this section (as we did, in fact, for our 2MRS analysis as well), but find that the results do not change appreciably as a result.




\begin{figure}[]
\begin{center}
\includegraphics[width=.48\textwidth]{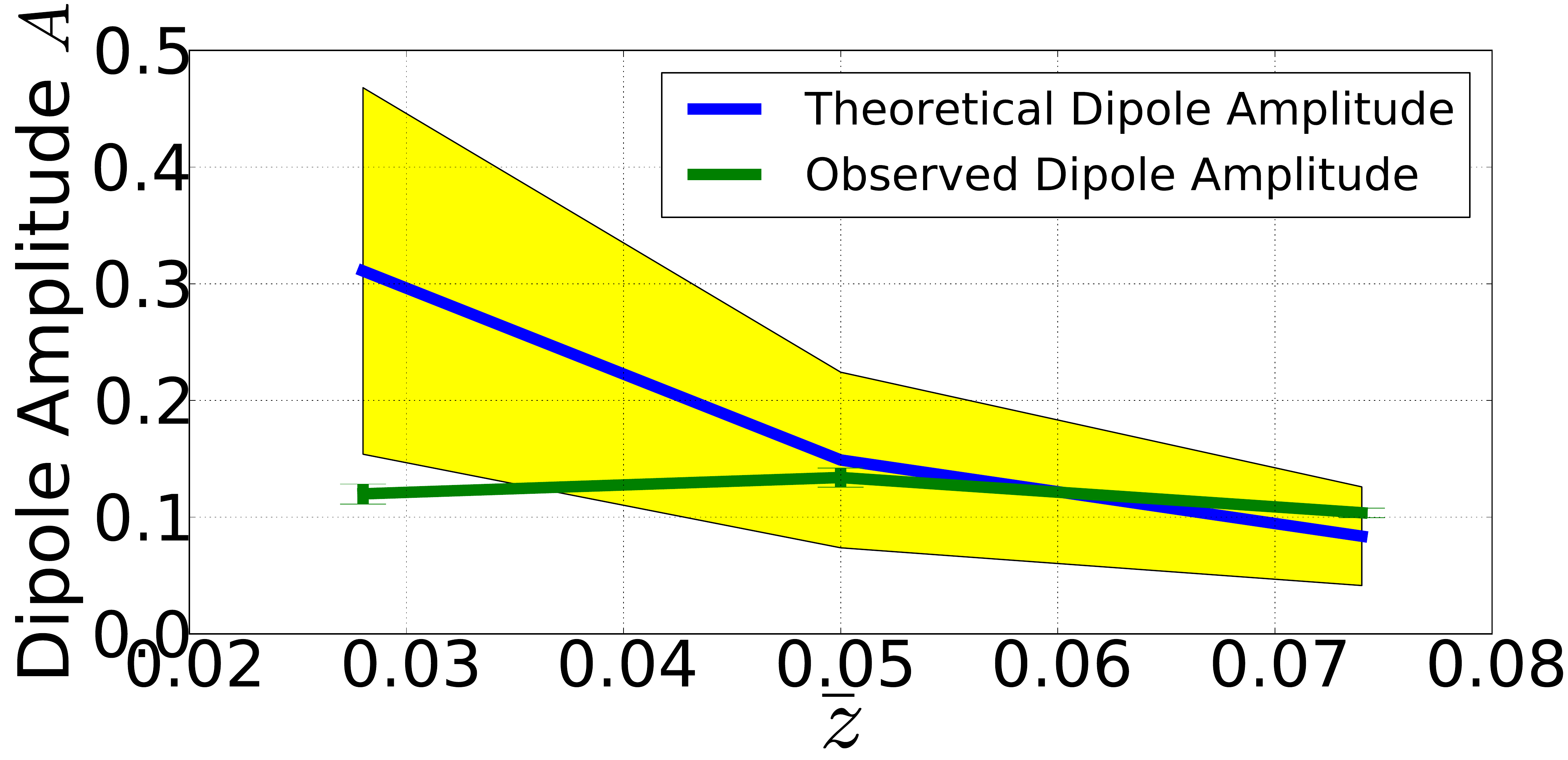}
\includegraphics[width=.48\textwidth]{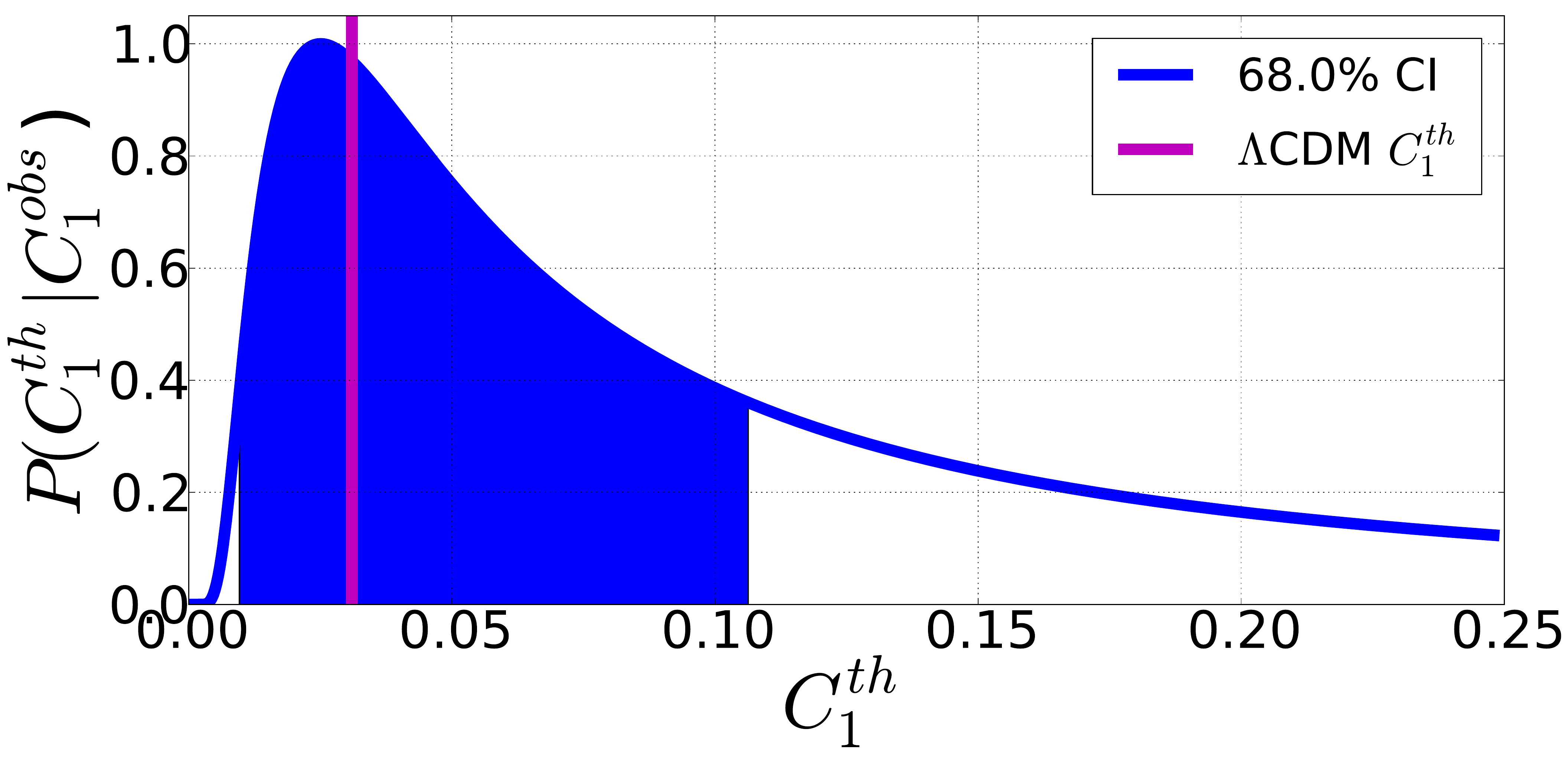}
\caption{\textit{Top panel:} Results for the dipole amplitude in the 2MASS
  survey, as a function of mean redshift $\zbar$ of the galaxy sample. The
  $\zbar=0.028$ sample corresponds to the 2MRS galaxies with $\blteight$ cut
  out; the $\zbar=0.050$ sample to $K_s<12.5$ with $\blttwenty$ cut out in the
  2MASS XSC; the $\zbar=0.074$ sample to $K_s<13.5$ with $\blttwenty$ again
  cut out in the 2MASS XSC. Cosmic variance (yellow band around the
  theoretical prediction) is shown in the most pessimistic $\blttwenty$ case
  for \textit{all} of the samples. \textit{Bottom panel:} Comparison of the
  $\Lambda$CDM value for $\Coneth$ with the expected distribution of $\Coneth$
  given the observed value for the $K_s<12.5$ sample. Both panels demonstrate
  sound agreement between theory and observation for all three subsamples of
  2MASS.}
\label{fig:amplitudezbar}
\end{center}
\end{figure}

\subsection{Systematic Checks: Sky Cut and Supergalactic Plane}
\label{sec4.3.2}

\begin{figure}[]
\begin{center}
\includegraphics[width=.48\textwidth]{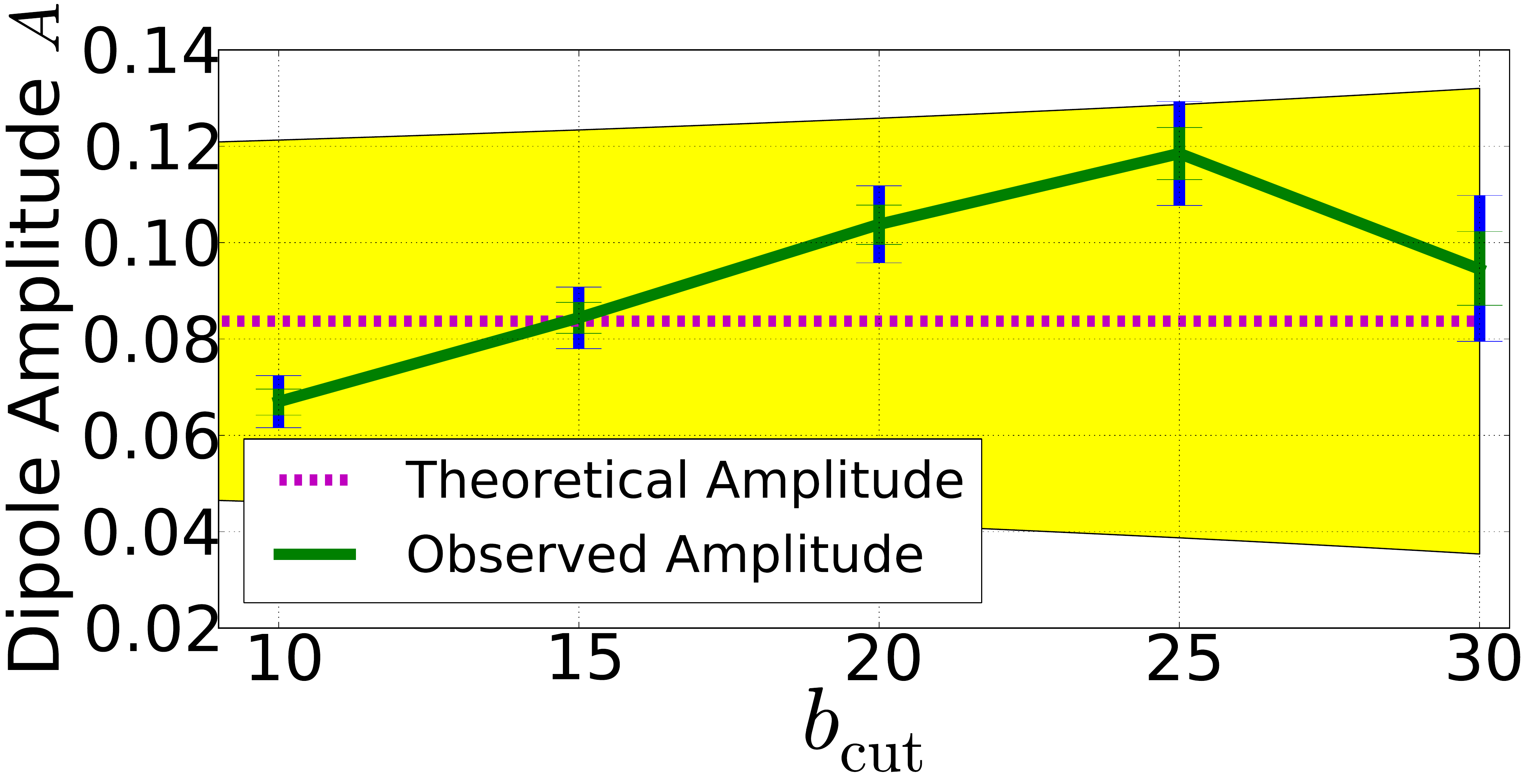}
\includegraphics[width=.48\textwidth]{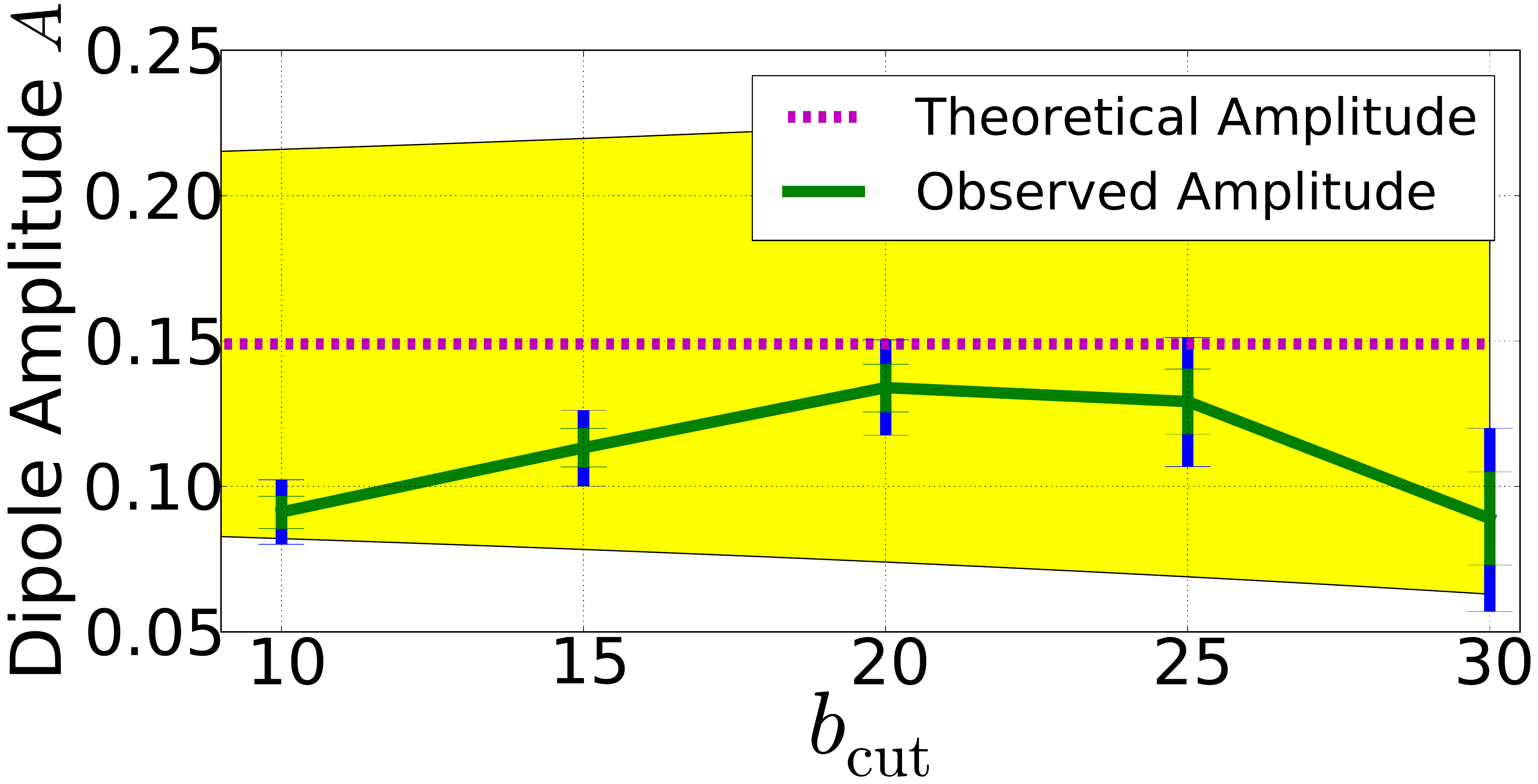}
\caption{ \textit{Top panel:} Dipole amplitude as a function of $\bcut$ for
  2MASS sources with $K_s<13.5$.  \textit{Bottom panel:} Same for 2MASS
  sources with $K_s<12.5$. See Fig.~\ref{fig:2mrsfunctionofbcut} for a fuller
  discussion of the significance of this type of plot.}
\label{fig:2massfunctionofbcut}
\end{center}
\end{figure}

We perform the same cuts in Galactic latitude as we did for 2MRS, but with the
expectation that the most reliable results will come for the $\blttwenty$ cut
rather than $\blteight$ as in the case of 2MRS. See Fig.~\ref{fig:2massfunctionofbcut} for a visual capture
of the results, with cosmic variance on the theoretical
prediction taken into account.

The basic conclusion here, as in the case of 2MRS, is that in no case are
results outside of the limits expected given cosmic variance. There is tension
between the $\bcut=10^\circ$ sample and other samples for both $K_s<13.5$ and
$K_s<12.5$, but that is to be expected given the much higher potential for
star-galaxy confusion when such low latitudes are included. The same could be
said of the $\bcut=15^\circ$ sample for $K_s<13.5$. In any case, cosmic
variance dominates the error budget for all cases, and dependence of results
on Galactic cut indicates no serious contamination from star-galaxy confusion
or other systematic effects that vary with Galactic latitude for samples with
Galactic $\blttwenty$, so we follow \citet{skrutskie2006two},
\citet{frith20052}, and others in taking this as our fiducial cut.

We also perform the same test as in the 2MRS case where we cut in
supergalactic latitude SGB. We create summary tables giving the key patterns,
as we did for 2MRS in Table \ref{tab4.4}. Results are presented for $K_s<13.5$
and $K_s<12.5$ in Table \ref{tab4.11}. The observations that we made for 2MRS
again (generally) hold here: in particular, $\fsources/\fsky$ decreases
monotonically as more and more of the supergalactic plane is excised, and the
ratio increases (almost) monotonically as more and more of the area around the
supergalactic poles is excised.

\begin{table*}[]
\footnotesize
\caption{Comparison of dipole parameters when performing various cuts in
  supergalactic coordinates, for 2MASS, for two different limiting $K$-band
  magnitudes. \vspace{0.1cm}}
\label{tab4.11}
\centering
\setlength{\tabcolsep}{0.7em} 
\begin{tabular}{| c | c | c | c | c | c || c | c | c | c | c | c |}
\hline
\multicolumn{12}{|c|}{\rule[-3mm]{0mm}{8mm} For objects with $K_s<13.5$:}\\
\hline
\rule[-3mm]{0mm}{8mm} $\sgbge$ & $\fsky$ & $N$ & $\fsources$ & $\frac{\fsources}{\fsky}$ & $A_{\rm peak}$ & $\sgblt$ & $\fsky$ & $N$ & $\fsources$ & $\frac{\fsources}{\fsky}$ & $A_{\rm peak}$ \\
\hline

\rule[-2mm]{0mm}{6mm} 0.0 & 0.65 & 386008 & 1.00 & 1.53 & 0.10 & --- & --- & --- & --- & --- & --- \\ \hline
\rule[-2mm]{0mm}{6mm} 2.0 & 0.63 & 368077 & 0.95 & 1.52 & 0.09 & 74.82 & 0.65 & 385740 & 1.00 & 1.53 & 0.10 \\ \hline
\rule[-2mm]{0mm}{6mm} 5.0 & 0.59 & 342390 & 0.89 & 1.51 & 0.08 & 65.90 & 0.64 & 378368 & 0.98 & 1.53 & 0.10 \\ \hline
\rule[-2mm]{0mm}{6mm} 10.0 & 0.52 & 301028 & 0.78 & 1.50 & 0.08 & 55.73 & 0.60 & 355464 & 0.92 & 1.53 & 0.07 \\ \hline
\rule[-2mm]{0mm}{6mm} 20.0 & 0.39 & 225502 & 0.58 & 1.49 & 0.13 & 41.15 & 0.49 & 295090 & 0.76 & 1.55 & 0.07 \\ \hline

\multicolumn{12}{|c|}{\rule[-3mm]{0mm}{8mm} For objects with $K_s<12.5$:}\\
\hline
\rule[-3mm]{0mm}{8mm} $\sgbge$ & $\fsky$ & $N$ & $\fsources$ & $\frac{\fsources}{\fsky}$ & $A_{\rm peak}$ & $\sgblt$ & $\fsky$ & $N$ & $\fsources$ & $\frac{\fsources}{\fsky}$ & $A_{\rm peak}$ \\
\hline

\rule[-2mm]{0mm}{6mm} 0.0 & 0.65 & 91008 & 1.00 & 1.53 & 0.13 & --- & --- & --- & --- & --- & --- \\ \hline
\rule[-2mm]{0mm}{6mm} 2.0 & 0.63 & 86657 & 0.95 & 1.52 & 0.11 & 74.82 & 0.65 & 90951 & 1.00 & 1.53 & 0.13 \\ \hline
\rule[-2mm]{0mm}{6mm} 5.0 & 0.59 & 80177 & 0.88 & 1.50 & 0.10 & 65.90 & 0.64 & 89203 & 0.98 & 1.53 & 0.12 \\ \hline
\rule[-2mm]{0mm}{6mm} 10.0 & 0.52 & 70212 & 0.77 & 1.48 & 0.10 & 55.73 & 0.60 & 83805 & 0.92 & 1.53 & 0.10 \\ \hline
\rule[-2mm]{0mm}{6mm} 20.0 & 0.39 & 52528 & 0.58 & 1.48 & 0.14 & 41.15 & 0.49 & 69923 & 0.77 & 1.55 & 0.10 \\ \hline

\hline
\end{tabular}
\end{table*}

\subsection{Dipole Amplitude as a Function of Redshift/Photometric Cuts}
\label{sec4.3.3}

Taking as most reliable the case with a cut for $\blttwenty$, and keeping the
SFD, quadrupole, and octopole templates in place, we proceed to compare
theoretical predictions with observational results for the dipole amplitude
for the two different magnitude cuts we have used, $K_s<13.5$ and $K_s<12.5$,
which correspond to $\zbar=0.074$ and 0.050, respectively (where, recall,
$\zbar$ is defined under Eq.~(\ref{eq:Nz_2MASS})). Again, the 2MRS sample
corresponds to $\zbar=0.028$, and we include this data point in our
comparisons as well.

See Fig.~\ref{fig:amplitudezbar} for results. Note that measurement errors are
once again tiny in comparison with cosmic-variance errors.

The magnitude of the dipole in 2MASS has, of course, been calculated
previously as part of computations of the entire power spectrum. In
particular, \citet{frith20052} give $C_1\approx 0.004$ for both $K_s<13.5$ and
$K_s<12.5$. This value converts to $A \approx 0.054$; cf.~our values of $A =
0.104 \pm 0.004$ and $A = 0.134 \pm 0.008$ (68 percent confidence) for
$K_s<13.5$ and $K_s<12.5$.  While both our measurement and the Frith et
al.\ measurement are in two-sigma cosmic-variance agreement with theory, the
\textit{mutual} discrepancy between these two observational results is
noteworthy. While we do not fully understand this discrepancy with the Frith
  et al.\ result, we note that our results are in better agreement if we do
  {\it not} marginalize over the dust map and the quadrupole and octopole templates.

\subsection{Dipole Direction as a Function of Redshift/Photometric Cuts}
\label{sec4.3.4}

\begin{figure}[]
\begin{center}
\includegraphics[width=.48\textwidth]{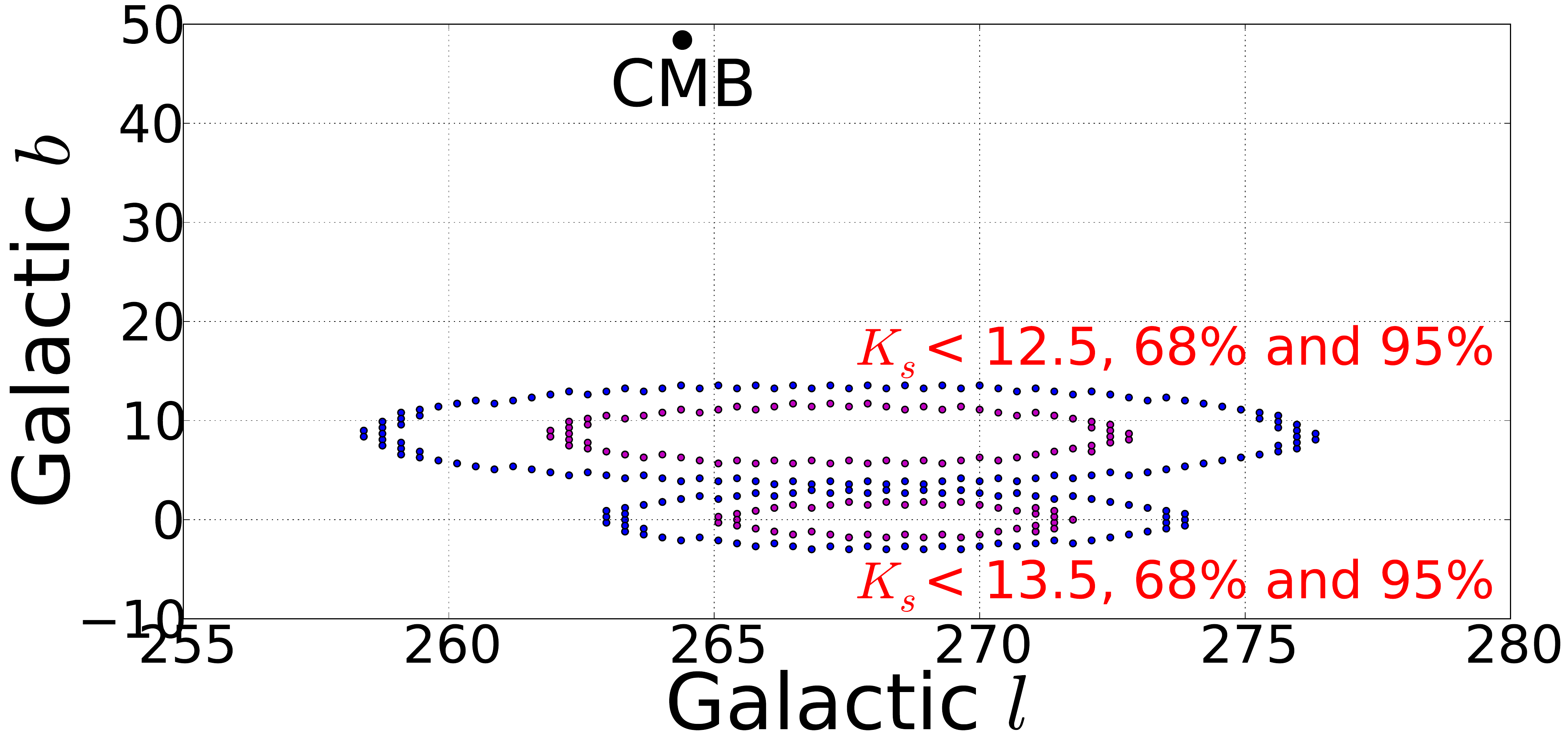}
\caption{Confidence intervals to go with results for the dipole direction in
  the 2MASS XSC, $K_s < 13.5$ (smaller circles) and $K_s<12.5$ (larger
  circles). In both cases we apply a cut eliminating $\blttwenty$, and apply
  the SFD map and quadrupole and octopole maps as systematic templates. The
  CMB kinematic dipole direction is indicated. Like 2MRS, 2MASS
    is too shallow to expect agreement between its dipole direction and the direction of the
    CMB dipole, so this is not an anomalous result.}
\label{fig:2massdirection}
\end{center}
\end{figure}

In Fig.~\ref{fig:2massdirection}, we present the results for the dipole
direction in both the $K_s<13.5$ and more conservative $K_s<12.5$ cases. Once
again, dipole amplitudes are on the order $10^{-1}$ while the kinematic dipole
is expected to be on the order $10^{-3}$, so no particular agreement with the
direction of the CMB dipole is expected. We do expect the $K_s<12.5$ sample to
give a direction relatively close to that of the 2MRS dipole, given the
overlap in the samples, and $K_s<13.5$ to give a result close to
$K_s<12.5$. In fact, we would regard it as anomalous if the results were not
all consistent with one another, \textit{if} the samples were genuinely
sampling the same population -- larger samples would simply have smaller error
bars than smaller samples. However, the populations being sampled are
different, given that the structure associated with $\zbar = 0.028, 0.050,$
and 0.074 are quite distinct from one another. So even internal
inconsistencies (between different values of $\zbar$) are tolerable. Indeed,
we find that the direction of the $K_s<13.5$ dipole is not fully consistent
with the direction of the $K_s<12.5$ dipole, but especially when we include
the quadrupole and octopole modes as systematics templates, as we do in
Fig.~\ref{fig:2massdirection}, the inconsistency is very mild.

\subsection{2MASS: Conclusion}
\label{2massconclusion}

We draw the basic conclusion that there are no anomalous results in applying
tests of dipole amplitude and direction to subsets of the 2MASS dataset and
comparing these results with theoretical predictions. We now proceed to more
critical tests using higher-redshift objects that might begin to probe the
kinematic dipole, as these objects exist at scales on which the
local-structure dipole should have become comparable to, or even smaller than,
the kinematic dipole.

\section{Difficulties With Using
     X-ray Surveys}
\label{sec5.1}

 We begin by considering surveys that detect very
high-energy photons (X-ray and gamma-ray), and then address the opposite end
of the spectrum (radio).

First, a brief note on X-ray data. Flux-weighted dipoles have been previously
calculated using the soft X-ray band ($<2$ keV) data from ROSAT and the hard
X-ray (2-10 keV) background data as observed by HEAO1-A2. For example,
\citet{plionis1999rosat} use the 1.5 keV ($\sim 0.8$ nm) ROSAT All-Sky Survey
(RASS) data to calculate a flux-weighted dipole, and \citet{scharf20002}
perform a similar analysis on HEAO1-A2. Contamination is a major issue for
both analyses. At the time of publication of these studies, only roughly
three-quarters of the unresolved X-ray flux in the soft and hard bands had
been accounted for (by extrapolation of objects resolved in deep fields)
\cite{scharf20002}, and theoretical modeling of populations contributing to
the unresolved flux (AGNs, starburst galaxies, hot IGM in rich clusters, etc.)
remained difficult. In the case of hard X-rays, Scharf et al.~argue that at
least a third of the structure in the data may be Galactic in origin
(associated especially with the bulge), and soft X-rays are even more strongly
contaminated by Galactic emission. \citet{treyer1998large} point out that in
the soft band, Galactic emission is present as a contaminant at all scales;
the hard band is better, but Treyer et al.~still rely on the Galactic
hard-band emission model of \citet{iwan1982large} to predict that for Galactic
latitude $>20^\circ$, the variations in flux due to Galactic emission are less
than 3 percent.

All told, it is very difficult to remove the foreground in X-ray all-sky
surveys successfully without some relatively uncertain modeling, without
fairly serious suspicion of contamination, and without removing a good deal of
the background too (especially in the soft band; \citet{plionis1999rosat}
estimate that Virgo contributes as much as 20 percent to the dipole amplitude
in RASS). Theoretical predictions would also be difficult to make without a
well-understood redshift distribution, especially given that the populations
contributing to the X-ray background are not especially well-modeled. Hence we
stay away from attempting to perform X-ray analyses here.

Some authors have searched for a dipole in the XRB to the extent that it is
possible to do so. The results that Scharf et al.~find in hard X-rays for the
flux-weighted dipole basically align with theoretical predictions of what they
refer to as the Compton-Getting effect (another name for the kinematic dipole,
following a paper by \citet{compton1935apparent} on the effect on cosmic-ray
intensity of Earth's motion through the Milky Way). More recently,
\citet{boughn2002large} analyzed the same (HEAO1-A2) dataset and found a limit
(95 percent confidence) on the amplitude of any intrinsic dipole at $5 \times
10^{-3}$. However, given the difficulty of definitively separating
extragalactic from foreground/Galactic emission in this dataset, and other
problems already noted, significant uncertainty attends any analysis in
X-rays, so we note all these results without making heavy use of them in the
remainder of this paper. Similarly, other populations of objects detected at
the very high-energy end of the spectrum, including blazars and clusters of
galaxies detected with gamma-ray satellites (see,
e.g.,~\citet{ando2007angular}), may be good targets for dipole searches in the
long term, especially once their bias is better-understood and future surveys
provide better statistics for the given target population; but we do not
pursue those here.

\section{Dipole in BATSE Gamma-Ray Bursts}
\label{sec5.2}

\begin{figure}[t]
\begin{center}
\includegraphics[width=.48\textwidth]{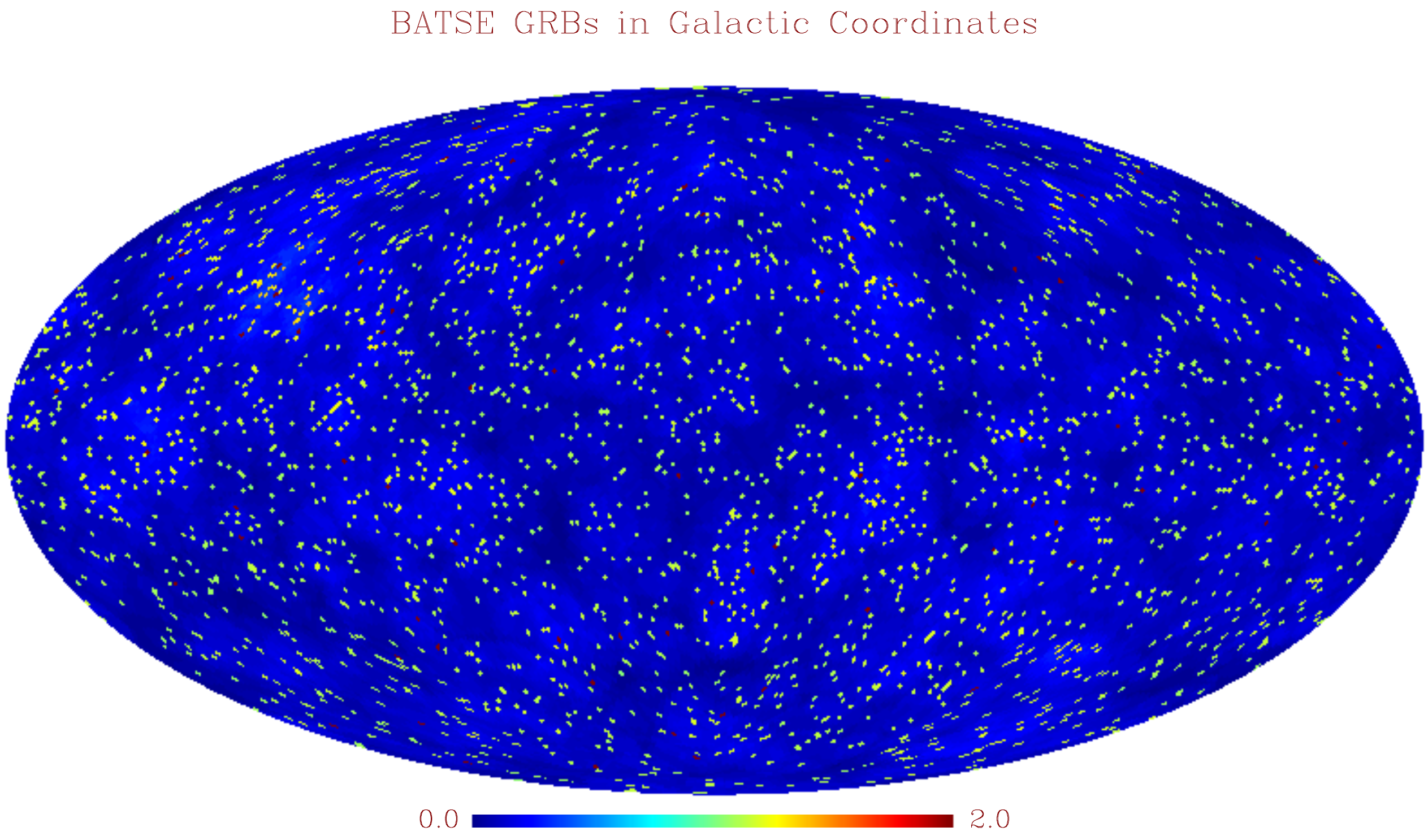}\\[0.2cm]
\includegraphics[width=.48\textwidth]{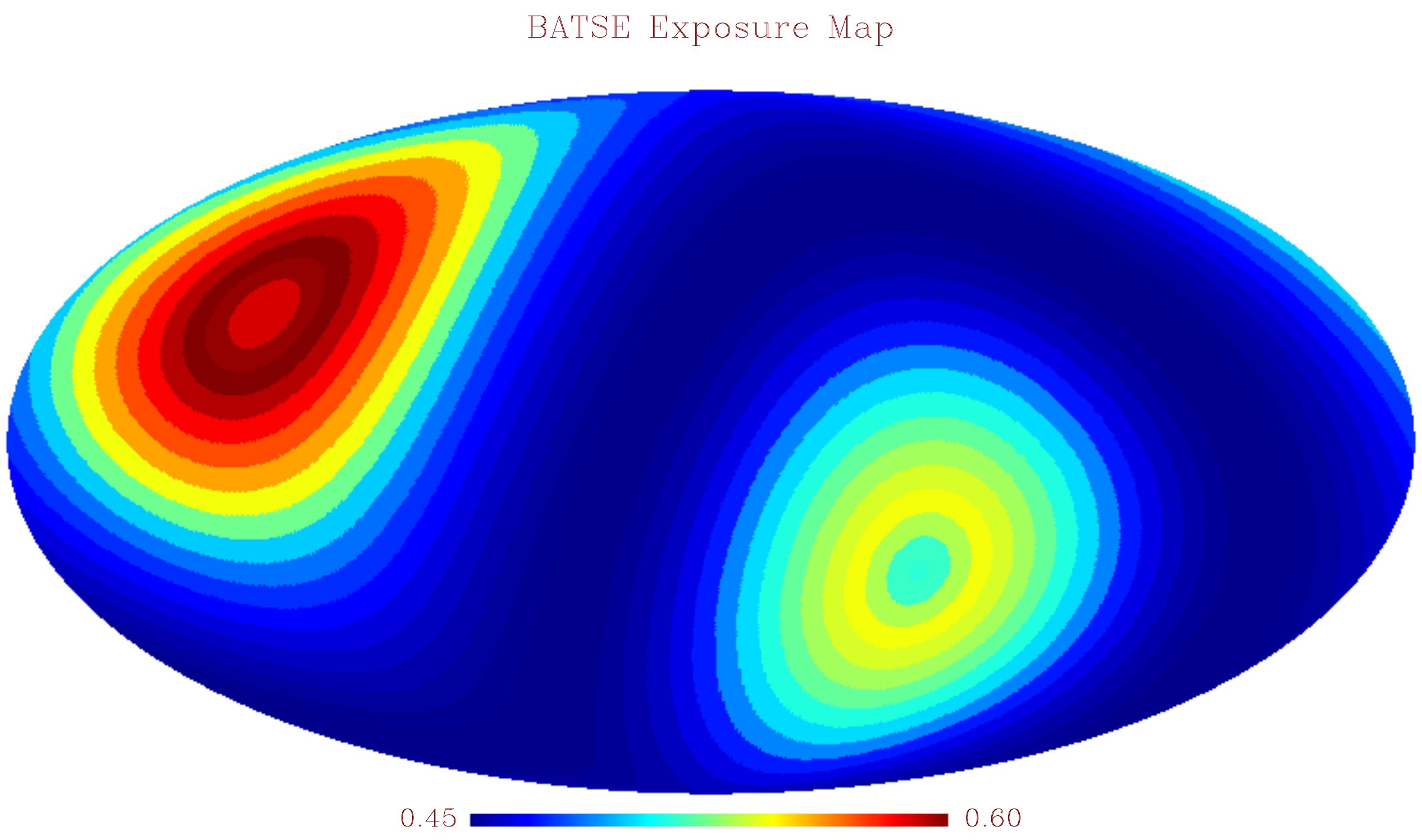}
\caption{\textit{Top panel:} GRB positions as recorded by BATSE, with error
  bars indicated as light circles/ovals around the GRBs; pixellized at
  NSIDE$=$64 (note that the dynamic range is limited to 2 even though a few
  pixels have 3 or 4 counts in them). \textit{Bottom panel:} The BATSE
  exposure function, which is proportional to the amount of time spent
  monitoring a given direction and varies with declination, in Galactic
  coordinates.}
\label{fig:grbpositions}
\end{center}
\end{figure}

For analysis of a 2D-projected dipole (i.e., not flux-weighted; see
Sec.~\ref{sec2.2}) in very high-energy surveys, we turn instead to gamma-ray
bursts (GRBs). GRBs are the most powerful explosions known in the universe, though
their exact nature and progenitor objects remain under some debate, and their
redshifts are difficult to measure since GRB observations are not
well-localized (see Fig.~\ref{fig:grbpositions}) and redshifts can only be
measured from their afterglows, which must be matched up with the position of
the original GRB, a highly nontrivial task given the error bars on the
position of a typical GRB. A review of previous research on the dipole in the
GRB distribution, as well as presentation of results using the formalism
outlined in Sec.~\ref{sec3.1} and applied to the BATSE
catalog, is presented in Sec.~\ref{sec5.2}. In Sec.~\ref{sec5.3}, we move to
the low-frequency end of the spectrum and present results from the NRAO VLA
Sky Survey (NVSS).

\subsection{Previous Work on the Isotropy of GRBs}
\label{sec5.2.1}

Up through the mid-1990s, there was a long history of assessing the isotropy
of gamma-ray bursts (GRBs) in an attempt to infer whether they were
cosmological or Galactic sources (or a combination thereof). For example,
\citet{maoz1993indications} argued in 1993 that gamma-ray bursts could be
shown to exist in an extended Galactic halo, some 130-270 kpc away from Earth,
by detecting slight but well-defined deviations from spherical symmetry
predicted for such a halo population. While his analysis did indeed suggest
that GRBs were nearby intergalactic objects, he argued that comparison with
more specific models would be necessary before considering the case closed.

In a similar spirit, \citet{briggs1995batse} argued persuasively that the
population of GRBs could not be Galactic, based on their observed
isotropy. This study found that the Galactic dipole and quadrupole moments
(calculated very straightforwardly as $\la\cos\theta\ra$ and $\la\sin^2 b -
1/3\ra$) did not differ significantly from those predicted for an isotropic
distribution. The majority of GRB models that assumed GRBs are a Galactic
population were found to be in $>2\sigma$ tension with the detected dipole and
quadrupole moments, and hence the conclusion of this research was that GRBs
are more isotropic than observed Galactic populations, suggesting either a
nearby intergalactic or, more likely, cosmological source.

\citet{scharf1995measurement} computed a fluence-weighted
dipole (where fluence is flux integrated over the timespan of the burst) in
analogy to the flux-weighted dipoles discussed in previous sections of this
paper. Combining fluence-weighted dipole information with straightforward
2D-projected dipole measurements (i.e., including photon count information)
better distinguishes a velocity dipole (due, as usual, to the Doppler effect
and relativistic aberration) from other possible sources of anisotropy. This
kind of test can be regarded as a supplement to the kinds of tests we perform here.

The current consensus that GRBs are cosmological is based not only on the
considerations discussed above and the absence of even a weak band
corresponding to the Milky Way in the GRB distribution
(\citet{tegmark1995angular}; see also \citet{Balasz}), but also (and
especially) on the observation of optical, X-ray, and radio counterparts to
GRBs that are clearly extragalactic (e.g.,~\cite{paciesas1999fourth,
  van1997discovery, fruchter1999hubble, metzger1997spectral}). Given the
extragalactic origins of GRBs, we should expect a GRB dipole sourced by the
same effects that give rise to a dipole in other sources we have analyzed. As
far back as the mid-1990s, \citet{maoz1994expected} predicted the dipole in
the clustering of GRBs, combining the effects of relativistic aberration and
the Doppler effect. These estimates are somewhat uncertain, but we too are
unsure how precisely GRBs trace large-scale structure (e.g., if they can be
described as a single population with a single bias, and so forth). Maoz finds
that the amplitude of this (kinematic) dipole is of order $A \sim 10^{-2}$ (to
within uncertainties of a factor of two), which is still an order of magnitude
larger than the CMB dipole, but much closer than the sources we dealt with in
the previous section. Maoz estimated that a large (of order $10^4$) sample of
GRBs would be necessary to detect the predicted dipole, and given that current
catalogs offer only on the order of $10^3$ bursts, we do not expect an
unequivocal detection. That said, we nevertheless run our tests, given that
useful constraints can be placed on the maximum possible dipole amplitude even
if we cannot confidently detect a dipole in currently available GRB catalogs.

\subsection{BATSE Data and Systematic Effects}

We perform our tests on GRBs in the BATSE catalog, data taken by the Compton
Gamma-Ray Observatory. The BATSE (Burst and Transient Source Experiment)
instrument onboard CGRO detected gamma-ray bursts within the nominal range of
50 to 300 keV. Other GRB datasets are available, including those from SWIFT
and Fermi (known previously as GLAST), but we use the BATSE catalog because it
has the most sources for an all-sky survey.

The most recent catalog of GRBs from BATSE is the ``current" catalog with 2702
sources. In this catalog, there are several complicating details that are
worth noting in regard to what bursts make it into the catalog and which do
not.

BATSE employed scintillators sensitive to gamma rays from $\sim 25$ to 2000
keV. A burst triggered the instrument when gamma-ray count rates exceeded some
minimum threshold relative to background in two or more of the eight detector
modules, within some energy range \cite{paciesas1999fourth}. Nominally,
that energy range was 50-300 keV, but on the order of 30 percent of BATSE's
observing time was spent with one of several trigger energy ranges different
from this. In addition, while the minimum detection threshold in count rates
relative to background was 5.5 sigma as a baseline, this value also changed
many times over the course of the experiment, and was not always the same for
different time intervals (BATSE tested count rates at 64, 256, and 1024 ms
intervals).

A different trigger energy range essentially represents a distinct burst
experiment, and a different detection threshold also changes the parameters of
the experiment in an important way. However, we argue that the time variation
in BATSE's ability to detect gamma-ray bursts is not sufficiently great to
affect our results on the dipole, especially given the lack of statistics for
the BATSE sample of 2702 bursts. \citet{kommers1997search} performed a search
for gamma-ray bursts and other gamma-ray transient phenomena with peak fluxes
below (by a factor of $\sim 2$) the flux necessary to count as a detection,
and also with energies outside the nominal 50-300 keV. They found that the
direction and intensity distributions of 91 likely GRB candidates \textit{not}
included in the final BATSE catalog imply that biases associated with the
trigger mechanism do not significantly affect the completeness of the
catalog. In addition to this result, there is no reason to expect that changes
in experimental parameters would have a particular effect on the dipole
quantity we investigate here. We note especially that changing trigger
criteria and energy ranges do not appreciably increase the chances that GRBs
will be confused with Galactic sources (e.g., soft gamma repeaters (SGRs)), so
contamination remains minimal; and any changes in trigger criteria apply
uniformly over the entire sky, so there is no obvious reason why this would
induce a dipole pattern. We therefore proceed to analyze the full catalog
without accounting for these changes.

However, one other experimental parameter is important for our dipole
analysis: sky exposure in BATSE varies as a function of declination (BATSE
spent different amounts of time looking at different declinations). We create
a template out of the exposure function (see Fig.~\ref{fig:grbpositions}) and
use this template as one of the systematic templates $t_i$ in the dipole
formalism outlined in Sec.~\ref{sec3.1}. This corresponds to weighting pixels
according to how much time the satellite spent observing a given area of the
sky: see the approach in, for example, \citet{scharf1995measurement}. The
choices we make here, to ignore changes in trigger criteria but take the
exposure function into account, corresponds to the choices made in the paper
by \citet{tegmark1995angular} calculating the angular power spectrum of the
BATSE 3B catalog, which found no evidence of deviations from isotropy on any
angular scale. Our present tests can be regarded as updates (since we use the
current catalog, which more than doubles the number of bursts in the 3B
catalog) of Tegmark's, with focus on the dipole (direction as well as
magnitude) using a real-space estimator that more naturally incorporates sky
cuts and systematic templates. (Note that Tegmark et al.~also impose a
weighted averaging scheme in harmonic space to account for the very large
position errors associated with GRBs, which are on the order of degrees,
orders of magnitude larger than typical position errors associated with
galaxies. This is unnecessary in our case given that the dipole probes scales
much larger than the uncertainties in GRB positions.)

The redshift selection function for GRBs is still only poorly understood,
though better statistics are consistently being built up. GRBs come from even
higher redshifts on average than NVSS sources (see Sec.~\ref{sec5.3})
(e.g.,~\citet{xiao2009estimating}), however, so we can confidently say that
regardless of the precise distribution, the local-structure dipole will be
subdominant in comparison with the kinematic dipole (given that it is already
subdominant for the NVSS sources discussed in the next section), and so we consider only the kinematic dipole as
a theoretical expectation below.


\subsection{Systematic Checks and Dipole Amplitude}
\label{sec5.2.2}

The positions of GRBs detected by BATSE are shown in
Fig.~\ref{fig:grbpositions}. They are not very well-localized; the
positions typically have error bars on the order of degrees. The GRBs do not
appear to cluster in any particular way by eye, but we apply our usual tests
to see whether this holds up rigorously.

In considering what systematic templates to put in place, maps of Galactic
foregrounds are unnecessary. In particular, inclusion of the SFD dust template
is unnecessary since gamma rays are highly penetrating and not subject to
appreciable dust extinction. We have explicitly verified that the difference
between the results including vs.~not including SFD template is
completely negligible.
In principle, there is the possibility of confusion with soft gamma repeaters
(SGRs) or other sources of gamma rays (pulsars, terrestrial gamma-ray flashes,
black holes, etc.). This is highly unlikely given that GRBs are easy to
distinguish from other gamma-ray sources based on spectral and time-domain
data. However, any foreground objects that might contaminate a pure GRB sample
are expected to vary with Galactic latitude, and since there is no reason not
to do so, we run our usual test of progressively excising the Galactic
plane. However, we expect no issues with astrophysical foregrounds given the
relatively clean nature of GRBs as a source; in fact, no sky cut at all should
be necessary.

In all tests below, the systematic template we do use, as alluded to above, is
the BATSE exposure function, which varies significantly with declination,
mimicking (partially) a dipole (see Fig.~\ref{fig:grbpositions}). Some of the
results presented below change both quantitatively and qualitatively (e.g.,
statements we would make about the supergalactic plane are different) if this
template is not taken into account. We also include quadrupole and octopole
templates, as usual. This does not affect the results apart from widening the
error bars.

\begin{figure}[]
\begin{center}
\includegraphics[width=.48\textwidth]{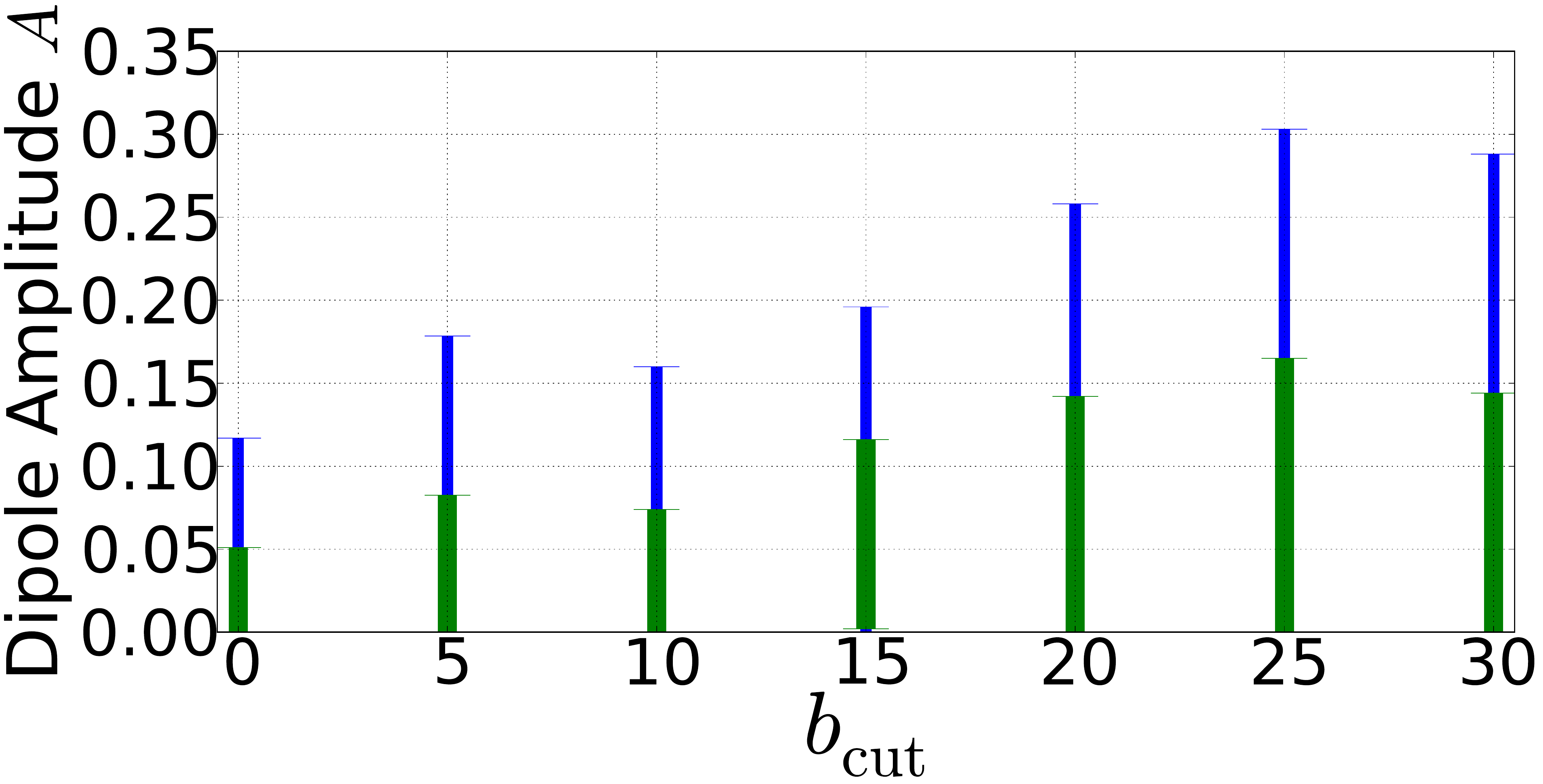}
\caption{Dipole amplitude in BATSE as a function of $\bcut$. Note that all
  measurements of the dipole amplitude for different Galactic cuts are
  mutually consistent, and are consistent with zero as well. Measurement
  errors in the form of 68 and 95 percent confidence intervals are shown; they
  tend to grow with the aggressiveness of the cut, as
  expected. }
\label{fig:batsefunctionofbcut}
\end{center}
\end{figure}

Given that cosmic variance should be much smaller on these scales than it was
for 2MASS, we expect that results at different values of minimum $|b|$ will be
consistent with each other within the \textit{measurement} error bars. Results
for varying Galactic $b_{\rm cut}$ are given in
Fig.~\ref{fig:batsefunctionofbcut}, from which it is clear that results at
different $b_{\rm cut}$ are indeed consistent with each other. As expected,
there is no detectable signal that varies with Galactic latitude.

We now proceed to run the usual test cutting the supergalactic plane. It
should be noted that for some cuts, a dipole is detected at marginal
significance in this series of tests \textit{if the BATSE exposure function is
  not taken into account}. Any kind of detection of the supergalactic plane in
gamma-ray bursts would be very surprising given the complete lack of any
association between GRBs and the local structure represented by the SGP (which
goes out to something on the order of $z \sim 0.02$ or 0.03, depending on
estimates; see \citet{lahav2000supergalactic}), but the result turns null when
we account for the exposure function. See Table \ref{tab5.3} for the results
of cutting in supergalactic latitude.

\begin{table*}[]
\footnotesize
\caption{Key patterns in cutting in supergalactic coordinates, for BATSE
  gamma-ray bursts.}
\label{tab5.3}
\centering
\setlength{\tabcolsep}{0.7em} 
\begin{tabular}{| c | c | c | c | c | c || c | c | c | c | c | c |}
\hline
\rule[-3mm]{0mm}{8mm} $\sgbge$ & $\fsky$ & $N$ & $\fsources$ & $\frac{\fsources}{\fsky}$ & $A_{\rm peak}$ & $\sgblt$ & $\fsky$ & $N$ & $\fsources$ & $\frac{\fsources}{\fsky}$ & $A_{\rm peak}$ \\
\hline

\rule[-2mm]{0mm}{6mm} 0.0 & 1.00 & 2702 & 1.00 & 1.00 & 0.00 & --- & --- & --- & --- & --- & --- \\ \hline
\rule[-2mm]{0mm}{6mm} 2.0 & 0.97 & 2606 & 0.96 & 1.00 & 0.00 & 74.82 & 0.97 & 2615 & 0.97 & 1.00 & 0.00 \\ \hline
\rule[-2mm]{0mm}{6mm} 5.0 & 0.91 & 2460 & 0.91 & 1.00 & 0.00 & 65.90 & 0.91 & 2479 & 0.92 & 1.01 & 0.00 \\ \hline
\rule[-2mm]{0mm}{6mm} 10.0 & 0.83 & 2184 & 0.81 & 0.98 & 0.00 & 55.73 & 0.83 & 2249 & 0.83 & 1.01 & 0.00 \\ \hline
\rule[-2mm]{0mm}{6mm} 20.0 & 0.66 & 1740 & 0.64 & 0.98 & 0.00 & 41.15 & 0.66 & 1840 & 0.68 & 1.03 & 0.00 \\ \hline

\hline
\end{tabular}
\end{table*}

We find another null result, which is good considering that detection of the
supergalactic plane in GRB data would be a highly unusual find, and would
almost certainly indicate something problematic about our analysis.

\subsection{Dipole Direction and Conclusion}
\label{sec5.2.3}


The still-relatively-small number of gamma-ray bursts (2702) in the BATSE
catalog places only the loosest of constraints on the direction of the
dipole. All but a very tiny patch of sky (centered around
$(l,b)=(273^\circ,31^\circ)$, and extending roughly 10 degrees in radius) is
within the 3-sigma confidence interval for the direction of the dipole, and
the CMB kinematic dipole direction (which is an appropriate direction of
comparison in this case, since the kinematic dipole should dominate
contributions to the dipole for GRBs) is only marginally outside the 2-sigma
confidence interval (which given the looseness of the constraints is not a
noteworthy result).

In all, the BATSE data places useful constraints on the dipole amplitude, but
not direction, in GRB data. Our results, while they are not yet strong enough
to allow for the detection of the expected kinematic dipole, do place
constraints on our ability to distinguish BATSE GRBs as a tracer of
large-scale structure, and also constrain any intrinsic dipole in the
large-scale structure that would manifest itself in the distribution of GRBs
on our sky: at 95 percent confidence, $A<0.117$ for the intrinsic dipole. 
Because the constraints on the dipole are relatively weak relative to
  constraints using other surveys employed in this paper, and because the
  radial distribution of BATSE sources $N(z)$ is poorly known, we have not attempted
  to produce a theoretical expectation for the BATSE local-structure dipole.

In the course of this analysis we have, however, effectively performed several
sanity checks on the BATSE dataset, showing that the supergalactic plane is
undetectable (as expected) using our analysis, that the Galaxy does not show
up at all in the data (which is in line with previous studies of the GRB
distribution), and that the BATSE exposure map must be taken into account in
order for these tests not to turn up anomalous. We would of course still like
to see constraints on the intrinsic dipole much better than those available
from GRBs, those constraints being on the order of $10^{-1}$. For this, we
turn to the radio survey NVSS.

\section{Dipole in NVSS}
\label{sec5.3}

For a long time, it was assumed that the distribution of radio sources was,
like that of GRBs, indistinguishable from isotropic and unclustered
(e.g.,~\citet{webster1976clustering}). In fact, even if the distribution of
radio sources was not intrinsically isotropic, radio sources have a large
range of intrinsic luminosities, and so structures would naturally wash out
when sources were projected onto the sky and radial information was removed
(\citet{baleisis1998searching}). However, recent results, especially using the
NRAO VLA Sky Survey (NVSS), have detected clustering in radio sources, and in
particular, a dipole. NVSS is a radio survey with nearly 1.8 million
extragalactic sources (\citet{condon1998nrao}). This survey presents an
excellent opportunity to actually test for the presence of the kinematic
dipole and possibly the intrinsic dipole in large-scale structure. NVSS has
more potentially non-negligible systematics to control for than the other
surveys we use, but it also has higher potential payoff because of its
combination of depth and sky coverage.

\subsection{Previous Work}
\label{sec5.3.1}

Several attempts at calculating the dipole in radio sources have been made in recent years.

\citet{baleisis1998searching} present theoretical predictions and
observational results for the dipole in the Green Bank 1987 and
Parkes-MIT-NRAO (PMN) catalogs. The combination of these catalogs gives $\sim
40,000$ sources with flux $> 50$ mJy at 4.85 GHz. They find that the magnitude
of the dipole is an order of magnitude larger than expected from the
contributions of large-scale structure (analogous to our local-structure
dipole) and the kinematic dipole. However, they are plagued by several
systematic errors. First, they find that the two catalogs they used have a
mismatch in flux. While they correct for this, it is hard to do so with high
precision and confidence. They also note that the radio sources in their
catalogs are likely drawn from multiple populations, though this is true of
any analysis that uses radio sources, and is not a crippling problem if the
redshift distribution is sufficiently well-understood.

\citet{blake2002detection} attempt to measure the kinematic dipole alone in
NVSS (see also \citet{blake2004angular} for analysis of the rest of the
angular power spectrum in NVSS). They make efforts to remove the contribution
of what they refer to as the ``clustering dipole," the dipole that when
flux-weighted gives a measure of the acceleration of the Local Group, and when
unweighted matches up with our local-structure dipole. They claim that the
clustering dipole should die away by $z < 0.03$ (based on results from the
\citet{rowan2000iras} analysis of the IRAS PSCz dipole, though the results of
\citet{erdogdu2006dipole} call this convergence into question) and contribute
roughly $A \sim 2 \times 10^{-3}$ to the total amplitude of the total dipole,
if it is not removed as they attempt to do. (Note that we have converted from
their peak-to-trough ``amplitude" $\delta$ to our peak-to-zero amplitude $A$).

Blake and Wall measure the remaining dipole -- which would ideally be a
kinematic dipole only, but which will in reality take contributions not only
from local structure beyond what IRAS observed, but also from more distant
large-scale structure, as we show below -- by expanding the angular
distribution of sources in spherical harmonics and measuring the harmonic
coefficients $a_{\ell m}$ for the dipole, quadrupole, and octopole, including
all $m$ values. Inclusion of higher harmonics is necessary because of the lack
of full-sky coverage. They find that a dipole model is a good fit by $\chi^2$,
and find good agreement with the direction of the CMB velocity dipole, which
they cite as $\theta = 97.2 \pm 0.1^\circ, \phi=168.0 \pm 0.1^\circ$, and which
converts to $(l,b)=264.3^\circ,48.1^\circ$).

We take the Blake and Wall results as the most reliable previous
result,\footnote{\citet{singal2011dy} has performed the most recent analysis
  in this vein. His results are suspicious, as he finds truly exorbitant
  speeds for the Local Group (on the order of 1700 km/s). This becomes more
  understandable given that the way in which he accounts for the sky cut,
  particularly the hole in NVSS at declination $<40^\circ$, is suspect (he
  simply cuts out dec $>40^\circ$ as well in order to counterbalance the hole
  at $<40^\circ$). Also, his method of detecting the dipole does not account
  for coupling between the dipole and other multipoles on the cut sky, and
  also neglects any contribution from the local-structure dipole to the
  results.} and compare our results to theirs. However, we do have reason to
expect that our estimate of the dipole will be more reliable than theirs; in
particular, we do a great deal more to take systematic effects into account
than they do in their analysis, using our real-space estimator. We also do not
attempt to remove local sources as they do, since our objective is to compare
our observational results to the full dipole signal expected from theory,
which includes both a local-structure and a kinematic contribution. Since we
are not flux-weighting, local sources do not contribute preferentially to the
dipole, and so we can afford to leave them in the analysis.

\subsection{Theoretical Predictions}
\label{sec5.3.2}

For NVSS, we must be careful to include in our theoretical predictions the
contributions of not only the local-structure dipole, but also the kinematic
dipole. It is no longer the case, as it was for 2MASS and 2MRS, that the
kinematic dipole is swamped by two orders of magnitude by the local-structure
dipole. Rather, the two are on the same order of magnitude, as also recognized
by \citet{baleisis1998searching} and \citet{blake2002detection}.

The local-structure contribution is calculated in the same way as always. This
part of the prediction will vary somewhat depending on what redshift
distribution $n(z)$ we use. \citet{dunlop1990redshift} derived $n(z)$ for
several different flux cutoffs, though the results for the dipole amplitude
and the redshift distribution itself are somewhat robust to changes in the
flux cutoff since radio galaxies display such a wide range of intrinsic
luminosities (\citet{baleisis1998searching}; \citet{blake2004angular} note
specifically that for the NVSS frequency of 1.4 GHz, the clustering of radio
galaxies is not strongly dependent on flux for fluxes between 3 mJy and 50
mJy). The redshift distribution developed by \citet{ho2008correlation} as a
best model for NVSS avoids several drawbacks of the Dunlop and Peacock
distribution, especially the assumption that bias is redshift-independent and
the heavy reliance on the functional form of the luminosity function rather
than data in constraining the redshift distribution. However, without
repeating Ho et al.'s rather detailed analysis,
we are left unable to calculate the redshift distribution for flux cuts
different than the 2.5 mJy cut that they use.


Fortunately, we can here make use of the fact that the predictions are not strongly tied to the flux
cutoff. We therefore follow Ho et al.~in modeling the NVSS redshift
distribution as follows:
\begin{equation}
W(z) = \frac{\alpha^{\alpha+1}}{\zstar^{\alpha+1}\Gamma(\alpha)} \beff z^\alpha e^{-\alpha z/\zstar}
\end{equation}
where $W(z) = b(z) n(z)$, where $b(z)$ is the bias as a function
of redshift and $n(z)$ ($\Pi(z)$ in Ho et al.) is the probability distribution
for the galaxy redshift. Ho et al.~give $\beff = 1.98$, $\zstar = 0.79$, and
$\alpha=1.18$ as best-fit parameters. Using all of this, we find that for this
distribution, the contribution of the local-structure dipole to the total
dipole in NVSS is $A = 9.8 \times 10^{-4} \simeq 0.0010$.

Meanwhile, the theoretically expected kinematic dipole may be
calculated as shown in Appendix \ref{appA}.
%
%
%
%
The predicted amplitude is
\begin{equation}
A = 2 \tilde \beta = 2 [1 + 1.25x(1-p)] \beta
\label{eq:Atildebeta}
\end{equation}
where the first term in brackets essentially represents the contribution of
relativistic aberration and the second term represents the contribution of the
Doppler effect. From the CMB, $\beta = v/c = 1.23 \times 10^{-3}$. Meanwhile,
$x$ and $p$ (exponents in the power laws for the intrinsic number counts of
galaxies as a function of limiting magnitude, and for the intrinsic flux
density of a galaxy as a function of frequency, respectively; see
Sec.~\ref{sec2.2.3}) are not known precisely, but can be estimated, as
  we do just below.

This expression is equivalent to that used by \citet{blake2002detection}
(converting from their $\delta$ to our $A=\delta/2$):
\begin{equation}
A = \left \lbrack 2 + x(1 + \alpha) \right \rbrack \beta
\end{equation}
with the substitution in Eq.~(\ref{eq:Atildebeta}) $x \rightarrow 2x/5$ and $p
\rightarrow -\alpha$. (The latter substitution is a straightforward matter of
notation; the first has to do with switching from magnitudes to
fluxes.\footnote{In our original notation, which follows
  \citet{itoh2010dipole}, the number of galaxies detected by the survey is
  proportional to $10^{x m_{\rm lim}}$, where $m_{\rm lim}$ is the limiting
  magnitude of the survey. In \citet{blake2002detection}, the number of
  galaxies is proportional to $S^{-x^\prime}$, where $S$ is flux, and we have
  added a prime to distinguish variables named $x$. Equating $S^{-x^\prime}$
  with $10^{x m_{\rm lim}}$ (up to a proportionality constant) and using the
  fact that $S \propto 10^{-2m_{\rm lim}/5}$, we have that $2 m_{\rm lim}
  x^\prime/5 = x m_{\rm lim}$, or $x = 2 x^\prime/5$. Blake and Wall give
  $x^\prime \approx 1$.}) We follow Blake and Wall (for NVSS) and also
Baleisis et al.~(who were not working with NVSS, but who did work with radio
catalogs) and take $x \approx 1$ and $\alpha \approx 0.75$, which yields a
kinematic dipole amplitude of $A_{\rm kin} \approx 0.0046$.  The full
theoretical value includes this kinematic dipole and the contribution from the
local-structure dipole ($A_{\rm loc\, str}=0.0010$). We can therefore write
the full theoretical prediction as
\begin{equation}
A=0.0046 \pm 0.0029\pm 0.0006,
\end{equation}
where the first uncertainty corresponds to the kinematic dipole and the second
to the local-structure dipole.
If we find a result that is outside these cosmic-variance errors from the central
value, and this is not a systematic effect, we might invoke the presence of an
intrinsic dipole as an explanation.

\subsection{Present Work}
\label{sec5.3.3}

\begin{figure}[t]
\begin{center}
\includegraphics[width=.48\textwidth]{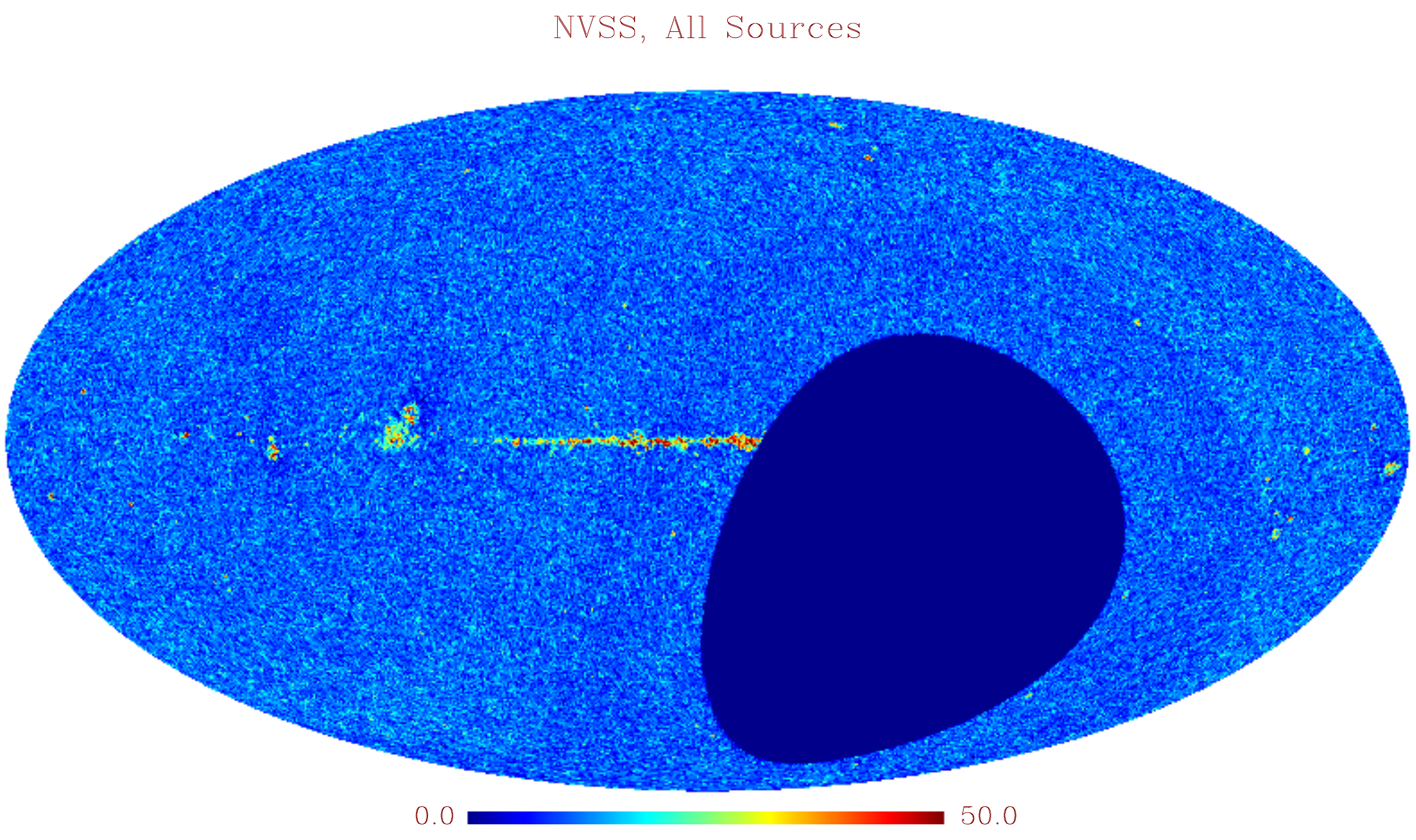}
\includegraphics[width=.48\textwidth]{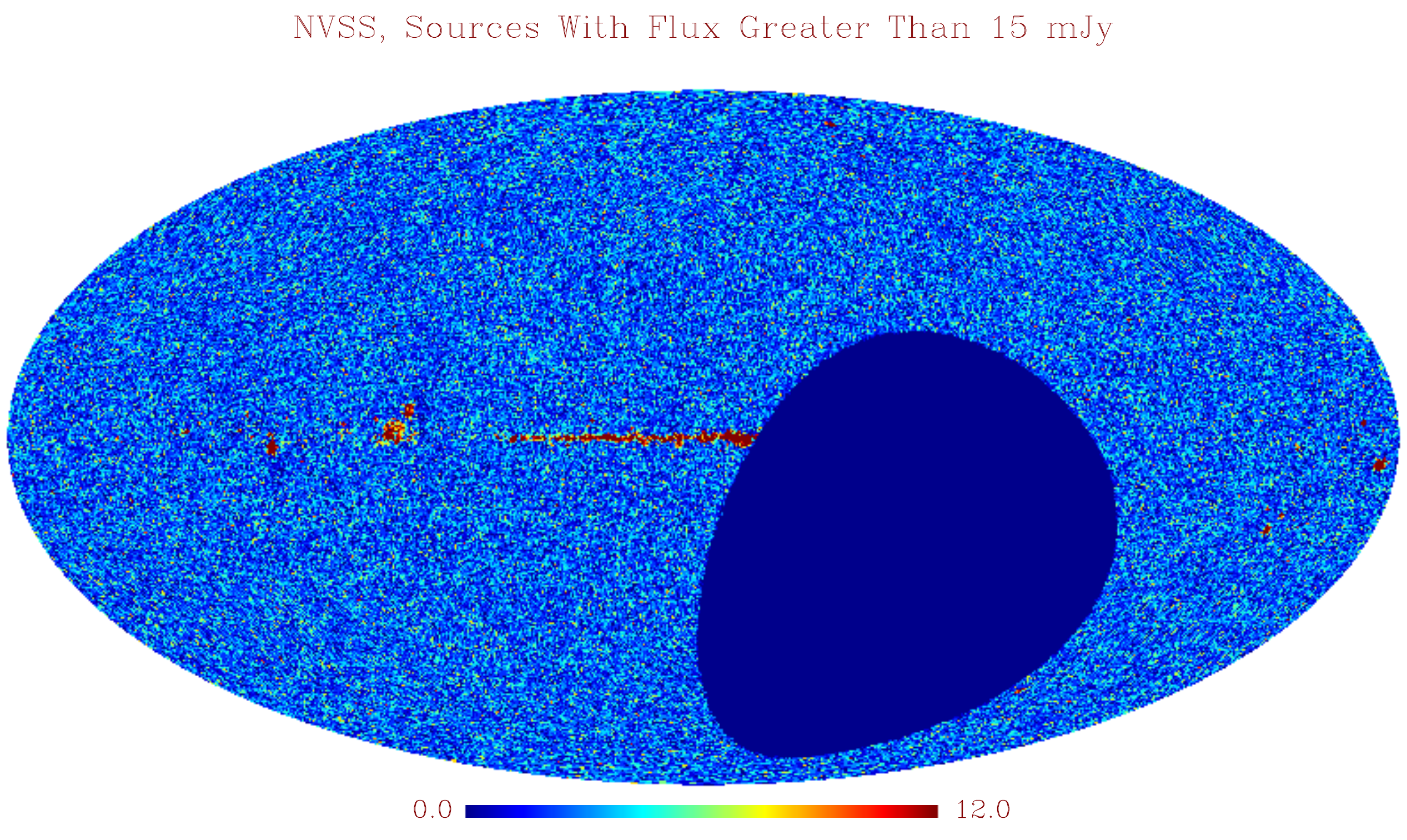}
\caption{{\it Top panel:} All sources in NVSS, in Galactic coordinates. Note the
  ``hole" in the data for declinations less than 40 degrees, and the
  declination-dependent striping (visible in this coordinate system as a
  series of ``wavy" stripes going outward from the pattern set by the
  declination-dependent hole in the data). {\it Bottom panel}: Sources with flux
  greater than 15 mJy. The spurious power goes largely away with this flux
  cut, and can no longer be seen by eye. (Dynamic ranges are restricted in
  both these maps so as to better show structure.)}
\label{fig:nvsssources}
\end{center}
\end{figure}

\begin{figure}[t]
\begin{center}
\includegraphics[width=.48\textwidth]{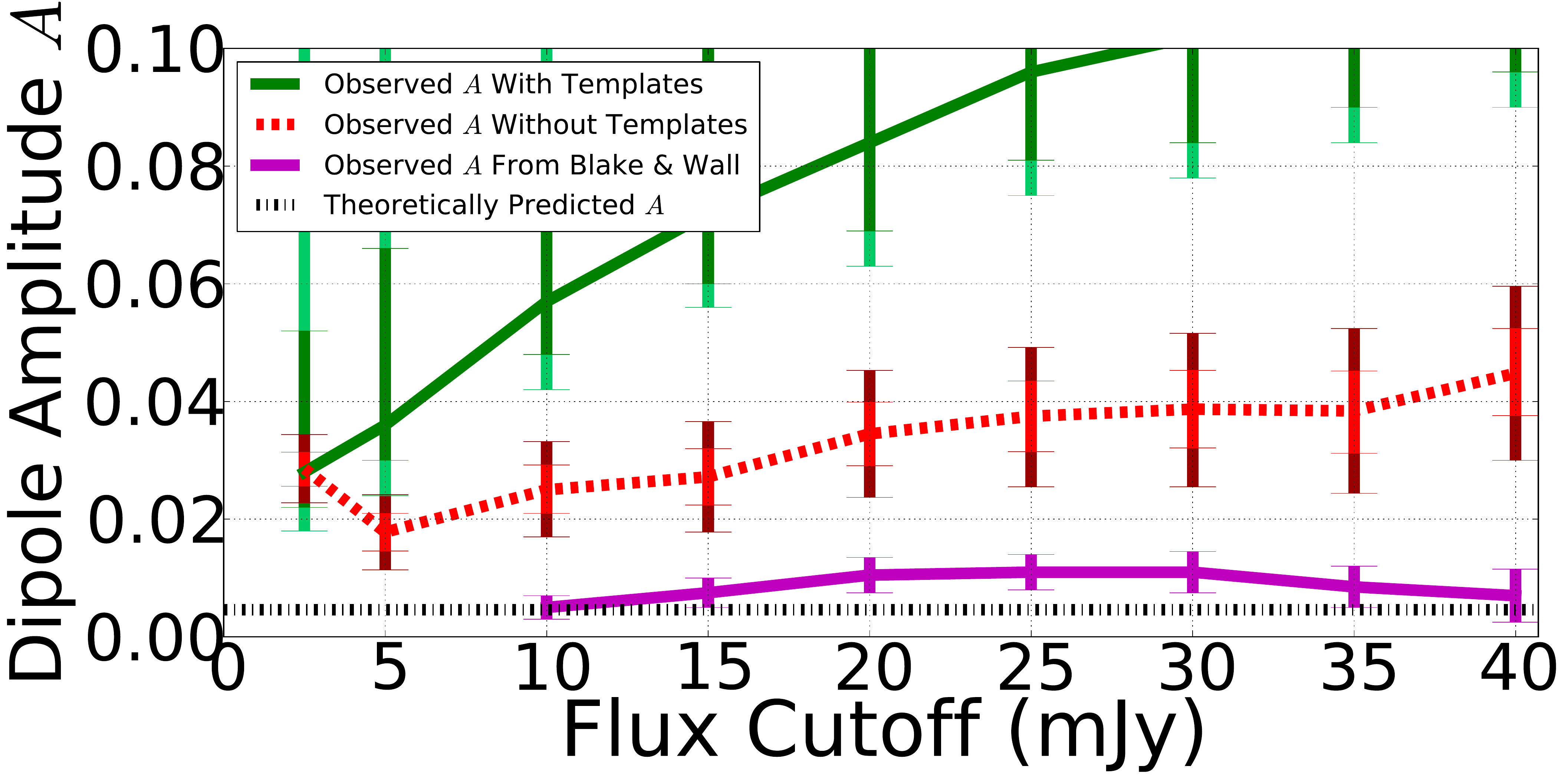}
\caption{Measured NVSS dipole amplitude as a function of flux cut, with error
  bars. We show both our results and those from
  \citet{blake2002detection}. For the dotted red curve, we include the Haslam
  et al.~map of 408 MHz Galactic emission as a systematics template, and
  remove $\bltfifteen$ and dec greater than 78 degrees/less than -10 degrees;
  for the solid green curve, we include sources with dec greater than -37
  degrees (thus
    effectively increasing $\fsky$ from 0.42 to 0.60), but also include
  quadrupole, octopole, and dec-dependent-striping templates. We suspect that
  the near-monotonic increase of the amplitude with the flux cut is due to
  spurious power in the NVSS map, and possibly the presence of local structure
  in the survey as well (since local structures preferentially have higher
  fluxes). The apparent agreement between theoretical predictions and
    the Blake and Wall results is partially misleading, as discussed in the
    text.}
\label{fig:plotoutputtablenvss}
\end{center}
\end{figure}

We turn first to examining the systematics that need to be accounted for in
the NVSS data. These systematics are illustrated in
    Fig.~\ref{fig:nvsssources}.

First, the survey did not observe below a declination of 40 degrees. For our
purposes, the pixellization around this ``hole" in declination is especially
important to pay attention to, as we find that a sky cut for pixels with
centers at dec $<40^\circ$ gives significantly different results for the dipole
amplitude and direction than a cut at dec $<37^\circ$ at low-resolution
pixellization (NSIDE$=$16). We work at much higher resolution (NSIDE$=$128),
where the effect is not as strong, but we still cut all pixels with dec
$<37^\circ$ just to be safe. We choose that particular number following
\citet{smith2010no}, and also because this appears to be a good conservative
choice if we want to completely avoid problems associated with the hole in
declination.

Second, faint sources very close to bright sources cannot be reliably
detected. We therefore mask out a $0.6^\circ$ radius around all sources
brighter than 2500 mJy, following \citet{ho2008correlation}. Blake and Wall do
not perform this same masking, although they do remove known local sources.

Third, we use as a systematic template the map of Galactic synchrotron
radiation of \citet{haslam1981408} and \citet{haslam1982408}. This is a 408
MHz radio continuum map of the entire sky that combines data from four
different surveys and is dominated by synchrotron emission. Information about
foreground emission is important when dealing with a radio survey such as
NVSS, and we can apply this map as a systematic template since NVSS could
plausibly pick up non-extragalactic signal from this emission.

Fourth, there is declination-dependent ``striping" in the maps, which appears
even by eye if no flux cut is imposed. This problem stems from the fact that
the NVSS observations were made using two different configurations of the VLA,
the D configuration for observations between declinations of $-10^\circ$ and
$+78^\circ$, and the DnC configuration for declinations between $-40^\circ$
and $-10^\circ$, and above $+78^\circ$. The striping is readily apparent by
eye for the full catalog, but is invisible by eye and largely absent even in
more rigorous tests for fluxes above $\sim 15$ mJy (see
\citet{blake2002detection}). We therefore begin by examining the stability of
our dipole results as a function of flux cut.

We find that neither the direction nor the amplitude of the dipole is stable
for different flux cuts. This is as expected for flux thresholds less than 15
mJy, since the striping artifact gradually dies away as the lower flux
threshold is increased from zero up to 15 mJy. But for flux thresholds above
15 mJy, the fact that the dipole remains unstable is a problem. Increases in
the dipole amplitude might be due to the influence of the local-structure
dipole, which could come into play more strongly for these brighter (and
therefore at least somewhat more local) sources. Fluctuations in the direction
(the best-fit $l$ ranges from 219.1 to 234.3, and the best-fit $b$ ranges from
11.9 down to -0.2) could also be the result of the local-structure dipole,
though this seems unlikely since the $b$ coordinate in particular moves
\textit{away} from the direction of the local-structure dipole that we have
seen in previous tests using 2MASS as we go to brighter and brighter sources
(higher and higher flux thresholds) in NVSS.

We implement two distinct and complementary strategies for dealing with these issues.

(1) The first strategy is to make a more aggressive cut in declination, which
makes use of only one subset of the NVSS maps, corresponding to the D
configuration of the VLA: that is, we remove all portions of the sky with
declination less than -10$^\circ$ or greater than 78$^\circ$. These results
show a great deal more stability in the direction of the dipole for flux cuts
above 5 mJy, and certainly above 15 mJy. The primary results can be seen in
Fig.~\ref{fig:plotoutputtablenvss}. We have also examined the stability of
results as a function of cut in Galactic $b$ for the fixed case of a flux cut
at 25 mJy. This provides our usual test for contamination that varies as a
function of Galactic latitude, and helps justify our choice of $\bltfifteen$
for our cut. We find that there is some fluctuation in the dipole signal as a
function of $\bcut$, likely an indication of remaining spurious power in the
map rather than genuine Galactic latitude-dependent contamination.



\begin{figure}[t]
\begin{center}
\includegraphics[width=.48\textwidth]{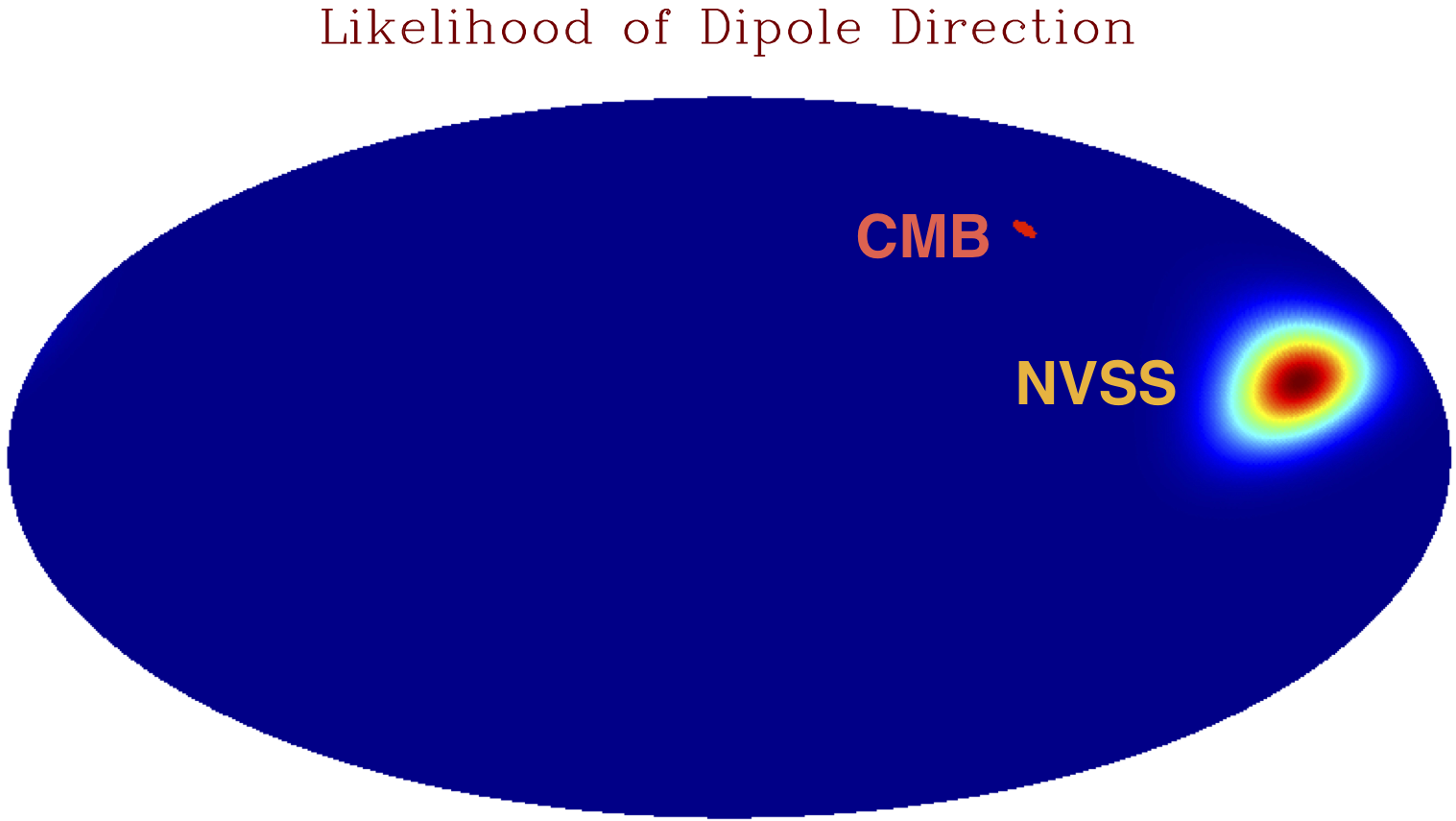}
\includegraphics[width=.48\textwidth]{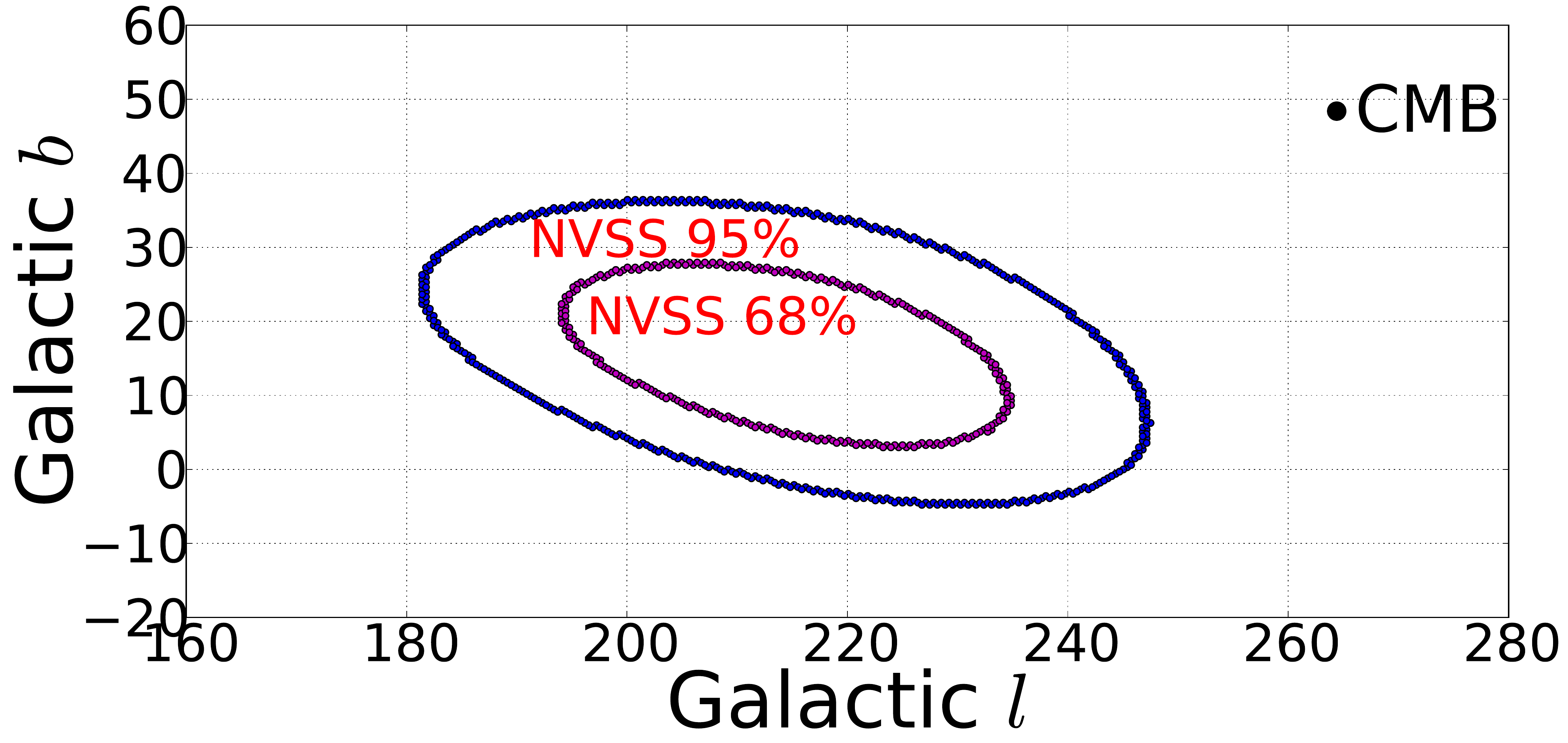}
\caption{{\it Top panel:} Likelihood of dipole direction in NVSS,
  marginalized over amplitude. {\it Bottom panel:} The CMB dipole direction
  is over 3 sigma away from the best-fit NVSS dipole direction. This is not a
  problem, however, since the NVSS dipole amplitude is $A \sim 10^{-2}$, and
  we expect the LSS dipole direction to match the CMB dipole direction only
  when amplitudes $A \sim 10^{-3}$ are probed. Both of these plots correspond to
  the case in which sources have flux greater than 15 mJy, only declinations
  between $-10^\circ$ and $78^\circ$ are kept, and $\bltfifteen$ is cut.}
\label{fig:nvssdirection}
\end{center}
\end{figure}

\begin{figure}[]
\begin{center}
\includegraphics[width=.20\textwidth]{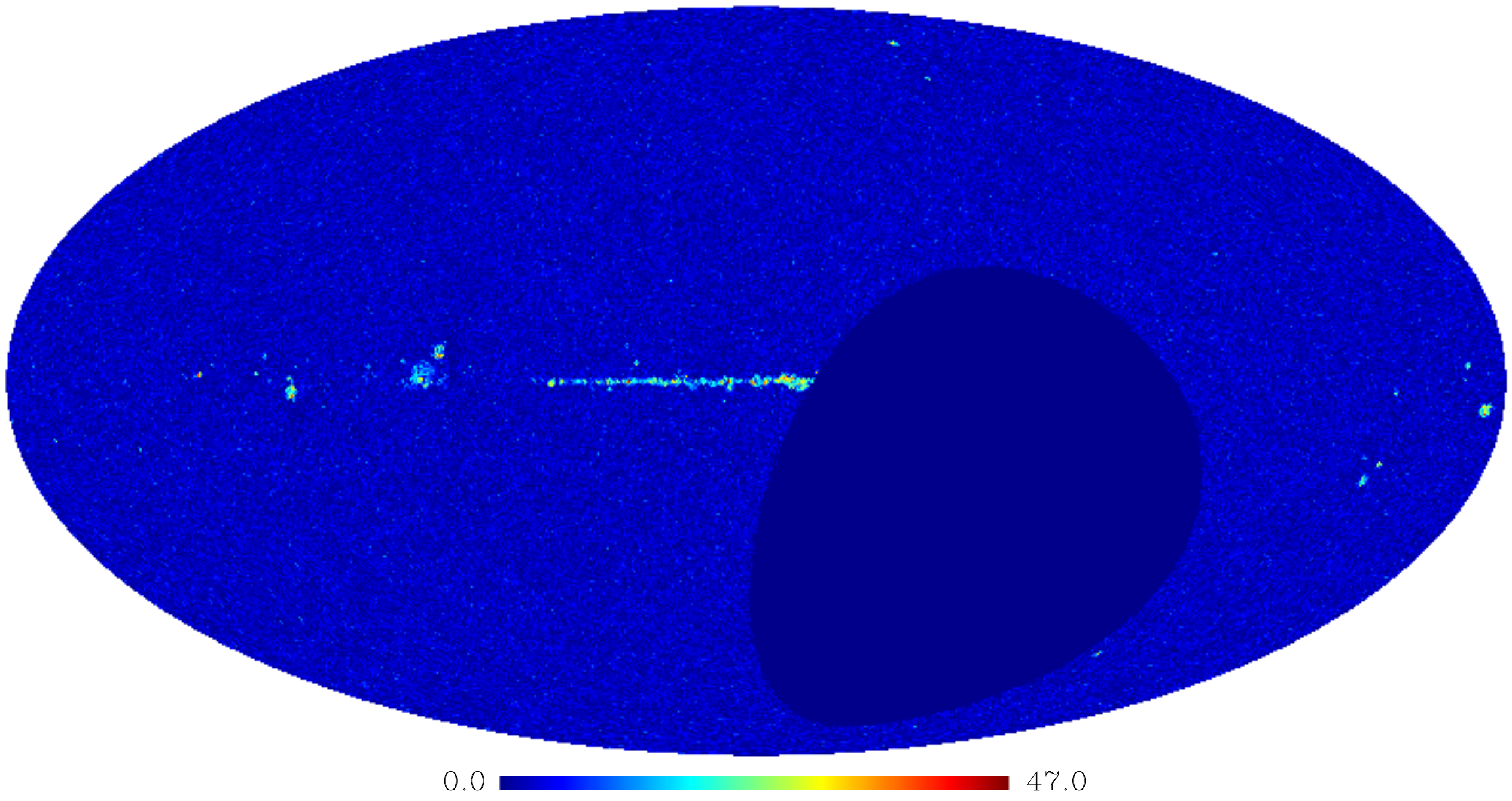}
\includegraphics[width=.20\textwidth]{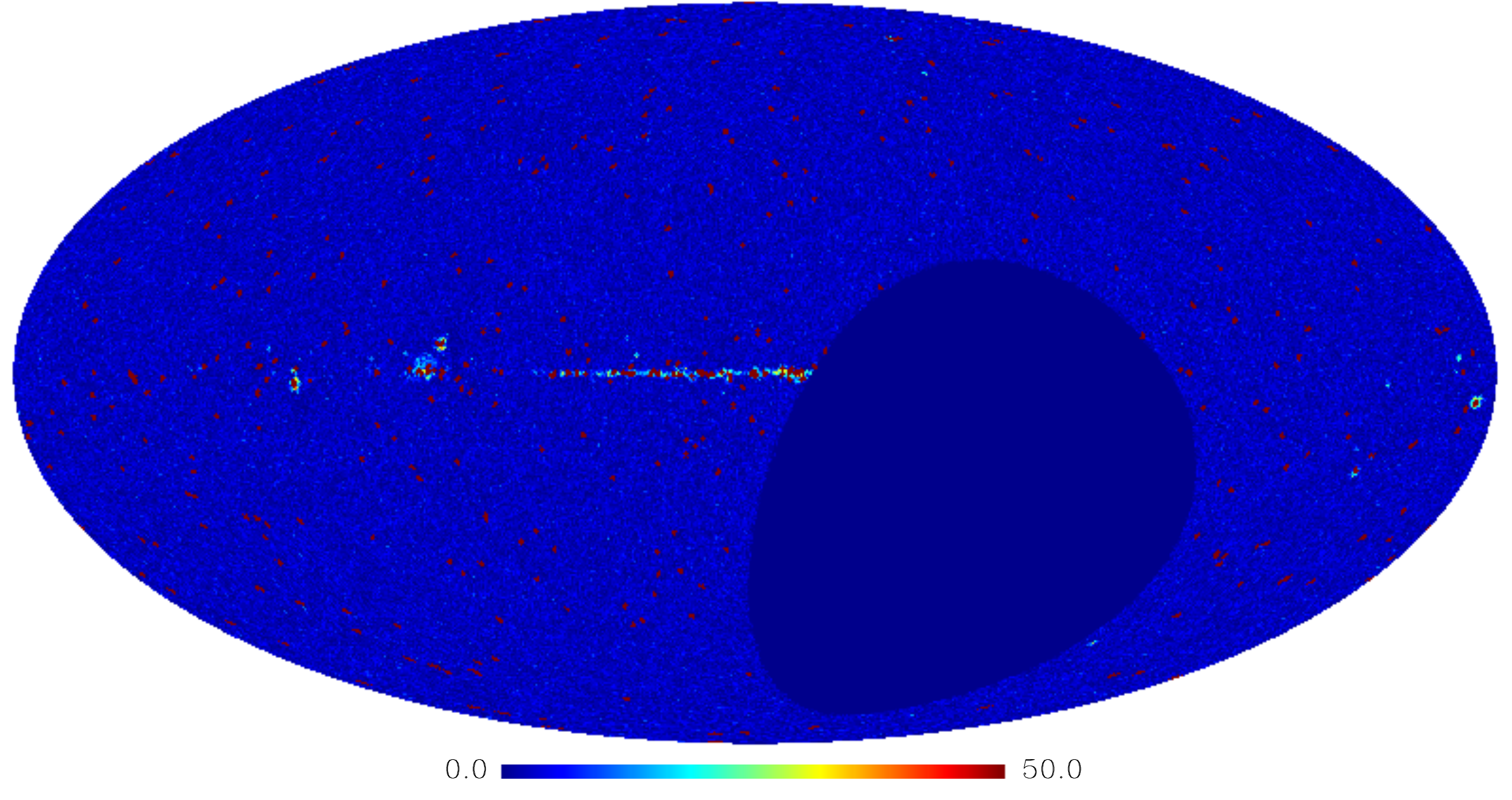}
\\[-1.0mm] no cut; only bright sources cut \\[1.0mm]
\includegraphics[width=.20\textwidth]{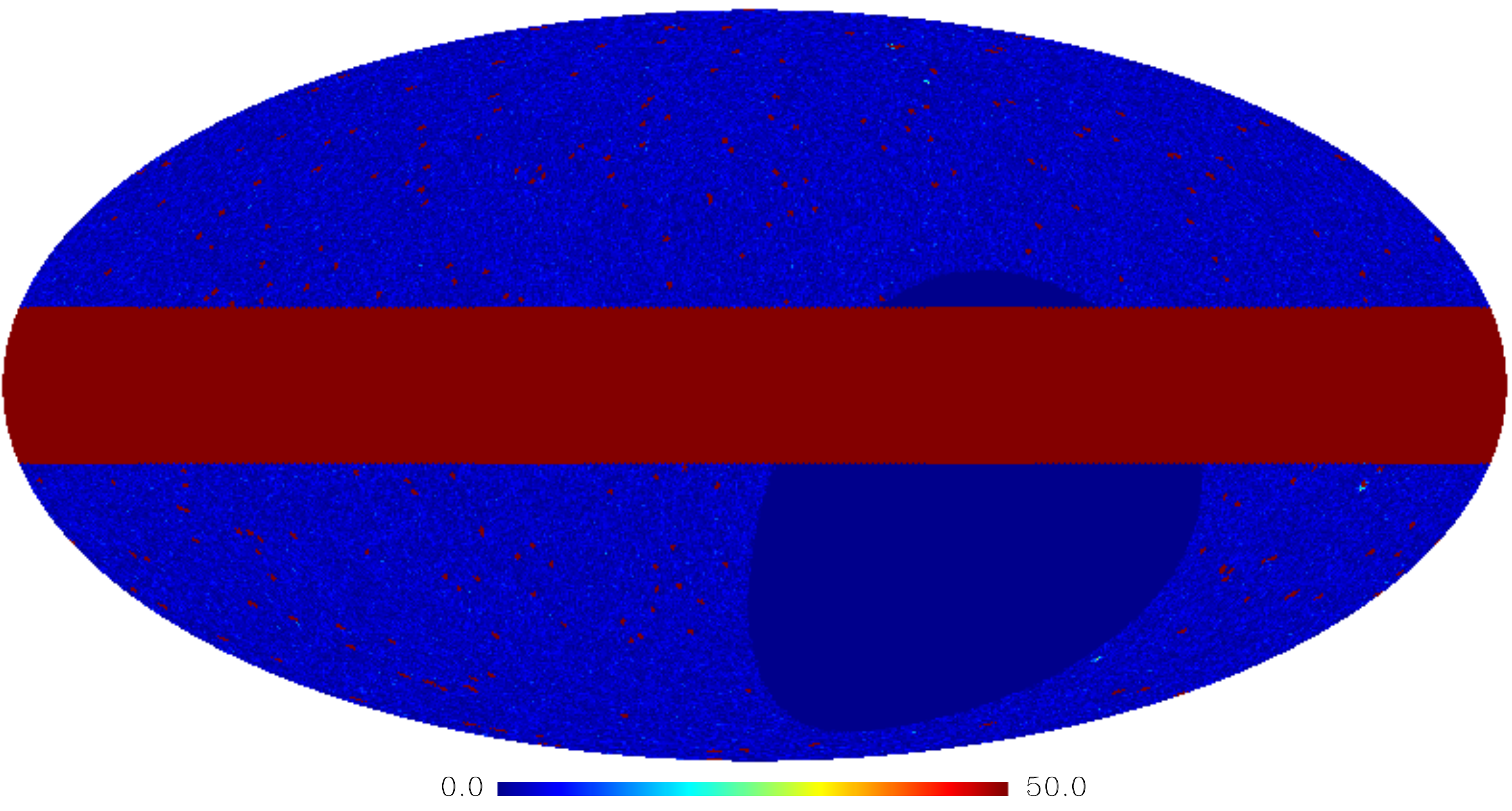}
\includegraphics[width=.20\textwidth]{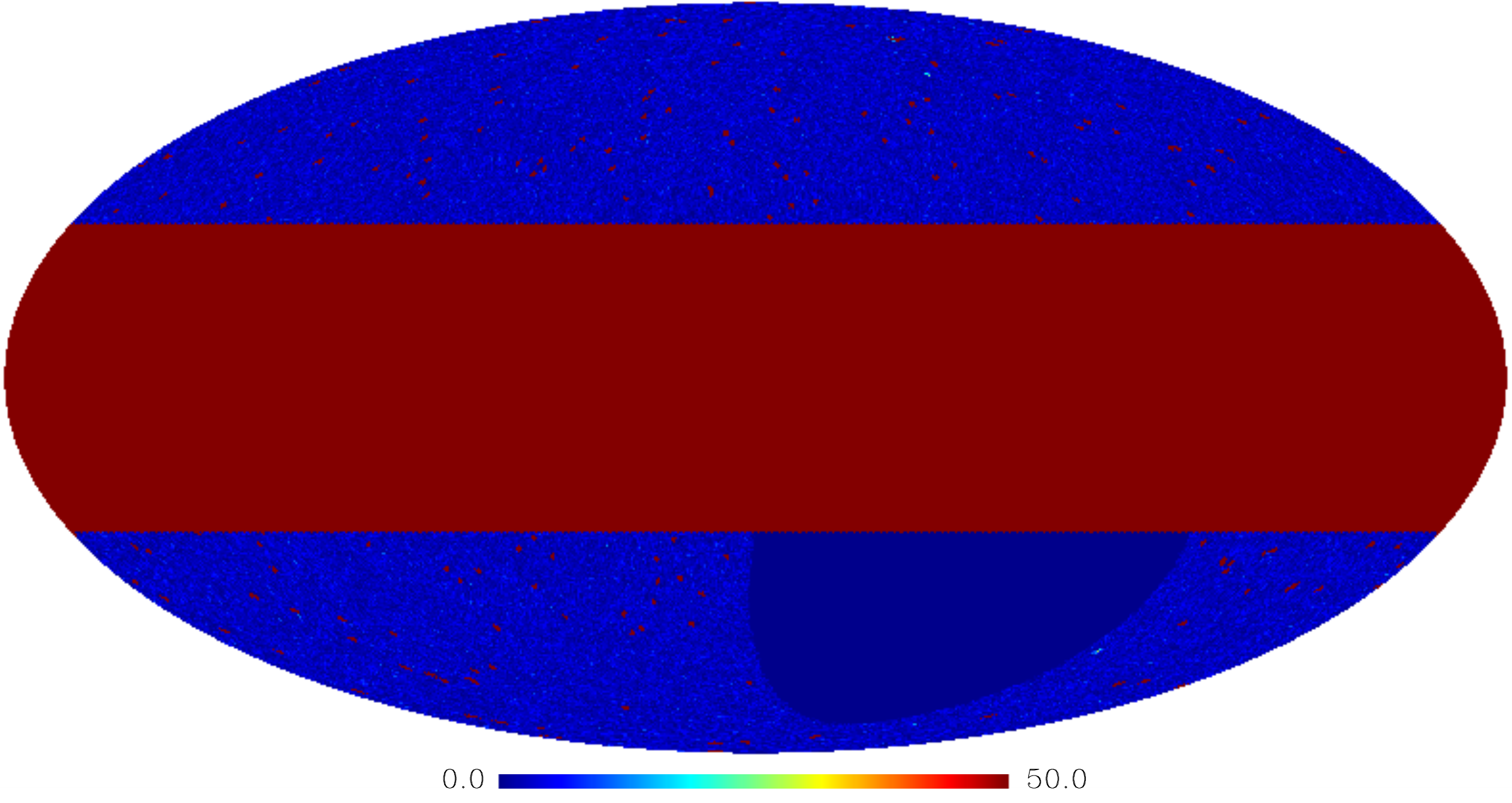}
\\[-1.0mm] $\bltfifteen$; $\bltthirty$ \\[1.0mm]
\includegraphics[width=.20\textwidth]{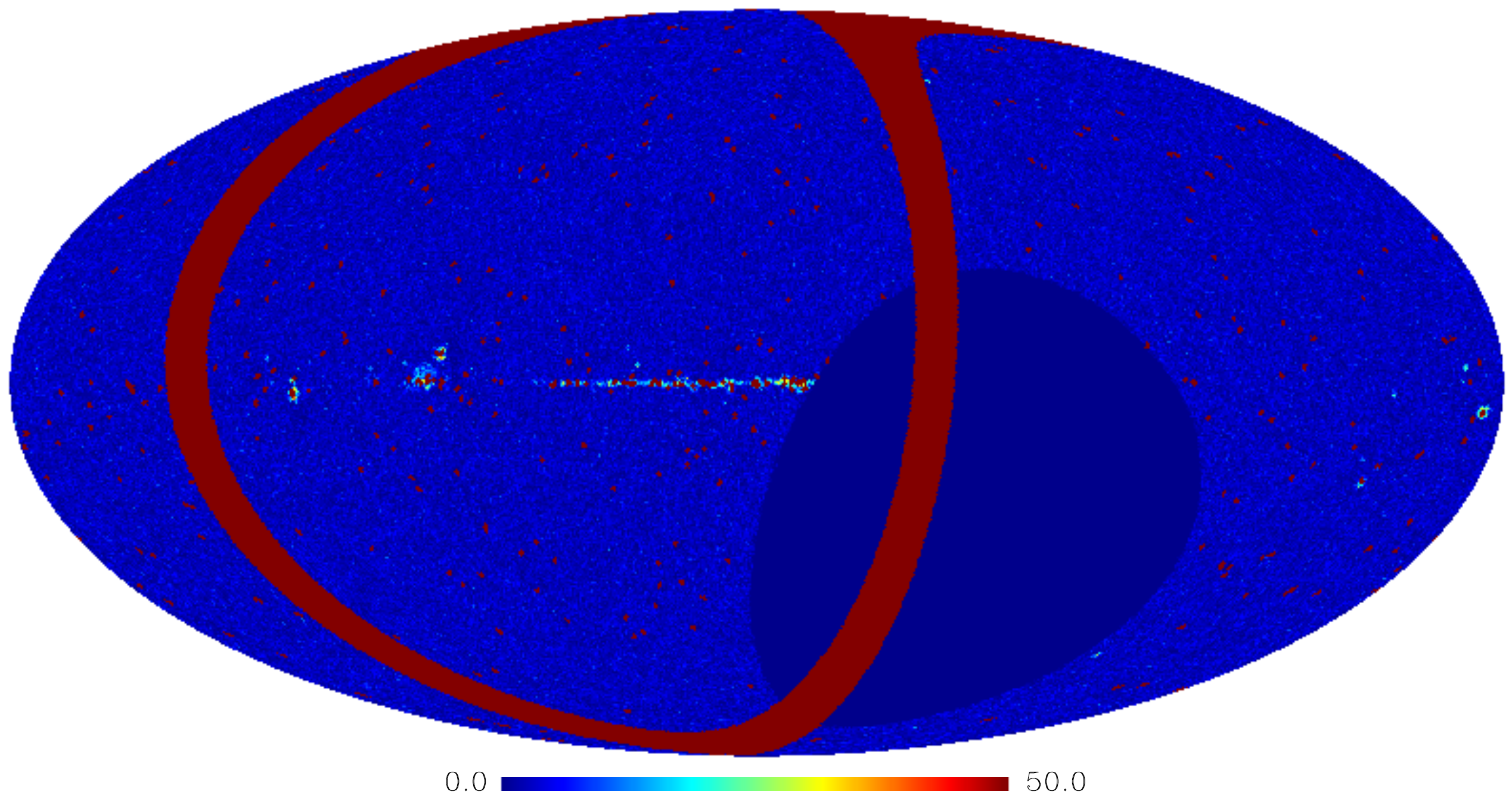}
\includegraphics[width=.20\textwidth]{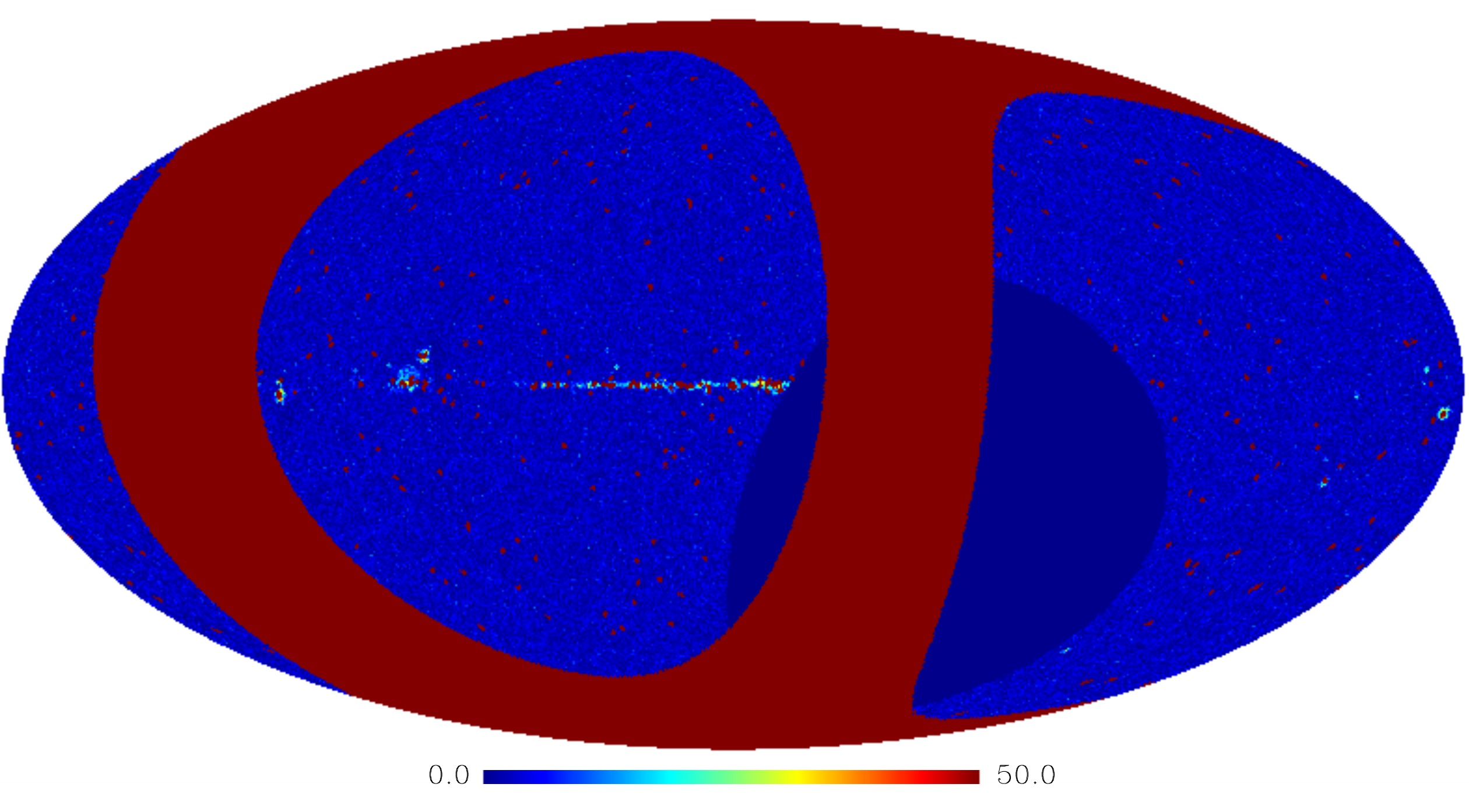}
\\[-1.0mm] $\sgblt 5^\circ$; $\sgblt 20^\circ$ \\[1.0mm]
\includegraphics[width=.20\textwidth]{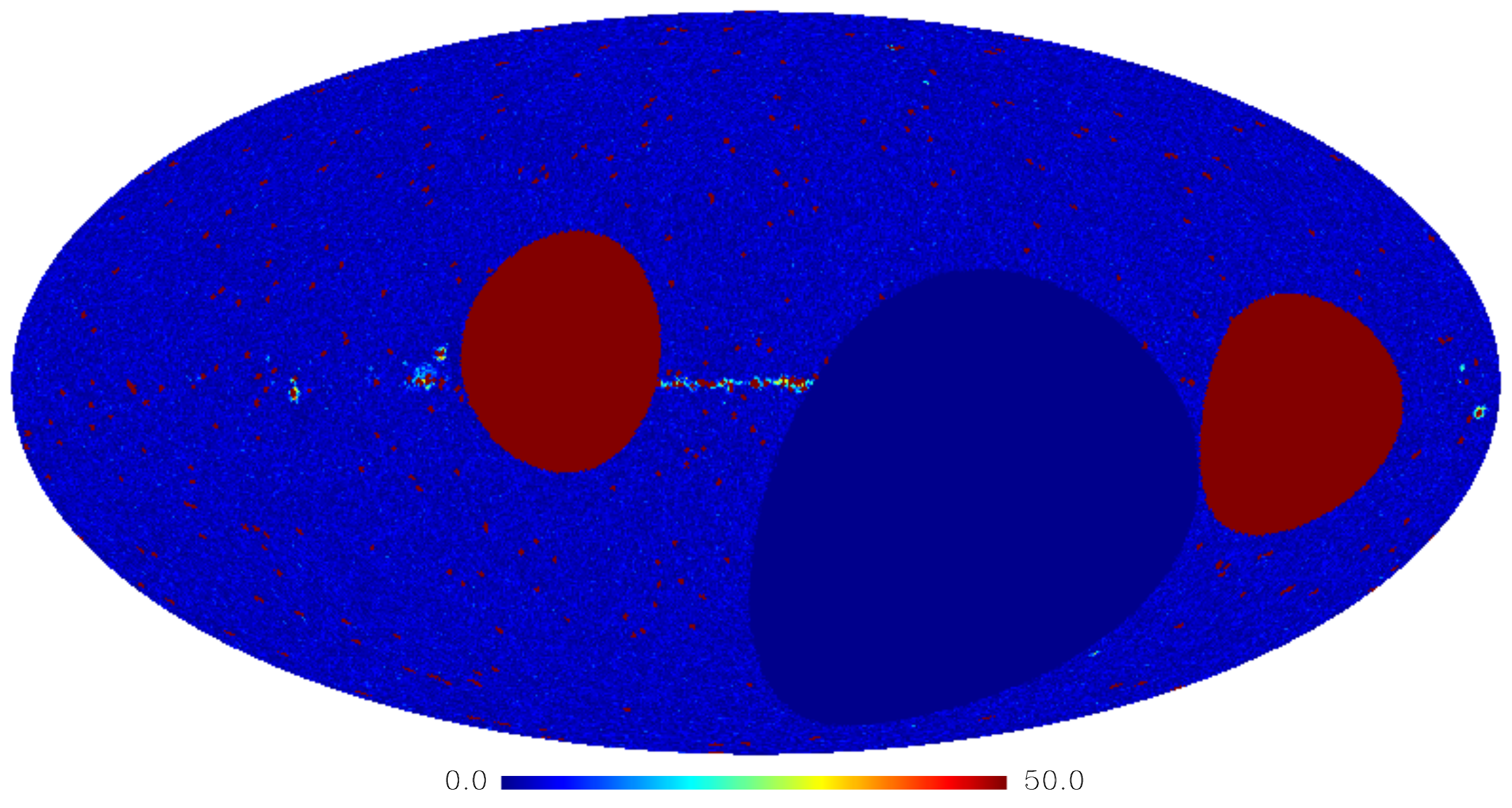}
\includegraphics[width=.20\textwidth]{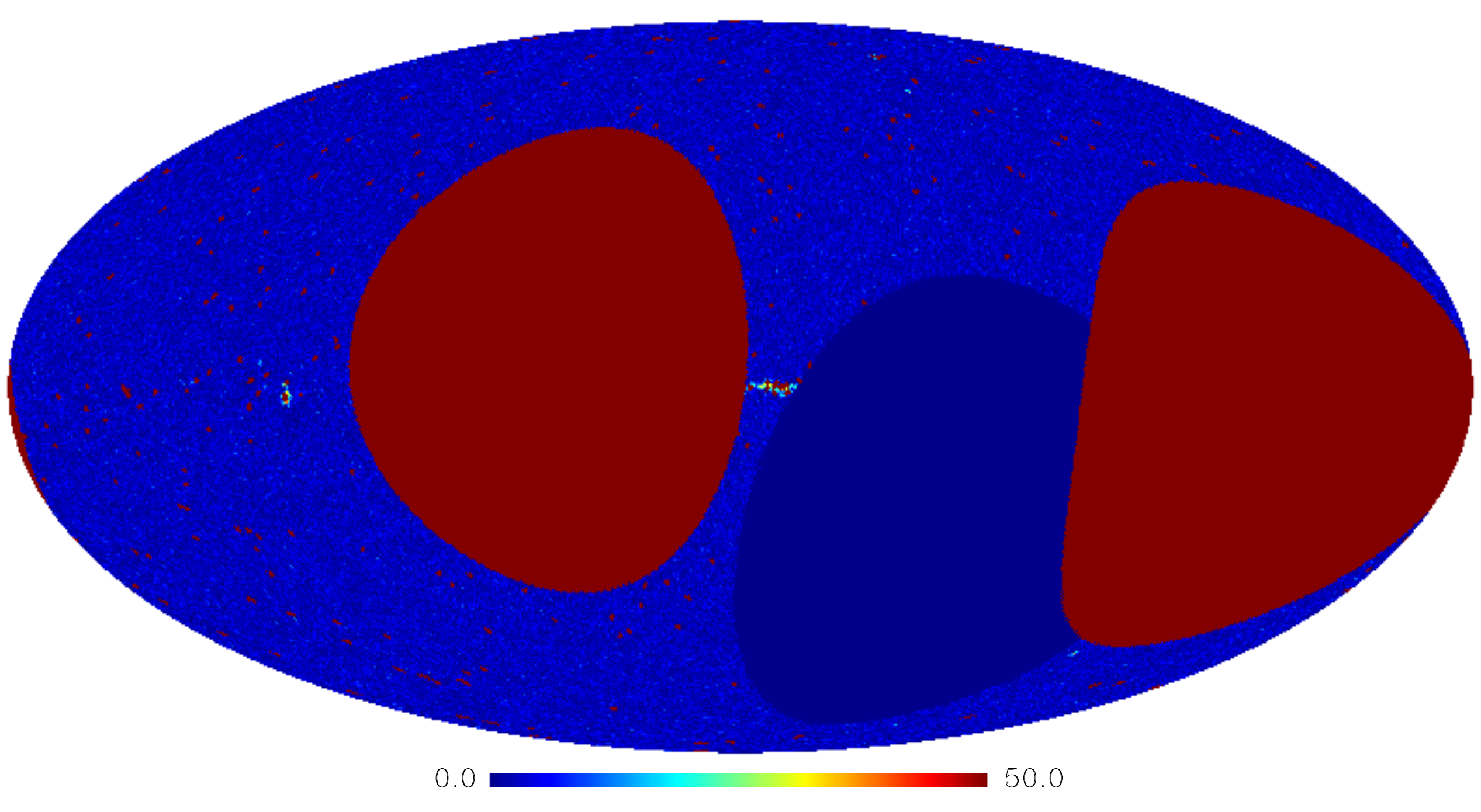}
\\[-1.0mm] $\sgbgt 65.90^\circ$; $\sgbgt 41.15^\circ$ \\[1.0mm]
\includegraphics[width=.20\textwidth]{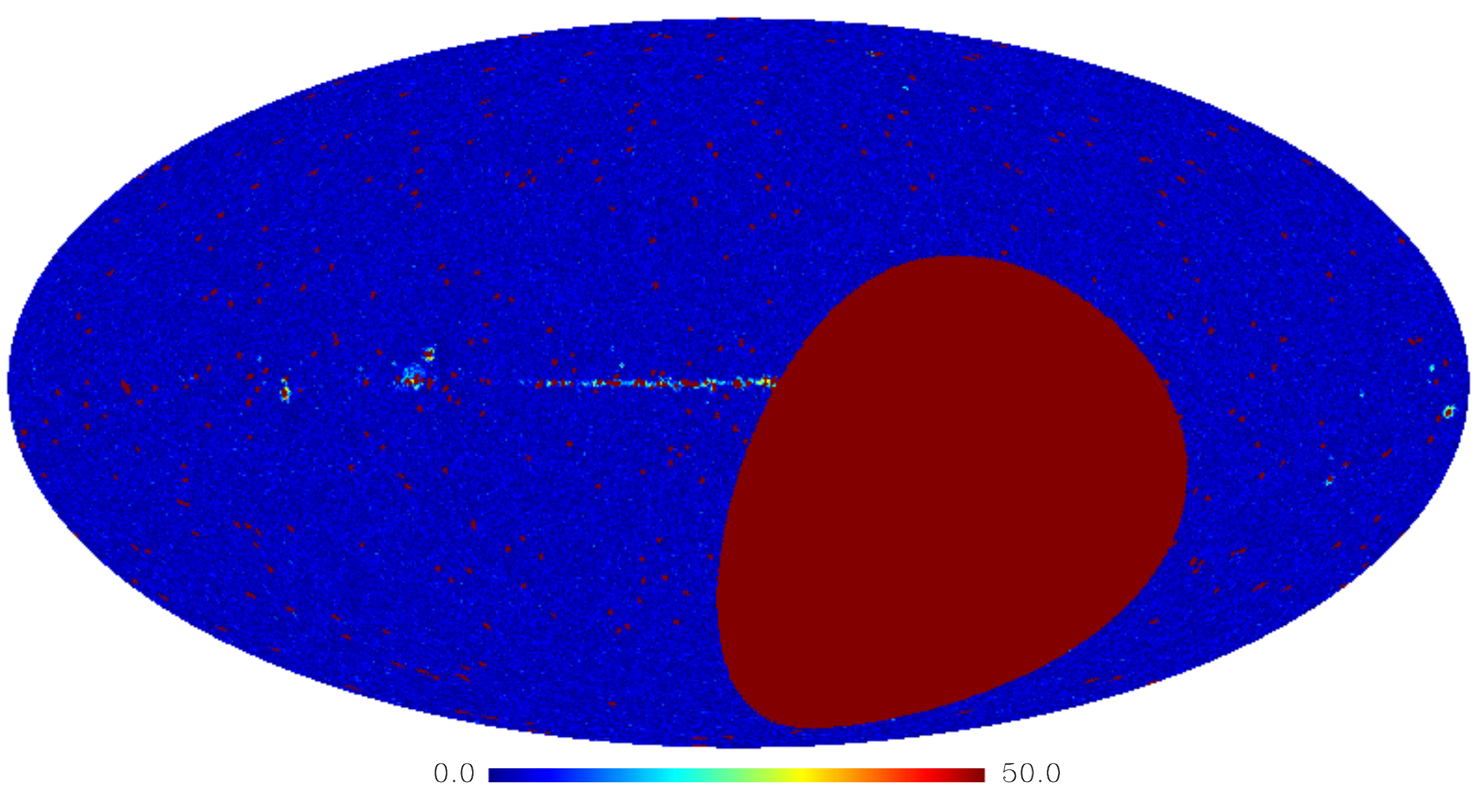}
\includegraphics[width=.20\textwidth]{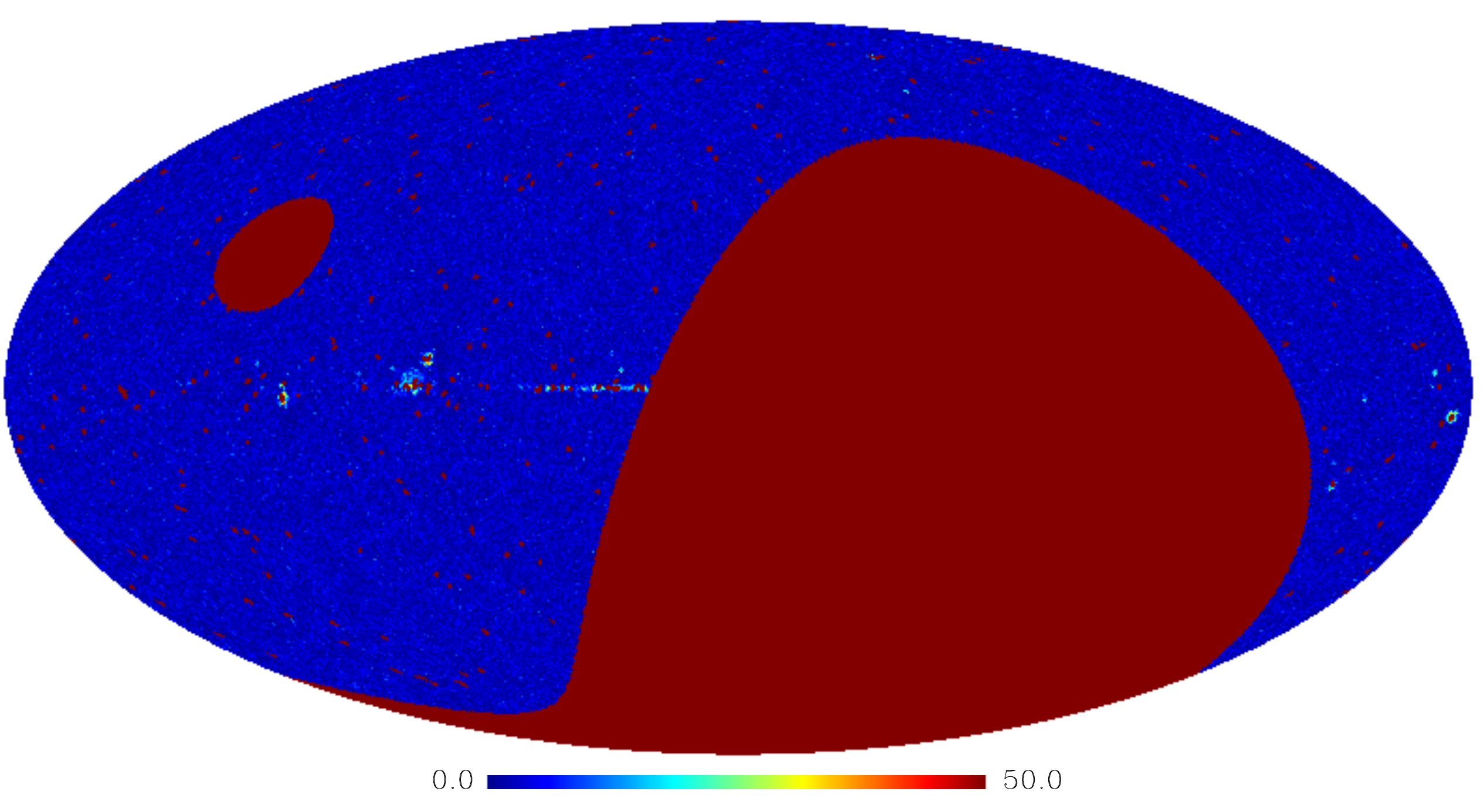}
\\[-1.0mm] $\declt -37^\circ$; $\declt -10^\circ$ or $\decgt 78^\circ$ \\[1.0mm]
\includegraphics[width=.20\textwidth]{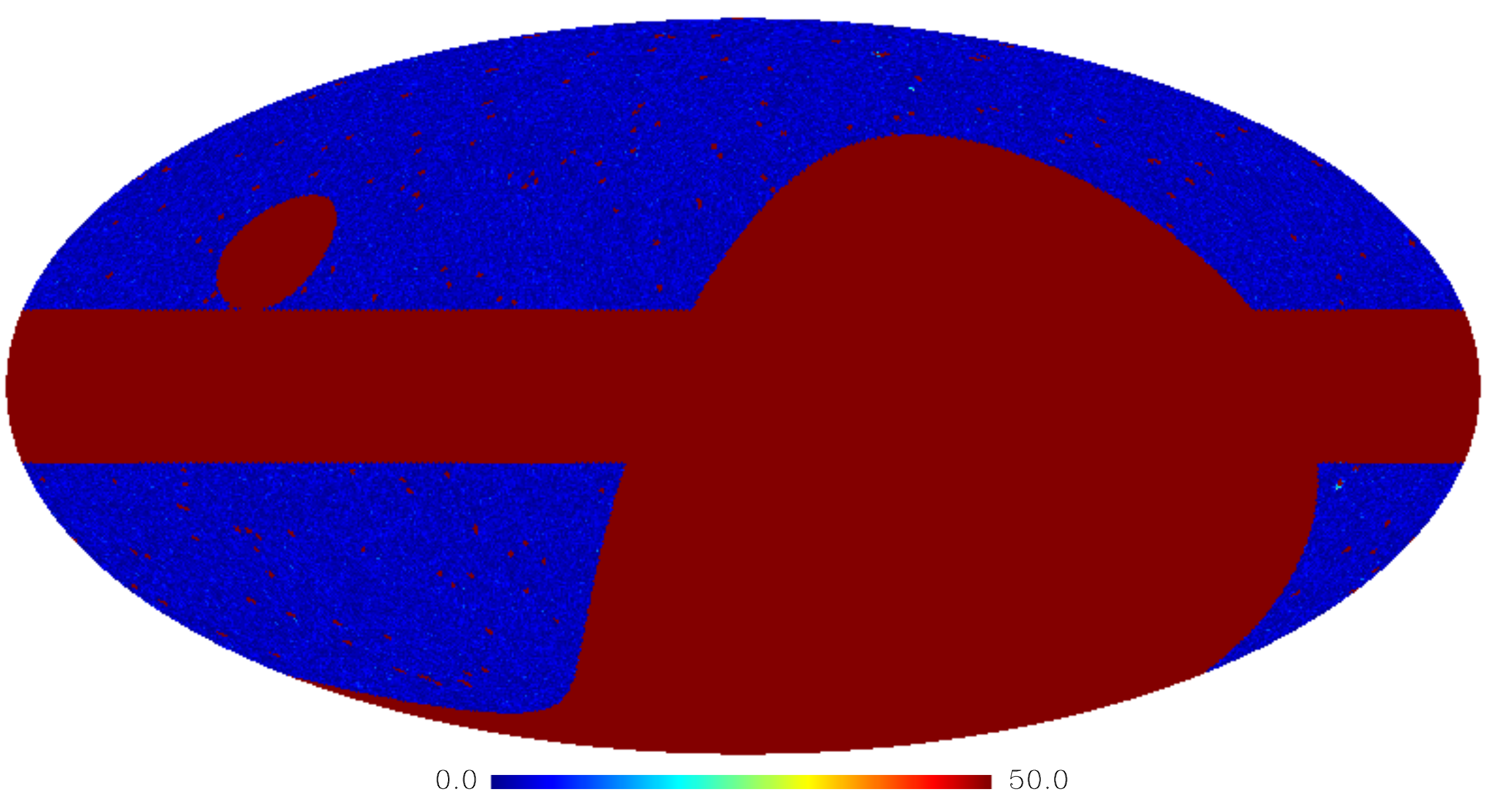}
\includegraphics[width=.20\textwidth]{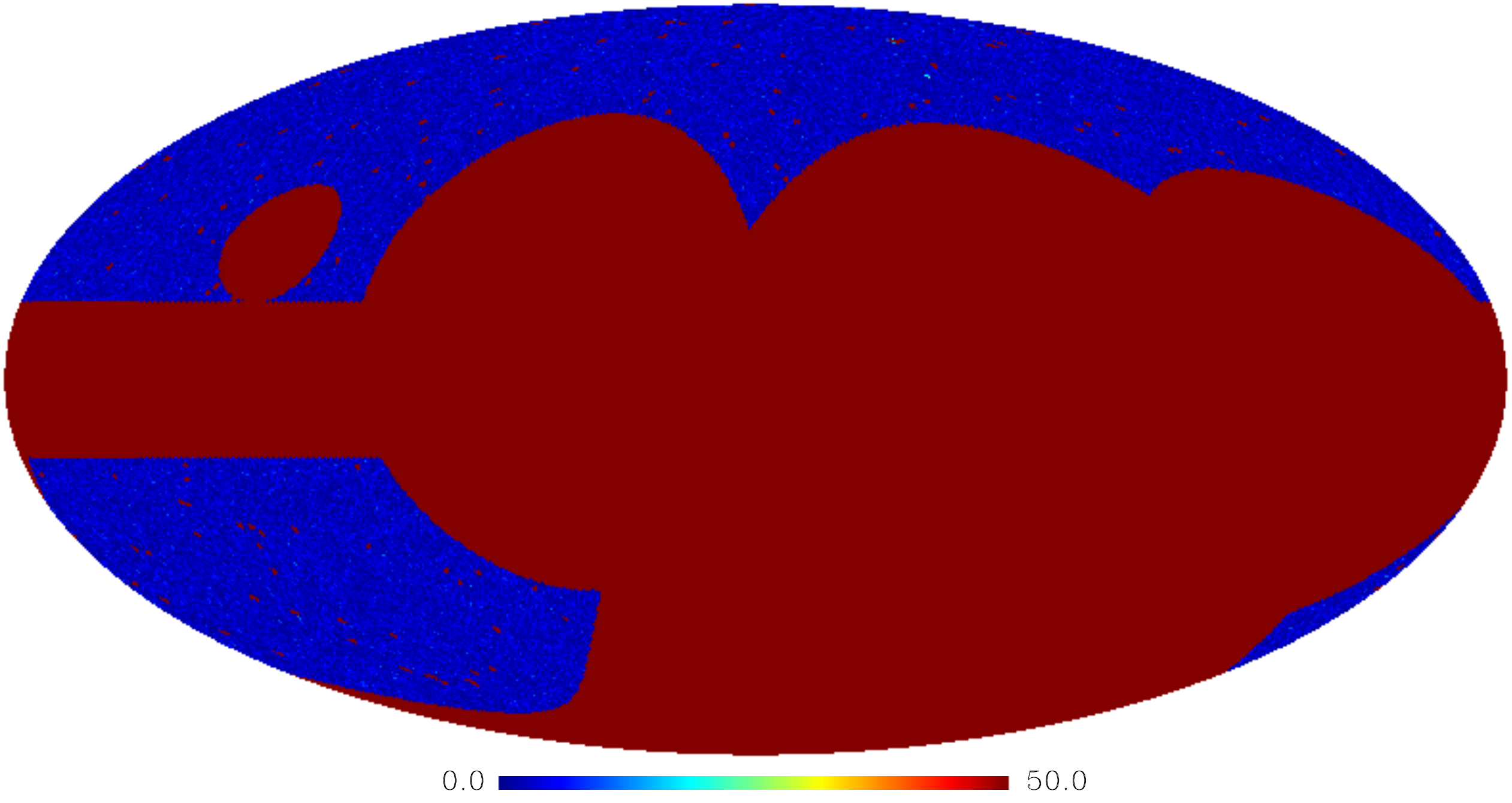}
\\[-1.0mm] same as immediately above but with $\bltfifteen$ and $\sgbgt 41.15^\circ$ also cut \\[0.02mm]
\caption{Various cuts employed in the analysis of the NVSS map. The background
  map is the NVSS map with all the sources that have flux greater than 15
  mJy.}
\label{fig:cuts}
\end{center}
\end{figure}

We regard the result with a flux cut at 15 mJy and $\bltfifteen$ as
paradigmatic, since it provides the best compromise between a large number of
sources (getting more sources requires getting into flux ranges where the
results are less trustworthy) and having stable results for the dipole.
Using this result as the fiducial, we perform our usual test of making cuts in
supergalactic latitude, and find that the presence of the supergalactic plane
is not visible in the results: the results are extremely stable as a function
of cut in SGB, with the ratio $\fsky/\fsources$ changing by less than 1
percent in all but one of the SGB cuts we study. We also compute our
likelihood distribution $P(\Coneth|\Coneobs)$ using the combination of a 15
mJy flux cut and $\bltfifteen$ cut. We calculate the direction of the dipole
for this same case, compare with the CMB kinematic dipole, and show the
results in Fig.~\ref{fig:nvssdirection}. The final plot of this section is
Fig.~\ref{fig:cuts}, which depicts the different cuts used in various portions
of the analysis for NVSS.

(2) The second strategy for attempting to control the various systematics in
NVSS is to include systematics templates that consist of degree-wide stripes
in declination: that is, rather than making highly selective cuts in
declination as before, we instead keep the entire NVSS map above $-37^\circ$
declination and then include systematics templates that have value 1 within
the declination range $(\theta, \theta+1], \theta = -37^\circ, -36^\circ, ...,
    88^\circ, 89^\circ$, and are zero elsewhere; this is a total of 127
      additional templates. This allows us to keep a much greater percentage
    of the sky in the analysis, and correspondingly more sources. This also
    allows us to marginalize over the quadrupole and octopole modes by
    incorporating them as systematics templates too (we avoided
    doing this with the more aggressive cut in declination since using so many
    templates on such a small fraction of the sky inflates the dipole
    amplitudes and error bars). The results of this test are also shown in
    Fig.~\ref{fig:plotoutputtablenvss}.

\begin{table*}[t]
\footnotesize
\caption{Summary of most reliable single results from each survey. From left
  to right in the table appear the name of the survey, the redshift range
  probed by the survey, the fraction of the sky covered, the number of sources
  available in the most reliable subset of the dataset, the observed dipole
  amplitude with error bar, the theoretical dipole amplitude (with
  cosmic-variance error bar if applicable), the direction of the best-fit
  observed dipole in Galactic coordinates $(l,b)$, and the most important
  systematic effect (in some cases, out of several candidates) that must be
  taken into account in attempting to detect a dipole in the dataset.}
  \vspace{3mm}
\label{tab6.1}
\centering
\setlength{\tabcolsep}{1em} 
\begin{normalsize} 
\begin{tabular}{| c | c | c | c | c | c | c | c |}
\hline
\rule[-3mm]{0mm}{8mm} Survey & Redshift & $\fsky$ & $N$ & $A_{\rm obs}$ & $A_{\rm th}$ & $(l,b)_{\rm obs}$ \\
\hline

\rule[-3mm]{0mm}{8mm} 2MRS & $0 < z < 0.1$ & 0.86 & 41,834 & $0.120 \pm 0.009$ & $0.311 \pm 0.122 $ & $(213.8^\circ, 35.2^\circ)$ \\ \hline
\rule[-3mm]{0mm}{8mm} 2MASS & $0 < z \lesssim 0.2$ & 0.65 & 386,008 & $0.104 \pm 0.004$ & $0.084 \pm 0.033 $ & $(268.4^\circ, 0.0^\circ)$ \\ \hline
\rule[-3mm]{0mm}{8mm} BATSE & $\bar z \gtrsim 2$ & 1.00 & 2702 & $<0.051$ (68\% CL) & [unc. prediction] & [weak constraints] \\ \hline
\rule[-3mm]{0mm}{8mm} NVSS & $\bar z \sim 1$ & 0.42 & 211,487 & $0.027 \pm 0.005$ & $0.0046 \pm 0.0035$ & $(214.5^\circ, 15.6^\circ)$ \\ \hline

\end{tabular}
\end{normalsize}
\end{table*}


In this case, the single most reliable value for the dipole amplitude is
$A=0.0280$, the value obtained from analyzing the sample with flux greater
than 2.5 mJy. While the dipole amplitude changes as a function of flux cut,
the direction is stable when we keep sources above $-37^\circ$ declination and
use templates to correct for the declination-dependent striping, and we no
longer have any clear reason to doubt the results drawn from the most
permissive flux cuts. The major caveats in this case are that (a) the use of
``stripe" templates may effectively eliminate actual signal as well as
spurious power, and (b) the direction we obtain for the dipole in case (2)
does not agree with that from case (1).

It is, however, at least somewhat encouraging that the ``best-case" observed
amplitude of the NVSS dipole is similar in both cases: $A=0.0272$ in case (1)
and $A=0.0280$ in case (2). The agreement between these values may or may not
be coincidental, but either way, the value is \textit{not} consistent with the
theoretical prediction of $A=0.0046$, at well over 99 percent confidence. The
reasonable implication is that the dipole we are measuring in the NVSS map is
partially spurious. There are several possible sources of this spurious
signal: the two most likely are that the declination-dependent striping may
not be wholly taken care of in the maps we test, even with our aggressive flux
cuts in case (1) and our templates in case (2) (observational error); or that
local sources contribute more strongly to the local-structure dipole than our
theoretical modeling allowed for, thus lifting the local-structure dipole's
contribution to the order of $10^{-2}$ rather than $10^{-3}$ (theoretical
error).

We strongly suspect that observational error plays the much greater
role. Theoretical error seems like a reasonable explanation in case (1), given
that the theoretical prediction for the dipole amplitude was computed using a
flux cut of 2.5 mJy, while the observational results were computed in case (1)
using a flux cut of 15.0 mJy. But it does not explain the discrepancy in case
(2), and in any case the theoretical predictions are ideally fairly robust to
flux cuts. (That said, for the most restrictive flux cuts, there is an
order-of-magnitude difference in flux between theory and observation, and this
is likely to motivate adjusting the theoretical
predictions at least somewhat, adding back in the local-structure dipole that
we expected to be subdominant.)

As a final note, we also include in Fig.~\ref{fig:plotoutputtablenvss} the
results that \citet{blake2002detection} found when doing a related dipole
analysis on NVSS. There are a handful of reasons why the lack of agreement
between our results and theirs is not unexpected. First, they remove local
structure in an effort to search for the kinematic dipole only; since we do
not do this, our results need not recover theirs, and in fact are expected to
give a higher signal. Also, our cut in declination is different than theirs,
leading to further potential discrepancies. The apparent agreement between
theoretical predictions and the Blake and Wall results is also partially
misleading, in that the theoretical prediction includes contributions from
both the kinematic dipole and the local-structure dipole at redshifts
  not excised by Blake and Wall ($z > 0.03$), but they
were attempting to measure only the kinematic dipole.

All this said, it is clear that radio surveys of this sort are an excellent
setting for the tests we perform, and we look forward to maps from, e.g.,
LOFAR and SKA to perform similar tests.



\section{Conclusion}
\label{secconclusion}

In this work, we have focused on what might be called the most straightforward
tests of statistical isotropy in large-scale structure -- looking for dipole
signals in existing surveys over a wide range of wavelengths. It turns out
that, despite the relative straightforwardness of the tests themselves
(Sec.~\ref{sec3.1}), the results must be carefully interpreted, as dipole
signals take contributions from several different sources. Some of these
sources, such as local-structure and kinematic dipoles, are theoretically
quite well-understood, while others, such as the intrinsic dipole, may be less
so; see Sec.~\ref{sec2.2}.


Observational results in infrared (2MRS/2MASS;
Sec.~\ref{sec2mrs} and \ref{sec4.3}), gamma rays (CGRO/BATSE;
Sec.~\ref{sec5.2}), and radio (NVSS; Sec.~\ref{sec5.3}) turn up no seriously
unexpected results, in either dipole amplitude or dipole direction. As long as
we are careful to take all sources of dipole signal into account in our
theoretical modeling, the observational results are in line with theoretical
predictions. 

Rigorous tests of this sort, while they ultimately turn up no
unexpected results, are valuable tests of current cosmological models, as they
add new wavebands in which rigorous tests of statistical isotropy have been
conducted, and ensure that statistical isotropy is probed at different epochs,
using different surveys with different systematics. Combined with similar
tests using maps of the cosmic microwave background, these measurements impose
interesting constraints on models of, and physical processes during,
cosmological inflation. We provide a summary of the most basic results, which
are elaborated upon heavily in the body of the paper, in Table \ref{tab6.1}.

One feature of these results is particularly worth highlighting: namely, that
they place constraints on the amplitude of any intrinsic dipole present in
large-scale structure. This is especially true of the BATSE and NVSS results,
since they are not expected to be dominated by the local-structure
dipole. BATSE places an upper bound on the intrinsic dipole amplitude at $1
\times 10^{-1}$ at 95 percent confidence, while our most optimistic NVSS
results places an upper bound at $4 \times 10^{-2}$ at 95 percent
confidence. As discussed in Sec.~\ref{sec5.3.3}, however, we cannot place as much
confidence in the NVSS result as we do in results from the other three surveys
analyzed in this paper. Detection of a dipole signal in any survey requires
that a great deal of attention be devoted to controlling for systematic errors
and spurious power in the survey, and while we have considerable confidence
that we have done this successfully for 2MRS, 2MASS, and BATSE, we outline in
Sec.~\ref{sec5.3.3} why acquiring and interpreting NVSS results presents the
greatest challenge of all.

Other surveys may provide interesting candidates for these same kinds of tests
in the future. Sloan luminous red galaxies (LRGs) are a sufficiently clean
dataset that these tests may be applicable and workable there (see, e.g.,
\citet{abate2012detected} for a related test). Also, survey results that have
yet to be released may be useful. The Wide-field Infrared Survey Explorer
(WISE), which produced a preliminary data release in April 2011 and has
full-survey results forthcoming now, covers more than 99 percent of the sky
(\citet{wright2010wide}). Tests performed on WISE would be similar to tests
performed on 2MASS, but would update 2MASS results with a more recent and
\textit{much} deeper survey. As the X-ray background becomes
better-understood, this may also serve as an increasingly valuable test of
statistical isotropy and setting in which to attempt to detect dipole
signals. The Dark Energy Survey (DES) will be useful in probing the
distribution of galaxies to high redshift, and will have sufficient sky
coverage to make the tests presented here useful. In microwaves, dipole
signals might be detectable in maps of the gravitational lensing of the CMB,
which provide a very good tracer of mass. Finally, new radio surveys such as
LOFAR and SKA will probe orders of magnitude more sources down to far lower
flux thresholds than NVSS (see, e.g., \citet{crawford2009detecting}), and
would provide very valuable updates to NVSS results on the dipole amplitude
and direction. The kinematic dipole, both direction and amplitude, should be
unambiguously detected in these surveys.

\acknowledgments

For help with the set of research projects that have gone into this paper,
thanks to Jim Condon, Sudeep Das, Jerry Fishman, Chris Hirata, Michael Hudson,
Tom Jarrett, Lucas Macri, Dominik Schwarz, and Weikang Zheng, all of whom were
very helpful at different points along the way. For useful comments in the
aftermath of initial arXiv posting, we thank Maciej Bilicki and Jim Zibin.  CG
would additionally like to thank the members of his dissertation committee,
all of whom provided valuable input: Philip Hughes, Jeff McMahon, Aaron
Pierce, and especially Tim McKay. DH would additionally like to thank Glenn
Starkman for sparking his interest in testing statistical isotropy with
large-scale structure.

This publication makes use of data products from the Two Micron All Sky
Survey, which is a joint project of the University of Massachusetts and the
Infrared Processing and Analysis Center/California Institute of Technology,
funded by the National Aeronautics and Space Administration and the National
Science Foundation.
We also acknowledge the use of the publicly available HEALPix
\cite{gorski2005healpix} package.

We have been supported by DOE grant under contract DE-FG02-95ER40899, NSF
under contract AST-0807564, and NASA under contract NNX09AC89G. DH
  thanks the Aspen Center for Physics, which is supported by NSF Grant
  No.\ 1066293, for hospitality.

\appendix

\section{Relativistic Aberration and Doppler Effect}
\label{appA}

We first address aberration, following the formalism of
\citet{burles2006detecting} (see also, e.g., \citet{calvao2005relativistic}),
who derive equations for aberration with the ultimate goal of showing that
aberration of the CMB temperature might be detectable statistically by Planck,
looking at shifts of CMB peaks. While this is not our goal, the formalism
still holds.

We define a spherical-coordinate system with the $z$-axis in the direction of
motion. If we take the ``unprimed" frame to be the CMB frame, and the
``primed" frame to be the frame of the Solar System barycenter, then the
azimuthal angle $\phi$ is unchanged between frames: $\phi =
\phi^\prime$. However, the polar angle $\theta$ is affected as follows:
\begin{equation}
\sin\theta = \frac{\sin\theta^\prime}{\gamma(1-\beta \cos\theta^\prime)}
\end{equation}
where $\beta$ is the relative velocity of the Solar System with respect to the
CMB, $\gamma = (1-\beta^2)^{-1/2}$ as usual, and $\theta=0$ corresponds to the
direction of forward motion. With the assumption that $\beta$ is small, which
is a good assumption given that CMB observations show it to be on the order of
$10^{-3}$, expansion in a Taylor series gives
\begin{equation}
\sin\theta = \sin\theta^\prime (1 + \beta \cos \theta^\prime).
\end{equation}
Finally, we take the arcsin of both sides and expand the arcsin function
assuming small $\beta$ to obtain
\begin{equation}
\theta  =  \theta^\prime + \beta \sin \theta^\prime.
\end{equation}
We are ultimately interested in calculating how areas (and volumes) on the
celestial sphere are stretched or compressed, and hence want the quantity $d
\Omega / d \Omega^\prime$. With that in mind, we compute
\begin{eqnarray}
\frac{d\theta}{d\theta^\prime} & = & 1+\beta \cos \theta^\prime \\
\frac{\sin \theta d \phi}{\sin \theta^\prime d \phi^\prime} & = & 1 + \beta \cos \theta^\prime 
\end{eqnarray}
and find that $d \Omega / d \Omega^\prime = (1 + \beta \cos
\theta^\prime)^2$. Hence areas and volumes on the sky, proportional to
$\sin(\theta)d\theta d\phi$, change as $(1 + \beta \cos \theta^\prime)^2
\approx 1 + 2 \beta \cos \theta^\prime$.

\citet{itoh2010dipole} provide a more complete derivation of this, including
both the Doppler effect (which changes frequencies and hence measured
magnitudes since we never measure bolometric magnitudes) and relativistic
aberration, and derive the following expression for the observed angular
number density of galaxies $n(\theta)$ given the limiting magnitude $m_{\rm
  lim}$:
\begin{equation}
n(\theta, m < m_{\rm lim}) = \bar n(m<m_{\rm lim}) \left \lbrack 1 + 2 \tilde \beta \cos \alpha \right \rbrack
\label{eq:n}
\end{equation}
where
\begin{equation}
\tilde \beta = [1 + 1.25x(1-p)] \beta.
\label{eq:betatilde}
\end{equation}
Here the intrinsic flux density of a galaxy is assumed to be a power law
$S_{\rm rest}(\nu) \propto \nu^p$, and the intrinsic number counts of galaxies
$\bar n$ is
\begin{equation}
\bar n(m < m_{\rm lim}) \propto 10^{x m_{\rm lim}}
\end{equation}
where $x$ is a numerical coefficient of order unity. The angle $\alpha$ is the
angle between the angular direction \textbf{$\theta$} and the angular
direction of the Earth's peculiar velocity \textbf{v} on the celestial sphere,
the same as $\theta^\prime$, but with more convenient notation. The factor of
2 in $2 \tilde \beta$ above comes from the same source as the square in $(1 +
\beta \cos \theta ')^2$ earlier. The correction for $\tilde \beta$ in
Eq.~(\ref{eq:betatilde}) is the contribution of the Doppler effect to the
overall kinematic dipole in observations of LSS.

The final formula for the combined effects of relativistic aberration and the
Doppler effect is $d\Omega/d\Omega'= 1 + 2 \beta \cos \theta^\prime$, so
that the predicted amplitude of the kinematic dipole is
\begin{equation}
A = 2 \tilde \beta = 2 [1 + 1.25x(1-p)] \beta .
\label{eq:Atildebeta}
\end{equation}

\section{Quadrupole and Octopole Templates}
\label{appB}

In Eq.~(\ref{eqn:masterequation}) and below, we show how the dipole formalism
from \citet{Hirata_hemispherical} can be used to separate out genuine dipole
signal in a map of large-scale structure from spurious signal due to
systematic effects or coupling with the monopole. In this Appendix, we detail
how we can guard against the possibility that some of the ``genuine" dipole
signal actually comes from higher multipoles.

The most straightforward way of doing this is simply to include the five
$\ell=2$ and seven $\ell=3$ modes as systematics templates in the analysis,
corresponding to the $\sum_i k_i t_i({\bf \hat n})$ term in
Eq.~(\ref{eqn:masterequation}). More specifically, the templates
  $t_i({\bf \hat n})$ are assigned to be $Y_{2m}(\nhat)$ for $-2\leq m\leq 2$ and
  $Y_{3m}(\nhat)$ for $-3\leq m \leq 3$. We expect that quadrupole and octopole modes
should not contribute to the dipole signal at all in the limit of full-sky
coverage. However, as the sky is cut, multipoles become coupled, and in the
limit of very small sky coverage, there is high degeneracy between the
quadrupole and octopole modes and the dipole mode, which is the only one of
interest.

\begin{figure}[]
\begin{center}
\includegraphics[width=.48\textwidth]{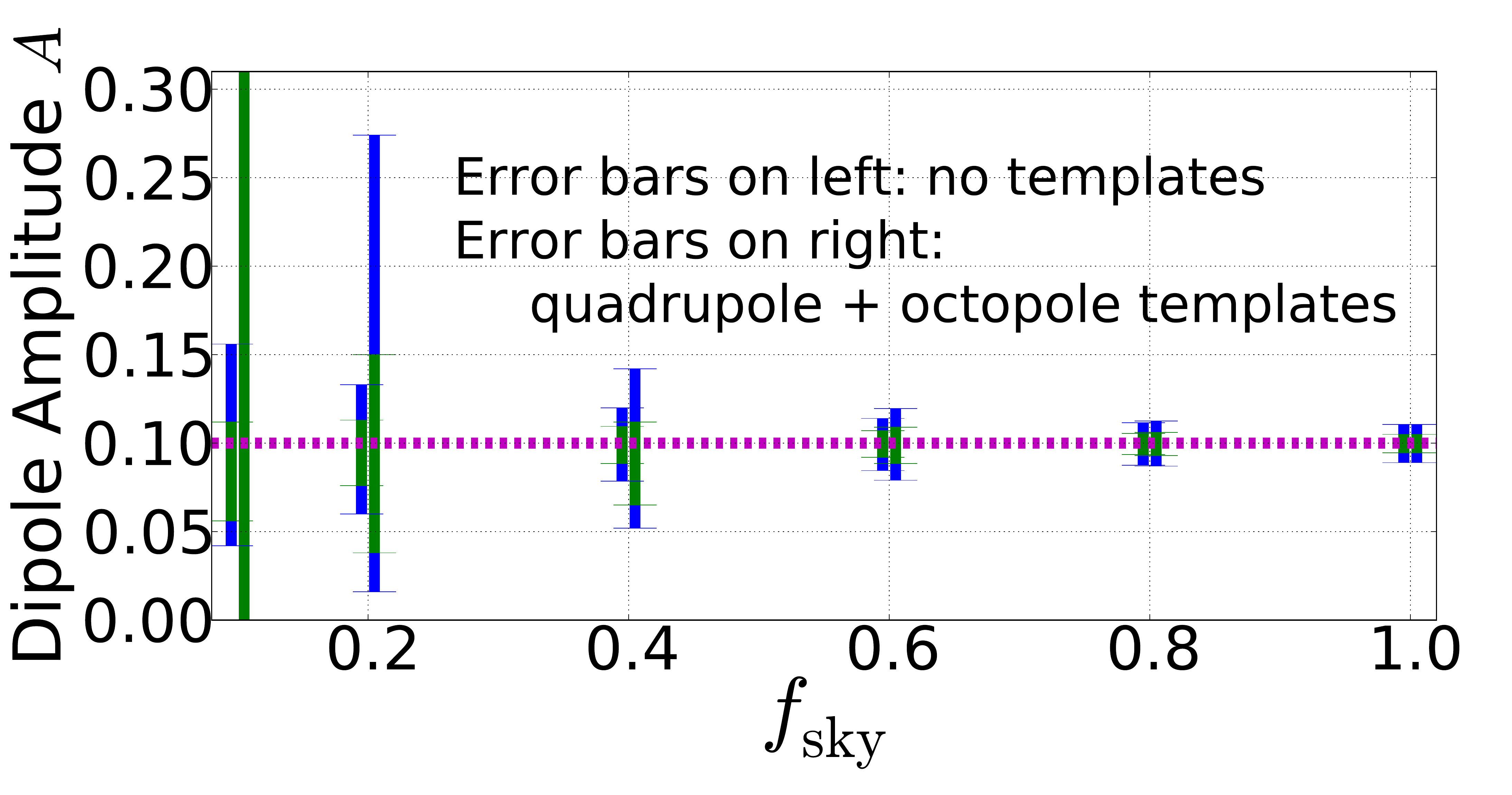}
\caption{Sensitivity of dipole amplitude measurements on marginalization over
  the quadrupole and octopole components, as a function of the fraction of the
  sky covered. We show the dependence of the detected dipole amplitude on
  $\fsky$ for the ideal case where the map being analyzed consists of a pure
  dipole. The dipole has amplitude $A=0.10$, indicated by the dotted magenta
  line. We find that while the correct value of the amplitude is, in all
  cases, recovered within the appropriate error bars, the inclusion of
  quadrupole and octopole templates does affect the size of the error
  bars. When $\fsky$ is more than about 0.7, the quadrupole and octopole
  templates make little difference, but below that value of $\fsky$, error
  bars start increasing noticeably as more of the sky is cut away.}
\label{fig:dipquadoct}
\end{center}
\end{figure}

To quantify this precisely, we run a simple test, the results of which are
shown in Fig.~\ref{fig:dipquadoct}. We take an artificially-generated map that
contains nothing but a pure dipole in a certain direction (in this case,
$l=61.4^\circ, b=33.4^\circ$). The dipole amplitude is $A=0.1$, so we expect that in the
full-sky limit, our dipole estimator should recover the result of $A=0.1$, and
the error bars should be the same regardless of whether we include quadrupole
and octopole templates -- there is, after all, no coupling between the dipole
and $\ell=2$/$\ell=3$ modes. However, as we make more and more aggressive cuts
(accomplished here by making cuts that are symmetric in Galactic latitude, as
we frequently do throughout our analyses of real datasets), we find that
including the quadrupole and octopole templates becomes more and more
important. While all results are consistent with the correct amplitude of
$A=0.1$, the error bars are much larger when $\ell=2$ and $\ell=3$ templates
are included when roughly half the sky or more is cut.

Given that the results depend only very weakly on inclusion of quadrupole and
octopole templates for small sky cuts, we do not always incorporate quadrupole
and octopole templates into our analysis in the limit of nearly full-sky
coverage. Meanwhile, especially in the case of NVSS where our analysis deals
almost exclusively with less than half the sky, we sometimes explicitly
compare the case where quadrupole and octopole templates are not included to
the case where they are, since inclusion of the templates substantially
weakens our results.

\bibliography{dipolessources}

\end{document}